\documentclass[amsmath,amssymb]{revtex4}
\usepackage{graphicx}% Include figure files
\usepackage{amscd,amsmath,amsthm,amsfonts,amssymb}
\def\[{\begin{equation}}
\def\]{\end{equation}}

\begin{document}
\title{Rogue wave patterns associated with Okamoto polynomial hierarchies}
\author{%%%% Author details
Bo Yang$^{1}$, Jianke Yang$^{2}$}
%%%%%%%%% Insert author address here
\address{$^{1}$ School of Mathematics and Statistics, Ningbo University, Ningbo 315211, China\\
$^{2}$ Department of Mathematics and Statistics, University of Vermont, Burlington, VT 05401, U.S.A}
\begin{abstract}
We show that new types of rogue wave patterns exist in integrable systems, and these rogue patterns are described by root structures of Okamoto polynomial hierarchies. These rogue patterns arise when the $\tau$ functions of rogue wave solutions are determinants of Schur polynomials with index jumps of three, and an internal free parameter in these rogue waves gets large. We demonstrate these new rogue patterns in the Manakov system and the three-wave resonant interaction system. For each system, we derive asymptotic predictions of its rogue patterns under a large internal parameter through Okamoto polynomial hierarchies. Unlike the previously reported rogue patterns associated with the Yablonskii-Vorob'ev hierarchy, a new feature in the present rogue patterns is that, the mapping from the root structure of Okamoto-hierarchy polynomials to the shape of the rogue pattern is linear only to the leading order, but becomes nonlinear to the next order. As a consequence, the current rogue patterns are often deformed, sometimes strongly deformed, from Okamoto root structures, unless the underlying free parameter is very large. Our analytical predictions of rogue patterns are compared to true solutions, and excellent agreement is observed, even when rogue patterns are strongly deformed from Okamoto root structures.
\end{abstract}
\maketitle

\section{Introduction}
Rogue waves are large and spontaneous nonlinear wave excitations that ``come from nowhere and disappear with no trace" \cite{Akhmediev2009}. Such waves were first studied in oceanography, since they posed a threat even to large ships \cite{Ocean_rogue_review,Pelinovsky_book}. Later, these waves were also investigated in optics and other physical fields due to their peculiar nature \cite{Solli_Nature,Wabnitz_book}. From a theoretical point of view, an important fact is that, many integrable nonlinear wave equations admit explicit rational solutions that correspond to rogue waves. This fact was first reported by Peregrine \cite{Peregrine}, who presented a simple rogue wave solution for the nonlinear Schr\"odinger (NLS) equation. Peregrine's solution was later generalized, and more intricate rogue wave solutions in the NLS equation were derived \cite{AAS2009,DGKM2010,KAAN2011,GLML2012,OhtaJY2012}. Since the NLS equation governs nonlinear wave packet evolution in a wide range of physical systems \cite{Benney,Ablowitz_book}, these theoretical rogue wave solutions of the NLS equation then motivated a lot of rogue-wave experiments, ranging from water waves to optical waves to acoustic waves \cite{Tank1,Tank2,He_Chabchoub,Plasma,Fiber1,Helium,PlasmaLin}. These combined theoretical and experimental studies significantly deepened our understanding of physical rogue wave events. Due to this success, rogue wave solutions have also been derived in many other physical integrable equations, such as the derivative NLS equations for circularly polarized nonlinear Alfv\'en waves in plasmas and short-pulse propagation in a frequency-doubling crystal
\cite{Kaup_Newell,KN_Alfven1,Wise2007,KN_rogue_2011,KN_rogue_2013,YangDNLS2019}, the Manakov equations for light transmission in randomly birefringent fibers and interaction between two incoherent light beams in crystals
\cite{Menyuk,BDCW2012,Stegeman_Manakov,Segev_coupled,ManakovDark,LingGuoZhaoCNLS2014,Chen_Shihua2015,ZhaoGuoLingCNLS2016}, the three-wave resonant interaction equations \cite{Ablowitz_book,BaroDegas2013,DegasLomba2013,ChenSCrespo2015,WangXChenY2015,ZhangYanWen2018}, and many others. Such theoretical rogue wave solutions have further stimulated experiments on optical rogue waves in randomly birefringent fibers \cite{Fiber2,Fiber3}. In addition to these rogue waves that arise from a uniform background, rogue waves that arise from a non-uniform background have also been predicted or observed in several wave systems \cite{Qin2015,Peli2019,Peli2020,He2021,YangYang3D}.

Pattern formation of rogue waves is an important question, because this information allows for the prediction of later rogue wave events from earlier wave forms. Rogue wave patterns were first studied for the NLS equation in \cite{AAS2009,DGKM2010,KAAN2011,GLML2012,OhtaJY2012,HeFokas,KAA2013,Akhmediev_triangular,AAS2009,AkhClark2010,He2017,Miller2020}, where the maximum-amplitude rogue waves and their near-field profiles were determined. In addition, clear geometric shapes of rogue waves on the spatial-temporal plane, such as triangles, pentagons, and heptagons, were numerically reported under certain parameters. Regarding the latter question of geometric shapes, significant progress was made recently in \cite{NLSRWs2021}, where an intimate connection between geometric shapes of NLS rogue waves and root structures of the Yablonskii-Vorob'ev polynomial hierarchy was revealed. Since roots of the Yablonskii-Vorob'ev polynomial hierarchy on the complex plane come in shapes such as triangles, pentagons and heptagons, previous numerical reports on this subject can then be analytically explained. Further significant progress in this direction was made in \cite{Yanguniversal}, where NLS rogue patterns associated with the Yablonskii-Vorob'ev hierarchy were shown to be universal in integrable systems, as long as rogue wave solutions of the integrable systems can be expressed by $\tau$ functions whose matrix elements are Schur polynomials with index jumps of two, as in the generalized derivative NLS equations, the Boussinesq equation, the Manakov system, and many others \cite{Yanguniversal,Junchao_Yajima,FengCSP,FengSasa}.

A natural next question is, are there other shapes of rogue wave patterns in integrable systems? If so, what special polynomials would be associated with such rogue patterns?

In this article, we show that there are indeed other shapes of rogue patterns in integrable systems. These new rogue patterns would arise if rogue solutions can be expressed by $\tau$ functions whose matrix elements are Schur polynomials with index jumps of three, which occurs in integrable systems such as the Manakov equations, the three-wave resonant interaction equations, and many others. Special polynomials associated with these new rogue patterns are the Okamoto polynomial hierarchies. We demonstrate these new rogue patterns in the Manakov system and the three-wave resonant interaction system. For each system, we derive asymptotic predictions of its rogue shapes through Okamoto polynomial hierarchies when one of the internal free parameters in the rogue waves gets large.
Since roots of Okamoto-hierarchy polynomials exhibit shapes such as double triangles, rhombuses, and squares, we then get new rogue patterns in similar shapes. However, unlike the previously reported rogue patterns associated with the Yablonskii-Vorob'ev hierarchy, a new feature in the present case is that, the mapping from the root structure of Okamoto hierarchies to the shape of the rogue pattern is linear only to the leading order, but becomes nonlinear to the next order. This nonlinear nature of the mapping then generates rogue shapes that are often deformed, sometimes strongly deformed, from Okamoto root structures, unless the underlying free parameter is very large so that the next-order nonlinear correction of the mapping becomes insignificant. Our analytical predictions of rogue patterns are compared to true solutions, and excellent agreement is observed, even when rogue shapes are strongly deformed from Okamoto root structures.

This paper is structured as follows. In Sec.~2, we describe some preliminary facts and results. We first introduce Okamoto polynomial hierarchies and study its root structures. Then, we provide bilinear rogue wave expressions in the Manakov and three-wave-interaction systems. In Sec.~3, we present our main results on rogue patterns in these two systems under a large parameter. In Sec.~4, we compare these analytical rogue-pattern predictions to true solutions in the two systems. In Sec.~5, we prove the analytical results stated in Sec.~3. Sec.~6 summarizes the paper. In the two appendices, we prove root-structure results of Okamoto-hierarchy polynomials, and derive bilinear rogue wave expression in the Manakov system, both of which are  stated in Sec.~2.

\section{Preliminaries} \label{sec:pre}

\subsection{Okamoto polynomials and their hierarchies}
Okamoto polynomials first arose in Okamoto's study of rational solutions to the Painlev\'e IV equation \cite{Okamoto1986}. He showed that a class of such rational solutions can be expressed as the logarithmic derivative of certain special polynomials, which are now called Okamoto polynomials. Later, determinant expressions of these polynomials were discovered by Kajiwara and Ohta \cite{KajiOhta1998PIV}. Let $p_{j}(z)$ be Schur polynomials defined by
\[ \label{defpkz}
\sum_{j=0}^{\infty}  p_{j}(z) \epsilon^j= \exp \left(z \epsilon + \epsilon^2 \right),
\]
with $p_{j}(z)=0$ for $j<0$. Then, the monic Okamoto polynomials $Q_{N}(z)$ and $R_{N}(z)$ with $N\geq1$ are defined as \cite{KajiOhta1998PIV,Clarkson2003PIV}
\[\label{OkamotoPoly1}
Q_{N}(z) = c_{N}
      \left| \begin{array}{cccc}
         p_{2}(z) & p_{1}(z) & \cdots &  p_{3-N}(z) \\
         p_{5}(z) & p_{4}(z) & \cdots &  p_{4-N}(z) \\
        \vdots& \vdots & \vdots & \vdots \\
         p_{3N-1}(z) & p_{3N-2}(z) & \cdots &  p_{2N}(z)
       \end{array}
 \right|,
\]
and
\[\label{OkamotoPoly2}
R_{N}(z) = d_{N}
\left| \begin{array}{cccc}
         p_{1}(z) & p_{0}(z) & \cdots &  p_{2-N}(z) \\
         p_{4}(z) & p_{3}(z) & \cdots &  p_{5-N}(z) \\
        \vdots& \vdots & \vdots & \vdots \\
         p_{3N-2}(z) & p_{3N-3}(z) & \cdots &  p_{2N-1}(z)
       \end{array}
 \right|,
\]
where
\[
c_{N}=3^{-\frac{1}{2}N(N-1)} \hspace{0.05cm} \frac{2! 5! \cdots (3N-1)!}{0! 1! \cdots (N-1)!},
\]
and
\[
d_{N}=3^{-\frac{1}{2}N(N-1)} \hspace{0.05cm} \frac{1! 4! \cdots (3N-2)!}{0! 1! \cdots (N-1)!}.
\]
Note that these two determinants are both Wronskians, because $p_{j+1}'(z)=p_j(z)$ from the definition of $p_j(z)$ in Eq.~(\ref{defpkz}), where the prime denotes differentiation. The first three $Q_N(z)$ and $R_N(z)$ polynomials are
\begin{eqnarray*}
&& Q_1(z)=z^2+2, \\
&& Q_2(z)=z^6+10 z^4+20 z^2+40, \\
&& Q_3(z)=z^{12}+28 z^{10}+ 260 z^8+ 1120 z^6+ 2800 z^4+ 11200 z^2+11200, \\
&& R_1(z)=z, \\
&& R_2(z)=z^4+4 z^2-4, \\
&& R_3(z)=z (z^8+ 16 z^6+ 56 z^4-560).
\end{eqnarray*}
Compared to the Okamoto polynomials introduced in \cite{Okamoto1986,KajiOhta1998PIV,Clarkson2003PIV}, the polynomials above are related to them through a simple scaling in $z$ and $(Q_N, R_N)$.

Like the Yablonskii-Vorob'ev polynomials \cite{Clarkson2003-II}, these Okamoto polynomials can also be generalized to hierarchies.
Let $p_{j}^{[m]}(z)$  be Schur polynomials defined by
\[\label{pkmz}
\sum_{j=0}^{\infty}  p_{j}^{[m]}(z) \epsilon^j= \exp \left(z \epsilon + \epsilon^m \right),
\]
where $m$ is a positive integer larger than one, and $p_{j}^{[m]}(z) \equiv 0$ if $j<0$. Then, the monic Okamoto polynomial hierarchies $R_{N}^{[m]}(z)$ and $Q_{N}^{[m]}(z)$ are defined by the Wronskians
\[\label{HeiTypePoly1}
Q_{N}^{[m]}(z) = c_{N}\left| \begin{array}{cccc}
         p^{[m]}_{2}(z) & p^{[m]}_{1}(z) & \cdots &  p^{[m]}_{3-N}(z) \\
         p^{[m]}_{5}(z) & p^{[m]}_{4}(z) & \cdots &  p^{[m]}_{4-N}(z) \\
        \vdots& \vdots & \vdots & \vdots \\
         p^{[m]}_{3N-1}(z) & p^{[m]}_{3N-2}(z) & \cdots &  p^{[m]}_{2N}(z)
       \end{array}
 \right|,
\]
and
\[\label{HeiTypePoly2}
R_{N}^{[m]}(z) = d_{N}
\left| \begin{array}{cccc}
         p^{[m]}_{1}(z) & p^{[m]}_{0}(z) & \cdots &  p^{[m]}_{2-N}(z) \\
         p^{[m]}_{4}(z) & p^{[m]}_{3}(z) & \cdots &  p^{[m]}_{5-N}(z) \\
        \vdots& \vdots & \vdots & \vdots \\
         p^{[m]}_{3N-2}(z) & p^{[m]}_{3N-3}(z) & \cdots &  p^{[m]}_{2N-1}(z)
       \end{array}
 \right|.
\]
If $m \hspace{0.1cm} \mbox{mod} \hspace{0.1cm} 3 = 0$, then $Q_{N}^{[m]}(z)=z^{N(N+1)}$ and $R_{N}^{[m]}(z)=z^{N^2}$. Such $m$ values turn out to be irrelevant to our rogue wave problem. Thus, in this article we require $m \hspace{0.1cm} \mbox{mod} \hspace{0.1cm} 3 \neq 0$, i.e., $m=2, 4, 5, 7, 8, 10, \cdots$. When $m=2$, $Q_{N}^{[2]}(z)$ and $R_{N}^{[2]}(z)$ are the Okamoto polynomials $Q_N(z)$ and $R_N(z)$. When $m>2$, $Q_{N}^{[m]}(z)$ and $R_{N}^{[m]}(z)$ give higher members of Okamoto hierarchies.

\subsection{Root structures of Okamoto polynomial hierarchies}
Root structures of Okamoto-hierarchy polynomials will play a key role in our analytical study of rogue wave patterns. For Okamoto polynomials $Q_N(z)$ and $R_N(z)$, their root structures have been investigated in \cite{Kametaka,Fukutani,Clarkson2003PIV}. It has been shown that for every positive integer $N$, $Q_N(z)$ and $R_N(z)$ have simple roots \cite{Kametaka,Fukutani}. In addition, graphs of root locations for many $Q_N(z)$ and $R_N(z)$ polynomials have been plotted, and double-triangle as well as rhombus-shape root structures have been observed \cite{Clarkson2003PIV}. But for higher members of Okamoto hierarchies, their root structures have not been studied yet to our knowledge.

In this subsection, we examine root structures of Okamoto hierarchies $Q_{N}^{[m]}(z)$ and $R_{N}^{[m]}(z)$. Defining integer $N_0$ as the remainder of $N$ divided by $m$, i.e.,
\[
N_0 \equiv N \hspace{0.1cm} \mbox{mod} \hspace{0.1cm} m,
\]
and denoting $[a]$ as the largest integer less than or equal to a real number $a$, then our results are summarized by the following two theorems.
\begin{quote}
\textbf{Theorem 1} \hspace{0.06cm} The Okamoto hierarchy polynomial $Q_{N}^{[m]}(z)$ is monic with degree $N(N+1)$, and is of the form
\[ \label{QNmform}
Q_{N}^{[m]}(z)=z^{N_Q}q_{N}^{[m]}(\zeta), \quad \zeta \equiv z^{m},
\]
where $q_{N}^{[m]}(\zeta)$ is a monic polynomial of $\zeta$ with all-real coefficients and a nonzero constant term. The non-negative integer $N_Q$ is the multiplicity of the zero root in $Q_{N}^{[m]}(z)$ and is given by
\[ \label{defNQ}
N_Q=N_{1Q}(N_{1Q}-N_{2Q}+1)+N_{2Q}^2,
\]
where $N_{1Q}$ and $N_{2Q}$ are non-negative integers. If $m>1$ and $m \hspace{0.1cm} \mbox{mod} \hspace{0.1cm} 3 =1$, these $(N_{1Q}, N_{2Q})$ values are
\[ \label{N12Qa}
(N_{1Q}, N_{2Q})=\left\{ \begin{array}{ll} (N_0, 0), & \mbox{when}\hspace{0.12cm} 0 \leq N_{0} \leq  \left[\frac{m}{3}\right],  \\
\left(\left[\frac{m}{3}\right], N_0-\left[\frac{m}{3}\right]\right), & \mbox{when}\hspace{0.12cm}  \left[\frac{m}{3}\right]+1 \leq  N_{0} \leq 2\left[\frac{m}{3}\right], \\
(m-1-N_0, \hspace{0.05cm} m-1-N_0), & \mbox{when}\hspace{0.12cm}  2\left[\frac{m}{3}\right]+1 \leq  N_{0} \leq m-1; \end{array} \right.
\]
and if $m \hspace{0.1cm} \mbox{mod} \hspace{0.1cm} 3 = 2$, these $(N_{1Q}, N_{2Q})$ values are
\[
(N_{1Q}, N_{2Q})=\left\{ \begin{array}{ll} (N_0, 0), & \mbox{when}\hspace{0.12cm} 0 \leq N_{0} \leq  \left[\frac{m}{3}\right],  \\
\left(N_0-\left[\frac{m}{3}\right]-1, \left[\frac{m}{3}\right]\right), & \mbox{when}\hspace{0.12cm}  \left[\frac{m}{3}\right]+1 \leq  N_{0} \leq 2\left[\frac{m}{3}\right], \\
(m-1-N_0, \hspace{0.05cm} m-1-N_0), & \mbox{when}\hspace{0.12cm}  2\left[\frac{m}{3}\right]+1 \leq  N_{0} \leq m-1. \end{array} \right.
\]
If $N_Q=0$, then zero is not a root of $Q_{N}^{[m]}(z)$.
\end{quote}
\begin{quote}
\textbf{Theorem 2} \hspace{0.06cm} The Okamoto hierarchy polynomial $R_{N}^{[m]}(z)$ is monic with degree $N^2$, and is of the form
\[ \label{RNmform}
R_{N}^{[m]}(z)=z^{N_R}r_{N}^{[m]}(\zeta), \quad \zeta \equiv z^{m},
\]
where $r_{N}^{[m]}(\zeta)$ is a monic polynomial of $\zeta$ with all-real coefficients and a nonzero constant term. The non-negative integer $N_R$ is the multiplicity of the zero root in $R_{N}^{[m]}(z)$ and is given by
\[ \label{defNR}
N_R=N_{1R}(N_{1R}-N_{2R}+1)+N_{2R}^2,
\]
where $N_{1R}$ and $N_{2R}$ are non-negative integers. If $m>1$ and $m \hspace{0.1cm} \mbox{mod} \hspace{0.1cm} 3 =1$, these $(N_{1R}, N_{2R})$ values are
\[
(N_{1R}, N_{2R})=\left\{ \begin{array}{ll} (0, N_0), & \mbox{when}\hspace{0.12cm} 0 \leq N_{0} \leq  \left[\frac{m}{3}\right],  \\
\left(\left[\frac{m}{3}\right]-1, N_0-1-\left[\frac{m}{3}\right]\right), & \mbox{when}\hspace{0.12cm}  \left[\frac{m}{3}\right]+1 \leq  N_{0} \leq 2\left[\frac{m}{3}\right], \\
(m-1-N_0, \hspace{0.05cm} m-N_0), & \mbox{when}\hspace{0.12cm}  2\left[\frac{m}{3}\right]+1 \leq  N_{0} \leq m-1; \end{array} \right.
\]
and if $m \hspace{0.1cm} \mbox{mod} \hspace{0.1cm} 3 = 2$, these $(N_{1R}, N_{2R})$ values are
\[
(N_{1R}, N_{2R})=\left\{ \begin{array}{ll} (0, N_0), & \mbox{when}\hspace{0.12cm} 0 \leq N_{0} \leq  \left[\frac{m}{3}\right],  \\
\left(N_0-1-\left[\frac{m}{3}\right], \left[\frac{m}{3}\right]+1 \right), & \mbox{when}\hspace{0.12cm}  \left[\frac{m}{3}\right]+1 \leq  N_{0} \leq 2\left[\frac{m}{3}\right], \\
(m-1-N_0, \hspace{0.05cm} m-N_0), & \mbox{when}\hspace{0.12cm}  2\left[\frac{m}{3}\right]+1 \leq  N_{0} \leq m-1. \end{array} \right.
\]
If $N_R=0$, then zero is not a root of $R_{N}^{[m]}(z)$.
\end{quote}

\vspace{0.08cm}
The proofs of these two theorems will be provided in Appendix A.

The most significant piece of information in these two theorems is the formulae for $N_Q$ and $N_R$, which give the multiplicities of the zero root in $Q_{N}^{[m]}(z)$ and $R_{N}^{[m]}(z)$ polynomials. These multiplicity formulae are particularly important for the analysis of rogue waves in the inner region under a large parameter (see later text). Compared to the multiplicity formula of the zero root in the Yablonskii-Vorob'ev polynomial hierarchy \cite{NLSRWs2021,Yanguniversal}, the present multiplicity formulae for Okamoto hierarchies are more involved, but their connection to the multiplicity formula of the Yablonskii-Vorob'ev hierarchy is still visible (see especially \cite{Yanguniversal}). For Okamoto polynomials $Q_N(z)$ and $R_N(z)$ (where $m=2$), these multiplicity formulae show that $N_Q=0$ for all $N$ values, and $N_R=0$ if $N$ is even and $N_R=1$ if $N$ is odd. This means that for any $N$, zero is not a root of $Q_N(z)$. In addition, for $R_N(z)$,
zero is not a root when $N$ is even and is a simple root when $N$ is odd.

Another piece of information from formulae (\ref{QNmform}) and (\ref{RNmform}) of these theorems is that, root structures of both $Q_{N}^{[m]}(z)$ and $R_{N}^{[m]}(z)$ polynomials are invariant under $2\pi/m$-angle rotation in the complex $z$ plane. This rotational symmetry of the root structures will have implications on shapes of rogue patterns away from the origin under a large parameter, as we will see later.

The only major piece of information missing from the above two theorems is multiplicities of \emph{nonzero} roots in these $Q_{N}^{[m]}(z)$ and $R_{N}^{[m]}(z)$ polynomials. For Okamoto polynomials $Q_N(z)$ and $R_N(z)$ (where $m=2$), it has been shown that all their roots are simple \cite{Kametaka,Fukutani}. For higher members of these polynomial hierarchies, their zero root clearly can be non-simple in view of the above $N_Q$ and $N_R$ formulae. However, it is unclear whether their nonzero roots can also be non-simple. We numerically studied this question for many particular polynomials of the two hierarchies, and found their nonzero roots to be always simple. Based on this numerical evidence, we propose the following conjecture.

\begin{quote}
\textbf{Conjecture 1}\hspace{0.06cm} Nonzero roots of Okamoto-hierarchy polynomials $Q_{N}^{[m]}(z)$ and $R_{N}^{[m]}(z)$ are all simple for arbitrary integers $N\ge 1$ and $m>1$.
\end{quote}

If this conjecture holds, then from Theorems 1-2, numbers of nonzero roots in $Q_{N}^{[m]}(z)$ and $R_{N}^{[m]}(z)$ would be
\[ \label{MQMR}
M_Q=N(N+1)-N_Q, \quad M_R=N^2-N_R,
\]
respectively, where $N_Q$ and $N_R$ are given in Eqs.~(\ref{defNQ}) and (\ref{defNR}).

\begin{figure}[htb]
\begin{center}
\includegraphics[scale=0.4, bb=500 000 385 630]{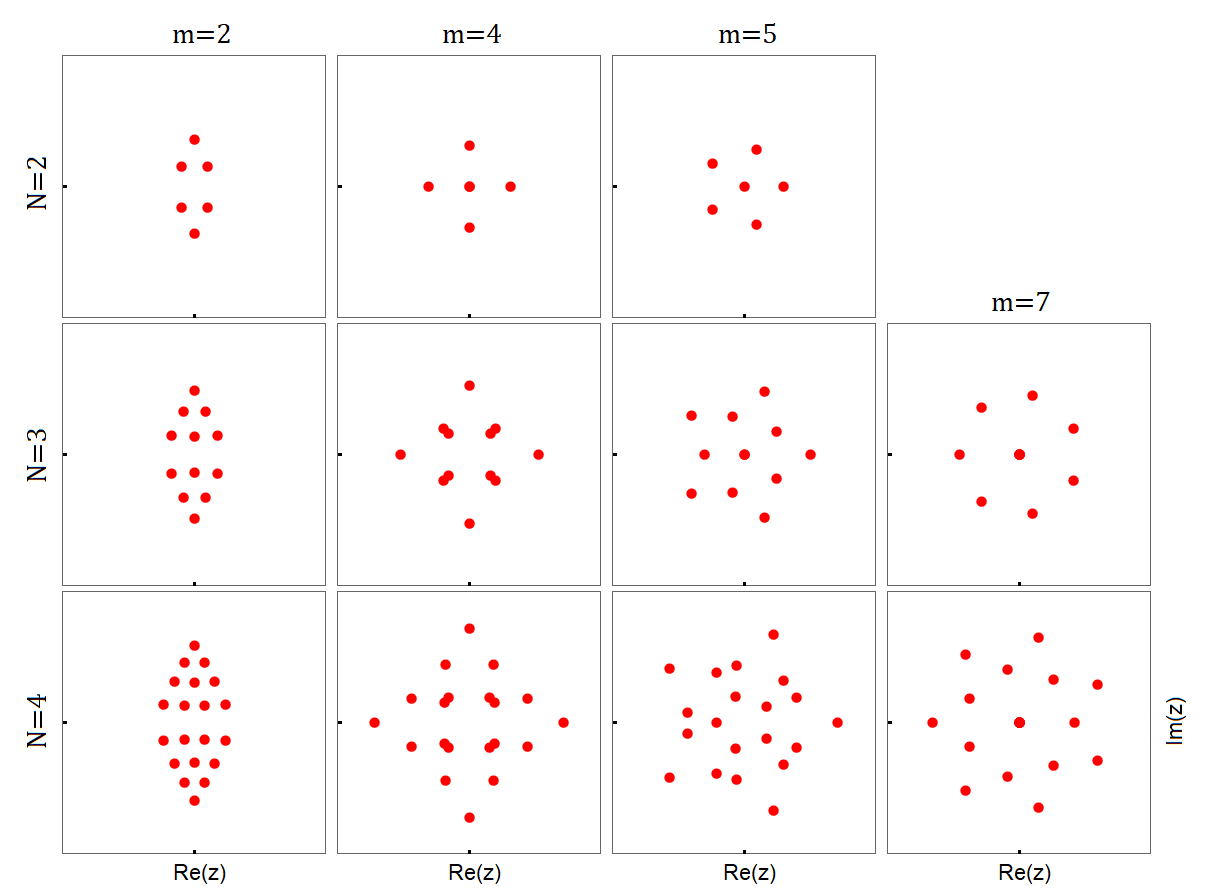}
\caption{Root structures of the $Q_{N}^{[m]}(z)$ polynomial hierarchy for $2\le N\le 4$ and $m=2, 4, 5, 7$. In all panels, $-8\le \mbox{Re}(z), \mbox{Im}(z)\le 8$.
\label{f:rootsQN}}
\end{center}
\end{figure}

To get a visual impression of root structures in Okamoto polynomial hierarchies, we plot in Fig.~\ref{f:rootsQN} roots of the $Q_{N}^{[m]}(z)$ hierarchy in the complex $z$ plane with $2\le N\le 4$ and $m=2,4,5,7$. The first column of this figure (with $m=2$), for roots of Okamoto polynomials $Q_N(z)$, exhibit ``double triangles" as reported in \cite{Clarkson2003PIV}. We caution the reader that sides of these double triangles are not exactly straight; thus our use of the term ``double triangles" is only in an approximate sense. The second column of this figure, for roots of $Q_{N}^{[4]}(z)$ polynomials, exhibit a ``square" shape with curved sides, intricate interiors, and some very close roots. The third column, for roots of $Q_{N}^{[5]}(z)$, exhibit a pentagon shape; while the fourth column, for roots of $Q_{N}^{[7]}(z)$, exhibit a heptagon shape. Compared to root shapes of the Yablonskii-Vorob'ev polynomial hierarchy, the present double-triangle and square shapes are new. The pentagon and heptagon shapes are not new, as they have appeared in the Yablonskii-Vorob'ev hierarchy before \cite{Clarkson2003-II,NLSRWs2021,Yanguniversal}. However, compared to pentagons and heptagons of Yablonskii-Vorob'ev roots, the current pentagons and heptagons of Okamoto roots have different interiors.

\begin{figure}[htb]
\begin{center}
\includegraphics[scale=0.4, bb=500 000 385 630]{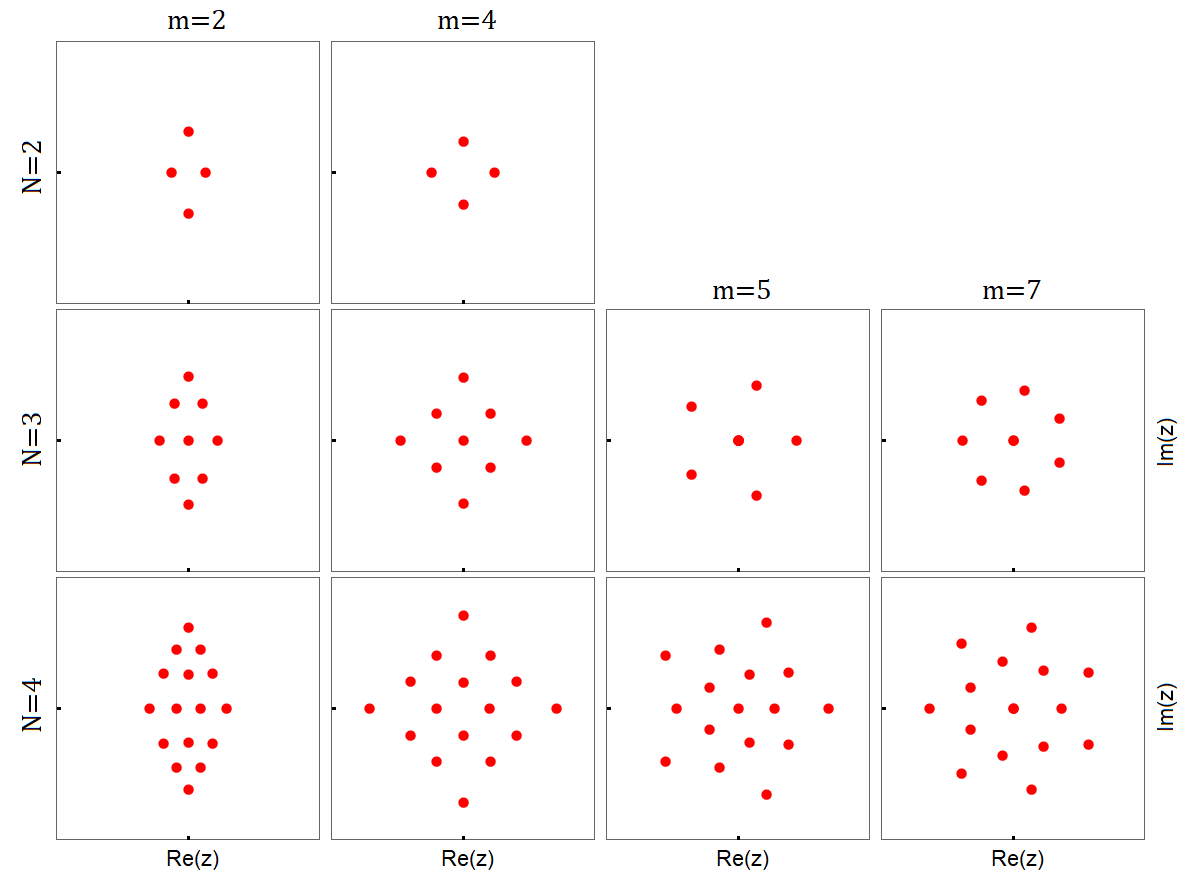}
\caption{Root structures of the $R_{N}^{[m]}(z)$ polynomial hierarchy for $2\le N\le 4$ and $m=2, 4, 5, 7$. In all panels, $-7\le \mbox{Re}(z), \mbox{Im}(z)\le 7$.
\label{f:rootsRN} }
\end{center}
\end{figure}

In Fig.~\ref{f:rootsRN}, we plot root structures of the $R_{N}^{[m]}(z)$ hierarchy in the complex plane with $2\le N\le 4$ and $m=2,4,5,7$. Shapes of these roots are somewhat similar to their counterparts for $Q_{N}^{[m]}(z)$ in the previous figure, but plenty of differences also exist between them. One difference is that, while the first column of Fig.~\ref{f:rootsQN} exhibit two separate triangles, the first column of the current figure exhibit two triangles that are joined together to form a rhombus. Another difference is that, interior roots in the second column of the current figure are more orderly than their counterparts in Fig.~\ref{f:rootsQN}. A third difference is that, even though shapes of roots in the fourth columns of the two figures are quite similar to each other, zero roots in corresponding panels actually have different multiplicities. For example, the zero root has multiplicity 5 in the upper panel of the fourth column of Fig.~1, but has multiplicity 2 in the corresponding panel of Fig.~2.

The aim of this paper is to show that, geometric shapes of certain types of rogue waves in integrable systems are closely related with root structures of Okamoto polynomial hierarchies. For this purpose, we present such rogue waves in two integrable systems below.

\subsection{Rogue waves in the Manakov system} \label{subsec_Manakov}

The Manakov system is \cite{Manakov}
\[ \label{ManakModel}
\begin{array}{l}
(\textrm{i} \partial_t+ \partial_x^2 )u_1+(\epsilon_{1}|u_{1}|^2+\epsilon_{2}|u_{2}|^2)u_{1}=0, \\
(\textrm{i} \partial_t+ \partial_x^2 )u_{2}+(\epsilon_{1}|u_{1}|^2+\epsilon_{2}|u_{2}|^2)u_{2}=0,
\end{array}
\]
where the nonlinear coefficients $\epsilon_{1}=\pm 1$ and $\epsilon_{2}=\pm 1$. These equations govern many physical processes such as the interaction of two incoherent light beams in crystals \cite{Kivshar_book,Stegeman_Manakov,Segev_coupled}, transmission of light in randomly birefringent optical fibers \cite{Menyuk, ManakovPMD1,Wabnitzexperiment1}, and evolution of two-component Bose-Einstein condensates \cite{BEC,BEC_Manakov_experiment}.

Rogue waves in the Manakov system are rational solutions that satisfy the following boundary conditions,
\[ \label{Manakovpw}
\begin{array}{l}
u_{1}(x,t)\rightarrow  \rho_{1}  e^{{\rm{i}} (k_{1}x + \omega_{1} t)},   \quad x, t \to \pm \infty, \\
u_{2}(x,t)\rightarrow  \rho_{2}  e^{{\rm{i}} (k_{2}x + \omega_{2} t)},   \quad x, t \to \pm \infty,
\end{array}
\]
where $(k_1, k_2)$ and $(\omega_1, \omega_2)$ are wavenumbers and frequencies of the two components in the plane-wave background, and $(\rho_1, \rho_2)$ are their amplitudes which will be set real positive using phase invariance of the system. Parameters of the background plane wave satisfy the following relations,
\[ \label{manaomega}
\begin{array}{l}
\omega_{1} = \epsilon_{1}\rho_{1}^2 +  \epsilon_{2}\rho_{2}^2 -k_{1}^2,  \\
\omega_{2} = \epsilon_{1}\rho_{1}^2 +  \epsilon_{2}\rho_{2}^2 -k_{2}^2.
\end{array}
\]
Due to Galilean invariance of the Manakov system, we can also set $k_1=-k_2$ without loss of generality. In this case, $\omega_1=\omega_2$.

Rogue waves in the Manakov system have been derived in Refs.~\cite{BDCW2012,ManakovDark,LingGuoZhaoCNLS2014,Chen_Shihua2015,ZhaoGuoLingCNLS2016,Yanguniversal} by various methods.
Some of those rogue waves are related to the Yablonski-Vorob'ev polynomial hierarchy \cite{Yanguniversal}. The ones that are related to Okamoto hierarchies turn out to be those studied in \cite{LingGuoZhaoCNLS2014,Chen_Shihua2015,ZhaoGuoLingCNLS2016} by Darboux transformation, when the characteristic equation of the underlying $3\times 3$ scattering matrix admits a triple eigenvalue. However, those rogue waves derived by Darboux transformation in \cite{LingGuoZhaoCNLS2014,Chen_Shihua2015,ZhaoGuoLingCNLS2016} were not general or explicit enough for asymptotic analysis. Thus, we will first present explicit and general expressions of those rogue waves by the bilinear method.

Before presenting our rogue-wave expressions, we need to introduce elementary Schur polynomials. These polynomials $S_j(\mbox{\boldmath $x$})$ with $ \emph{\textbf{x}}=\left( x_{1}, x_{2}, \ldots \right)$ are defined by the generating function
\begin{equation}\label{Elemgenefunc}
\sum_{j=0}^{\infty}S_j(\mbox{\boldmath $x$}) \epsilon^j
=\exp\left(\sum_{j=1}^{\infty}x_j \epsilon^j\right),
\end{equation}
or more explicitly,
\begin{equation}
S_{0} (\emph{\textbf{x}})=1,\quad  S_1(\mbox{\boldmath $x$})=x_1,
\quad S_2(\mbox{\boldmath $x$})=\frac{1}{2}x_1^2+x_2, \quad \cdots, \quad
S_{j}(\mbox{\boldmath $x$}) =\sum_{l_{1}+2l_{2}+\cdots+ml_{m}=j} \left( \ \prod _{j=1}^{m} \frac{x_{j}^{l_{j}}}{l_{j}!}\right).
\end{equation}
In addition, we define $S_{j} (\emph{\textbf{x}})=0$ when $j<0$.

Through Schur polynomials, our Manakov rogue waves related to Okamoto polynomial hierarchies are given by the following theorem.

\begin{quote}
\textbf{Theorem 3} \hspace{0.05cm} When the nonlinear coefficients in the Manakov system (\ref{ManakModel}) are $\epsilon_1=\epsilon_2=1$, and background amplitudes and wave numbers in the boundary conditions (\ref{Manakovpw}) satisfy the following constraints
%\[
%\epsilon_{1} \rho_{1}^2= \epsilon_{2}\rho_{2}^2=2 \left(k_1-k_2\right)^2,  \quad k_1\ne k_2,
%\]
\[ \label{CubicCondition}
\rho_1=\rho_2=\sqrt{2}\hspace{0.03cm} \left|k_1-k_2\right|, \quad k_1\ne k_2,
\]
the algebraic equation
\[ \label{QurticeqQ1dp}
\mathcal{F}'_{1}(p)= 0,
\]
where
\[\label{Q1polynomial}
\mathcal{F}_{1}(p)= \frac{\rho_{1}^2}{p-{\rm{i}}\hspace{0.03cm} k_1} + \frac{\rho_{1}^2}{p-{\rm{i}}\hspace{0.03cm}k_2}+ 2p,
\]
would admit a non-imaginary double root
\[\label{p0Mana}
p_0=\frac{\sqrt{3}}{2}(k_1-k_2)+\frac{{\rm{i}}}{2}(k_1+k_2).
\]
In this case, the Manakov system (\ref{ManakModel}) under boundary conditions (\ref{Manakovpw}) would admit nonsingular $(N_1, N_2)$-th order rogue wave solutions
\begin{eqnarray}
&& u_{1,N_1, N_2}(x,t)= \rho_{1}\frac{g_{1,N_1, N_2}}{f_{N_1, N_2}} e^{{\rm{i}} (k_1x+\omega_{1} t)}, \label{Schpolysolu1} \\
&& u_{2,N_1, N_2}(x,t)= \rho_{1}\frac{g_{2,N_1, N_2}}{f_{N_1, N_2}} e^{{\rm{i}} (k_{2}x + \omega_{2} t)}, \label{Schpolysolu2}
\end{eqnarray}
where $N_1$ and $N_2$ are arbitrary non-negative integers,
\[ \label{SchpolysolufN}
f_{N_1, N_2}=\sigma_{0,0}, \quad g_{1,N_1, N_2}=\sigma_{1,0}, \quad  g_{2,N_1, N_2}=\sigma_{0,1},
\]
$\sigma_{n,k}$ is given by the following $2\times 2$ block determinant
\[ \label{cubicrwstype3}
\sigma_{n,k}=
\det \left(
\begin{array}{cc}
  \sigma^{[1,1]}_{n,k} & \sigma^{[1,2]}_{n,k} \\
  \sigma^{[2,1]}_{n,k} & \sigma^{[2,2]}_{n,k}
\end{array}
\right),
\]
\[\label{Blockmatrix}
\sigma^{[I, J]}_{n,k}=
\left(
\phi_{3i-I, \, 3j-J}^{(n,k, \hspace{0.04cm} I, J)}
\right)_{1\leq i \leq N_{I}, \, 1\leq j \leq N_{J}},
\]
the matrix elements in $\sigma^{[I, J]}_{n,k}$ are defined by
\[ \label{Schmatrimnij}
\phi_{i,j}^{(n,k,I,J)}=\sum_{\nu=0}^{\min(i,j)} \left[ \frac{|p_{1}|^2 }{(p_{0}+p_{0}^*)^2}  \right]^{\nu} \hspace{0.06cm} S_{i-\nu}(\textbf{\emph{x}}_I^{+}(n,k) +\nu \textbf{\emph{s}})  \hspace{0.06cm} S_{j-\nu}(\textbf{\emph{x}}_J^{-}(n,k) + \nu \textbf{\emph{s}}^*),
\]
vectors $\textbf{\emph{x}}_I^{+}(n,k)=\left( x_{1,I}^{+}, x_{2,I}^{+},\cdots \right)$ and $\textbf{\emph{x}}_J^{-}(n,k)=\left( x_{1,J}^{-}, x_{2,J}^{-},\cdots \right)$ are defined by
\begin{eqnarray}
&&x_{r,I}^{+}(n,k)= p_r x +  \left(\sum _{l=0}^r p_l p_{r-l}\right)  (\textrm{i} t) +  n \theta_{r} + k \lambda_{r}+a_{r,I},  \quad \hspace{0.15cm} \mbox{if} \hspace{0.08cm} r \hspace{0.08cm} \mbox{mod} \hspace{0.05cm} 3 \ne 0,
\label{defxrp} \\
&&x_{r,J}^{-}(n,k)= p_r^* x - \left( \sum _{l=0}^r p_l^* p_{r-l}^* \right) (\textrm{i} t)  - n \theta_{r}^* -  k \lambda_{r}^* +a_{r,J}^*, \quad \mbox{if} \hspace{0.08cm} r \hspace{0.08cm} \mbox{mod} \hspace{0.05cm} 3 \ne 0, \label{defxrm} \\
&& x_{r,I}^{+}(n,k)=x_{r,J}^{-}(n,k)=0, \quad \mbox{if} \hspace{0.08cm} r \hspace{0.08cm} \mbox{mod} \hspace{0.05cm} 3 = 0,  \label{defxrpm}
\end{eqnarray}
the asterisk `*' represents complex conjugation, $\theta_{r}$, $\lambda_{r}$ and $s_{r}$ are coefficients from the expansions
\begin{eqnarray}
&&  \ln \left[\frac{ p \left( \kappa \right)-{\rm{i}}\hspace{0.03cm} k_1}{p_{0}-{\rm{i}}\hspace{0.03cm} k_1}\right]  =\sum_{r=1}^{\infty} \theta_{r}\kappa^{r},  \quad
  \ln \left[\frac{ p \left( \kappa \right)-{\rm{i}}\hspace{0.03cm} k_2}{p_{0}-{\rm{i}}\hspace{0.03cm} k_2}\right]  =\sum_{r=1}^{\infty} \lambda_{r}\kappa^{r}, \label{schucoeflambda} \\
&& \ln \left[\frac{1}{\kappa} \left(\frac{p_{0}+p_{0}^*}{p_{1}} \right) \left( \frac{ p \left( \kappa \right)-p_{0}}{p \left( \kappa \right)+p_{0}^*} \right)  \right] = \sum_{r=1}^{\infty}s_{r} \kappa^r,  \label{schurcoeffsr}
\end{eqnarray}
the function $p \left(\kappa\right)$ is defined by the equation
\[\label{defpk}
\mathcal{F}_{1}\left[p \left( \kappa \right)\right] = \frac{\mathcal{F}_{1}(p_{0})}{3} \left[ e^{\kappa} +2 e^{-\kappa/2}
\cos\left(\frac{\sqrt{3}}{2} \kappa \right) \right],
\]
$p_r= p^{(r)}(0)/r!$, with the superscript `$(r)$' denoting the $r$-th derivative of $p(\kappa)$, and
\[
(a_{1, 1}, a_{2,1}, a_{4,1}, a_{5, 1}, \dots, a_{3N_1-1,\hspace{0.05cm} 1}), \quad (a_{1, 2}, a_{2,2}, a_{4,2}, a_{5, 2}, \dots, a_{3N_2-2, \hspace{0.05cm} 2})
\]
are free complex constants.
\end{quote}

The proof of this theorem will be provided in Appendix B.

\textbf{Remark 1} \hspace{0.04cm} Regarding the polynomial degree of $\sigma_{n,k}$ in the above theorem, we can show, by rewriting $\sigma_{n,k}$ into a larger determinant similar to what was done in Ref.~\cite{OhtaJY2012}, that
\[  \label{sigmadegree}
\mbox{deg}(\sigma_{n,k})=2\left[N_{1}(N_1-N_2+1)+N_{2}^2\right]
\]
in both $x$ and $t$.

\textbf{Remark 2} \hspace{0.04cm} The algebraic equation (\ref{QurticeqQ1dp}) is a quartic equation. Under parameter conditions (\ref{CubicCondition}), this quartic equation admits two double roots, one being $p^{(1)}_{0}=p_0$, and the other being $p^{(2)}_{0}=-p_0^*$. If we replace $p_0$ by $p^{(2)}_{0}$ in Eqs.~(\ref{Schmatrimnij})-(\ref{defpk}), the resulting functions (\ref{Schpolysolu1})-(\ref{Schpolysolu2}) are still Manakov rogue waves. However, these other Manakov rogue waves are equivalent to those given in Theorem 3 when parameters $(a_{r, 1}, a_{r, 2})$ in them are properly related. See Remark 2 in Ref.~\cite{YangYang3waves} for details.

\textbf{Remark 3} \hspace{0.04cm} Regarding coefficients $s_r$ defined in the expansion (\ref{schurcoeffsr}), we can show that the left-side function in this expansion is independent of background parameters $(k_1, k_2, \rho_1, \rho_2)$. Thus, $s_r$ are specific constants. Our calculation of these constants gives
\[ \label{svalueMan}
s_1=s_2=0, \hspace{0.1cm} s_3=-0.025, \hspace{0.1cm} s_4=s_5=0, \hspace{0.1cm} s_6\approx 0.00092, \hspace{0.1cm} s_7=s_8=0, \hspace{0.1cm} s_9\approx -0.000045, \hspace{0.1cm} s_{10}=s_{11}=0, \hspace{0.1cm} \cdots.
\]
We believe that this pattern of $s_{r}=0$ for $r \hspace{0.07cm} \mbox{mod}\hspace{0.07cm} 3 \neq 0$ holds for all higher $r$ indices as well. But our analysis will not assume this ``fact".

\textbf{Remark 4} \hspace{0.04cm} There are three functions of $p(\kappa)$ that satisfy Eq.~(\ref{defpk}), and these three functions are related as $p(\kappa e^{{\rm{i}}2j\pi/3})$, where $j=0, 1, 2$. We can choose any one of these three functions in the above theorem and keep complex parameters $a_{r,I}$ free without loss of generality. See Remark 3 in Ref.~\cite{YangYang3waves} for details. The choice of these three $p(\kappa)$ functions is made by the choice of the $p_1$ value in $p(\kappa)$'s Taylor expansion. This $p_1$ is any one of the three cubic roots of a certain constant. After $p_1$ is picked, $p(\kappa)$ will be uniquely determined.

\subsection{Rogue waves in the three-wave resonant interaction system}

The (1+1)-dimensional three-wave resonant interaction system is
\begin{eqnarray}
&& \left(\partial_t+c_1 \partial_x\right)u_1= \epsilon_{1} u_{2}^* u_{3}^*, \nonumber \\
&& \left(\partial_t+c_2 \partial_x\right)u_2= \epsilon_{2} u_{1}^* u_{3}^*,  \label{3WRIModel}\\
&& \left(\partial_t+c_3 \partial_x\right)u_3= \epsilon_{3} u_{1}^* u_{2}^*, \nonumber
\end{eqnarray}
where $(c_1, c_2, c_3)$ are group velocities of the three waves, and $(\epsilon_1, \epsilon_2, \epsilon_3)$ are real-valued nonlinear coefficients. To remove ambiguity, we order the three group velocities as $c_{1} > c_{2} > c_{3}$, and make $c_{3}=0$ by choosing a coordinate system that moves with velocity $c_3$. The nonlinear coefficients $\epsilon_n$ can be normalized to $\pm 1$ by variable scalings. In addition, we can fix $\epsilon_{1}=1$ without loss of generality.

Rogue waves in this three-wave interaction system are rational solutions which approach plane-wave solutions as $x, t\to \pm \infty$, i.e.,
\begin{eqnarray}
&& u_{1}(x,t)\rightarrow  \rho_{1}  e^{{\rm{i}} (k_{1}x + \omega_{1} t)},  \hspace{1.8cm}  x, t\to \pm \infty, \nonumber \\
&& u_{2}(x,t)\rightarrow  \rho_{2}  e^{{\rm{i}} (k_{2}x + \omega_{2} t)},  \hspace{1.8cm}  x, t\to \pm \infty, \label{BoundaryCond}\\
&& u_{3}(x,t)\rightarrow  \textrm{i} \hspace{0.04cm} \rho_{3}  e^{-{\rm{i}} [ (k_1+k_2)x +(\omega_1+\omega_2) t]}, \hspace{0.4cm} x, t\to \pm \infty,   \nonumber
\end{eqnarray}
where $(k_1, k_2)$ and $(\omega_1, \omega_2)$ are wavenumbers and frequencies of the first two waves, and $(\rho_1, \rho_2, \rho_3)$ are the complex amplitudes of the three waves. Parameters of these plane waves satisfy the following relations,
\begin{eqnarray}
&& \rho_{1} \left( \omega_{1} + c_{1}k_{1} \right) = -\epsilon_{1} \rho_{2}^* \rho_{3}^*,  \nonumber \\
&& \rho_{2} \left( \omega_{2} + c_{2}k_{2} \right) = -\epsilon_{2} \rho_{1}^* \rho_{3}^*,  \label{Pararlation}\\
&& \rho_{3} \left( \omega_{1} +\omega_{2} \right) = \epsilon_{3} \rho_{1}^* \rho_{2}^*.  \nonumber
\end{eqnarray}
In this article, we assume $\rho_1$, $\rho_2$, and $\rho_3$ are all nonzero. In view of the phase invariance, we can normalize $\rho_1$ and $\rho_2$ to be real. Then the above relations show that $\rho_3$ is real as well. In addition, the gauge invariance  allows us to impose a restriction on the four parameters $(k_1, k_2, \omega_1, \omega_2)$, such as fixing one of them as zero, or equating $k_1=k_2$, or equating $\omega_1=\omega_2$, without any loss of generality. Under such a restriction, wavenumber and frequency parameters $(k_1, k_2, \omega_1, \omega_2)$ would be fully determined from the three real background-amplitude parameters $(\rho_1, \rho_2, \rho_3)$ through equations (\ref{Pararlation}).

General rogue-wave solutions in the three-wave interaction system (\ref{3WRIModel}) have been derived by the bilinear method in~\cite{YangYang3waves}. To present these rogue waves, we introduce notations
\[ \label{gamma123}
\gamma_{1} \equiv \epsilon_1 \frac{ \rho_{2} \rho_{3} }{ \rho_{1}} ,\  \gamma_{2} \equiv \epsilon_2 \frac{\rho_{1} \rho_{3} }{ \rho_{2}}, \
\gamma_3\equiv \epsilon_3 \frac{ \rho_{1} \rho_{2} }{ \rho_{3}},
\]
and
\[\label{Q2polynomial}
\mathcal{F}_{2}(p)= \left( \frac{  \gamma_{1} c_{2}}{\gamma_3(c_{2}-c_{1})} \right) \frac{1 }{p} - \left( \frac{\gamma_{2} c_{1}}{\gamma_3(c_{2}-c_{1})}  \right) \frac{1 }{p-{\rm i}} - p.
\]
Then, rogue waves in three-wave interactions that are related to Okamoto polynomial hierarchies are given by the following theorem.

\begin{quote}
\textbf{Theorem 4} \cite{YangYang3waves} \hspace{0.05cm} In the soliton-exchange case where $(\epsilon_1, \epsilon_2, \epsilon_3)=(1, -1, 1)$, and under parameter conditions
\[ \label{CubicRestrict}
\rho_{2}=\pm \sqrt{\frac{c_{1}}{c_{2}}}\rho_{1},\ \ \ \rho_{3}= \pm \sqrt{\frac{c_{1}-c_{2}}{c_{2}}}\rho_{1},
\]
the algebraic equation
\[ \label{QurticeqQ1dp2}
\mathcal{F}'_{2}(p)= 0
\]
admits a non-imaginary double root $p_{0}=(\sqrt{3}+\rm{i})/2$. In this case, the three-wave interaction system (\ref{3WRIModel}) under boundary conditions (\ref{BoundaryCond}) admits nonsingular $(N_1, N_2)$-th order rogue-wave solutions
\begin{eqnarray}
  && u_{1, N_1,N_2}(x,t)= \rho_{1}\frac{g_{1, N_1,N_2}}{f_{N_1,N_2}} e^{{\rm{i}} (k_1x+\omega_{1} t)}, \label{Schpolysolu1b} \\
  && u_{2, N_1,N_2}(x,t)= \rho_{2}\frac{g_{2, N_1,N_2}}{f_{N_1,N_2}} e^{{\rm{i}} (k_{2}x + \omega_{2} t)}, \label{Schpolysolu2b} \\
  && u_{3, N_1,N_2}(x,t)= {\rm{i}} \hspace{0.04cm} \rho_{3}\frac{g_{3, N_1,N_2}}{f_{N_1,N_2}} e^{-{\rm{i}} [(k_1+k_{2})x + (\omega_{1}+\omega_{2}) t]},  \label{Schpolysolu3b}
\end{eqnarray}
where $N_1$ and $N_2$ are arbitrary non-negative integers,
\[ \label{SchpolysolufN2}
f_{N_1,N_2}=\sigma_{0,0}, \quad g_{1, N_1,N_2}=\sigma_{1,0}, \quad  g_{2, N_1,N_2}=\sigma_{0,-1}, \quad g_{3, N_1,N_2}=\sigma_{-1,1},
\]
$\sigma_{n,k}$ is given by the following $2\times 2$ block determinant
\[ \label{cubicrwstype3b}
\sigma_{n,k}=
\det \left(
\begin{array}{cc}
  \sigma^{[1,1]}_{n,k} & \sigma^{[1,2]}_{n,k} \\
  \sigma^{[2,1]}_{n,k} & \sigma^{[2,2]}_{n,k}
\end{array}
\right),
\]
\[\label{Blockmatrix2}
\sigma^{[I, J]}_{n,k}=
\left(
\phi_{3i-I, \, 3j-J}^{(n,k, \hspace{0.04cm} I, J)}
\right)_{1\leq i \leq N_{I}, \, 1\leq j \leq N_{J}},
\]
the matrix elements in $\sigma^{[I, J]}_{n,k}$ are defined by
\[ \label{Schmatrimnij9a}
\phi_{i,j}^{(n,k, I, J)}=\sum_{\nu=0}^{\min(i,j)} \left[ \frac{|p_{1}|^2 }{(p_{0}+p_{0}^*)^2}  \right]^{\nu} \hspace{0.06cm} S_{i-\nu}(\textbf{\emph{x}}_I^{+}(n,k) +\nu \textbf{\emph{s}})  \hspace{0.06cm} S_{j-\nu}(\textbf{\emph{x}}_J^{-}(n,k) + \nu \textbf{\emph{s}}^*),
\]
vectors $\textbf{\emph{x}}_I^{+}(n,k)=\left( x_{1,I}^{+}, x_{2,I}^{+},\cdots \right)$ and $\textbf{\emph{x}}_J^{-}(n,k)=\left( x_{1,J}^{-}, x_{2,J}^{-},\cdots \right)$ are defined by
\begin{eqnarray}
&&x_{r,I}^{+}(n,k)= \left( \alpha_{r} - \beta_{r} \right) x +\left( c_{1}\beta_{r}-c_{2}\alpha_{r} \right)t + n \theta_{r} + k \lambda_{r} + a_{r,I},  \quad \hspace{0.17cm} \mbox{if} \hspace{0.08cm} r \hspace{0.08cm} \mbox{mod} \hspace{0.05cm} 3 \ne 0, \label{defxrp2} \\
&&x_{r,J}^{-}(n,k)=   \left( \alpha^*_{r} - \beta^*_{r} \right) x +\left( c_{1}\beta^*_{r}-c_{2}\alpha^*_{r} \right)t - n \theta^*_{r} - k \lambda^*_{r} +a^*_{r,J},
\quad \mbox{if} \hspace{0.08cm} r \hspace{0.08cm} \mbox{mod} \hspace{0.05cm} 3 \ne 0, \label{defxrm2} \\
&& x_{r,I}^{+}(n,k)=x_{r,J}^{-}(n,k)=0, \quad \mbox{if} \hspace{0.08cm} r \hspace{0.08cm} \mbox{mod} \hspace{0.05cm} 3 = 0,
\end{eqnarray}
$\alpha_{r}$, $\beta_{r}$, $\theta_{r}$, $\lambda_{r}$ and the vector $\textbf{\emph{s}}=(s_1, s_2, \cdots)$ are defined through the expansions
\begin{eqnarray}
&& \frac{ \gamma_{1}}{c_{1}-c_{2}} \left(\frac{1 }{p \left( \kappa \right)} -\frac{1}{p_{0}} \right)= \sum_{r=1}^{\infty} \alpha_{r}\kappa^{r}, \label{schucoefalpha2} \\
&& \frac{ \gamma_{2}}{c_{1}-c_{2}} \left(\frac{1}{p \left( \kappa \right)-{\rm{i}}} -\frac{1}{p_{0}-{\rm{i}}} \right)= \sum_{r=1}^{\infty} \beta_{r}\kappa^{r},  \label{schucoefbeta2}
\\
&& \ln \frac{ p \left( \kappa \right)}{p_{0}}  =\sum_{r=1}^{\infty} \lambda_{r}\kappa^{r},  \quad \ln \frac{ p \left( \kappa \right)-{\rm{i}}}{p_{0}-{\rm{i}}}  =\sum_{r=1}^{\infty} \theta_{r}\kappa^{r},   \label{schucoeflambda2} \\
&& \ln \left[\frac{1}{\kappa} \left(\frac{p_{0}+p_{0}^*}{p_{1}} \right) \left( \frac{ p \left( \kappa \right)-p_{0}}{p \left( \kappa \right)+p_{0}^*} \right)  \right] = \sum_{r=1}^{\infty}s_{r} \kappa^r,  \label{schurcoeffsr2}
\end{eqnarray}
the function $p\left( \kappa \right)$ which appears in Eqs. (\ref{schucoefalpha2})-(\ref{schurcoeffsr2}) is defined by the equation
\[  \label{Q1ptriple}
\mathcal{F}_{2}\left[p \left( \kappa \right)\right]= \frac{\mathcal{F}_{2}(p_{0})}{3} \left[ e^{\kappa} +2 e^{-\kappa/2}
\cos\left(\frac{\sqrt{3}}{2} \kappa \right) \right],
\]
$\mathcal{F}_{2}(p)$ is given by Eq.~(\ref{Q2polynomial}), or equivalently,
\[ \label{Q2ptriple9}
\mathcal{F}_{2}(p)=-\left(\frac{1}{p}+\frac{1}{p-{\rm{i}}}+p \right)
\]
in view of the parameter constraints (\ref{CubicRestrict}), $p_1\equiv (dp/d\kappa)|_{\kappa=0}$,
and
\[
(a_{1, 1}, a_{2,1}, a_{4,1}, a_{5, 1}, \dots, a_{3N_1-1,\hspace{0.05cm} 1}), \quad (a_{1, 2}, a_{2,2}, a_{4,2}, a_{5, 2}, \dots, a_{3N_2-2, \hspace{0.05cm} 2})
\]
are free complex constants.
\end{quote}

We note that the above rogue expressions are a bit simpler than those presented in Ref.~\cite{YangYang3waves}, since $x_{r,I}^{+}(n,k)$ and $x_{r,J}^{-}(n,k)$ for $r$-indices of
$r \hspace{0.08cm} \mbox{mod} \hspace{0.05cm} 3 = 0$ have been set as zero here. The reason for this simplification is analogous to that explained in Appendix A of Ref.~\cite{NLSRWs2021} in a different setting. We also note that there was a typo in the $\beta_r$ expansion equation (35) of Ref.~\cite{YangYang3waves}, where $c_2-c_1$ there should have been $c_1-c_2$. That typo has been fixed in our current $\beta_r$ expansion equation~(\ref{schucoefbeta2}).

As in the Manakov case, the quartic equation (\ref{QurticeqQ1dp2}) under parameter conditions (\ref{CubicRestrict}) also admits another double root $-p_{0}^*=(-\sqrt{3}+\rm{i})/2$, but this other double root does not lead to new rogue solutions \cite{YangYang3waves}. In addition, from the expansion (\ref{schurcoeffsr2}) for $s_r$, we find numerically that
\[ \label{svalue3wave}
s_1=s_2=0, \hspace{0.1cm} s_3\approx -0.025 - 0.0053{\rm{i}}, \hspace{0.1cm} s_4=s_5=0, \hspace{0.1cm} s_6\approx 0.00088+0.00040 {\rm{i}}., \hspace{0.1cm} s_7=s_8=0, \hspace{0.1cm} \cdots.
\]
So, it seems that $s_{r}=0$ when $r \ \mbox{mod}\ 3 \neq 0$ for the three-wave system as well. Furthermore, the previous Remark 4 holds here too.

\subsection{Special rogue solutions to be studied}

Rogue waves in Theorems 3 and 4 for the Manakov and three-wave-interaction systems contain a wide variety of solutions that exhibit different wave patterns. In this article, we will only study rogue waves in these two theorems where $N_1=0$ or $N_2=0$. In these cases, the $2\times 2$ block determinants in Eqs.~(\ref{cubicrwstype3}) and (\ref{cubicrwstype3b}) reduce to a single-block determinant, which makes our analysis a little simpler. For convenience, we introduce the terminology:
\begin{itemize}
\item \emph{Q-type $N$-th order rogue waves}: rogue waves in Theorems 3 and 4 where $N_1=N \hspace{0.05cm} (>0)$ and $N_2=0$;

\vspace{-0.2cm}
\item \emph{R-type $N$-th order rogue waves}: rogue waves in Theorems 3 and 4 where $N_1=0$ and $N_2=N \hspace{0.05cm}(>0)$.
\end{itemize}
The reason for the word choices of `Q-type' and `R-type' here is that the underlying rogue waves will be related to $Q_{N}^{[m]}(z)$
and $R_{N}^{[m]}(z)$ Okamoto polynomials respectively, as we will show later.

For Q-type $N$-th order rogue waves, $\sigma_{n,k}$ in Eqs.~(\ref{cubicrwstype3}) and (\ref{cubicrwstype3b}) becomes
\[ \label{sigma1}
\sigma_{n,k}^{(Q)}=\left(\phi_{3i-1, \, 3j-1}^{(n,k)}\right)_{1\leq i, j \leq N},
\]
and for R-type $N$-th order rogue waves, $\sigma_{n,k}$ in Eqs.~(\ref{cubicrwstype3}) and (\ref{cubicrwstype3b}) is
\[  \label{sigma2}
\sigma_{n,k}^{(R)}=\left(\phi_{3i-2, \, 3j-2}^{(n,k)}\right)_{1\leq i, j \leq N},
\]
where $\phi_{i,j}^{(n,k)}$ is given by Eq.~(\ref{Schmatrimnij}) for the Manakov system and (\ref{Schmatrimnij9a}) for the three-wave system, but with indices $I$ and $J$ removed. Internal parameters are $(a_1, a_2, a_4, a_5, \cdots, a_{3N-1})$ for Q-type waves, and are $(a_1, a_2, a_4, a_5, \cdots, a_{3N-2})$ for R-type waves. We normalize $a_1=0$ by a shift of the $(x, t)$ axes. Then, internal complex parameters in these rogue waves are $(a_{2}, a_{4}, a_{5}, \cdots, a_{3N-1})$ for Q-type, and $(a_{2}, a_{4}, a_{5}, \cdots, a_{3N-2})$ for R-type.

\section{Rogue wave patterns under a large parameter in the Manakov and three-wave systems} \label{sectheorems}

Now, we consider solution patterns of Q-type and R-type rogue waves in the Manakov and three-wave interaction systems when a single complex internal parameter $a_m$ in them is large (in magnitude), where $2\le m \le 3N-1$ for Q-type, and $2\le m \le 3N-2$ for R-type. In both cases, $m \hspace{0.08cm} \mbox{mod} \hspace{0.05cm} 3 \ne 0$.

\subsection{Rogue-pattern results in the Manakov system}
Our results on patterns of Q-type and R-type rogue waves in the Manakov system under a large internal parameter are summarized in the following two theorems.

\begin{quote}
\textbf{Theorem 5} \hspace{0.06cm} For Q-type $N$-th order rogue waves $[u_{1,N, 0}(x,t), u_{2,N, 0}(x,t)]$ in the Manakov system, suppose $|a_m|\gg 1$ and all other internal parameters are $O(1)$. In addition, suppose all nonzero roots of $Q_{N}^{[m]}(z)$ are simple. Then, the following asymptotics for these rogue waves holds.

\begin{enumerate}
\item
In the outer region on the $(x, t)$ plane, where $\sqrt{x^2+t^2}=O\left(|a_{m}|^{1/m}\right)$, this rogue wave asymptotically separates into $M_{Q}$ fundamental rogue waves, where $M_Q$ is given in Eq.~(\ref{MQMR}). These fundamental rogue waves are $[\hat{u}_{1}(x-\hat{x}_{0}, t-\hat{t}_{0})\hspace{0.06cm} e^{{\rm{i}} (k_1x+\omega_{1} t)}, \hspace{0.07cm} \hat{u}_{2}(x-\hat{x}_{0}, t-\hat{t}_{0})\hspace{0.06cm} e^{{\rm{i}} (k_{2}x + \omega_{2} t)}]$, where functions
$[\hat{u}_{1}(x, t), \hat{u}_{2}(x, t)]$ are obtained from $(\rho_1 g_1/f, \rho_1 g_2/f)$ in Eq.~(\ref{SchpolysolufN}) of Theorem 3 by setting $N_1=0, N_2=1$ and $a_{2,1}=0$, or more explicitly,
\[\label{u11Mana}
\hat{u}_{1}(x,t)=\rho_1 \frac{\left[ p_{1} x+ 2 p_{0} p_{1}\left(  \textrm{i}t  \right) + \theta_{1}\right] \left[ p_{1}^* x- 2 p_{0}^* p_{1}^*\left(\textrm{i}t  \right) - \theta_{1}^* \right]+\zeta_{0}}{\left| p_{1} x+ 2 p_{0} p_{1}\left(  \textrm{i}t  \right) \right|^2+\zeta_{0}},
\]
\[ \label{u21Mana}
\hat{u}_{2}(x,t)=\rho_1 \frac{\left[ p_{1} x+ 2 p_{0} p_{1}\left(  \textrm{i}t  \right) + \lambda_{1}\right] \left[ p_{1}^* x- 2 p_{0}^* p_{1}^*\left(\textrm{i}t  \right)  -  \lambda_{1}^* \right]+\zeta_{0}}{\left| p_{1} x+ 2 p_{0} p_{1}\left(  \textrm{i}t  \right) \right|^2+\zeta_{0}},
\]
\[
\theta_{1}=\frac{p_{1}}{p_{0}-\textrm{i} k_{1}}, \ \lambda_{1}=\frac{p_{1}}{p_{0}-\textrm{i} k_{2}}, \ \zeta_{0} =\frac{|p_{1}|^2}{(p_{0}+p_{0}^*)^2},
\]
their positions $(\hat{x}_{0}, \hat{t}_{0})$ are given by
\begin{eqnarray}
&& \hat{x}_{0}= \frac{1}{\Re(p_0)}\Re\left[ \frac{p_{0}^*}{p_{1}}\left(z_{0}a_m^{1/m}-\Delta_{Q}\right) \right],  \label{x0ManaQ} \\
&& \hat{t}_{0}=\frac{1}{2 \Re(p_0)} \Im\left[ \frac{1}{p_{1}}\left(z_{0}a_m^{1/m}- \Delta_{Q}\right)\right],  \label{t0ManaQ}
\end{eqnarray}
$\Re$ and $\Im$ represent the real and imaginary parts of a complex number, $z_0$ is any of the $M_Q$ nonzero simple roots of $Q_{N}^{[m]}(z)$, and $\Delta_{Q}$ is a $z_0$-dependent $O(1)$ quantity whose formula will be given by Eq.~(\ref{DeltaQform}) in later text. The error of this fundamental rogue wave approximation is $O(|a_{m}|^{-1/m})$. Expressed mathematically, when $|a_{m}|\gg 1$ and $(x-\hat{x}_{0})^2+(t-\hat{t}_{0})^2=O(1)$, we have the following solution asymptotics
\[ \label{uNManaQ1}
\begin{array}{l}
u_{1,N, 0}(x,t) = \hat{u}_{1}(x-\hat{x}_{0},t-\hat{t}_{0}) \hspace{0.06cm} e^{{\rm{i}} (k_1x+\omega_{1} t)} + O\left(|a_{m}|^{-1/m}\right),    \\
u_{2,N, 0}(x,t) = \hat{u}_{2}(x-\hat{x}_{0},t-\hat{t}_{0}) \hspace{0.06cm} e^{{\rm{i}} (k_2x+\omega_{2} t)} + O\left(|a_{m}|^{-1/m}\right).
\end{array}
\]
\item
If zero is a root of the Okamoto-hierarchy polynomial $Q_{N}^{[m]}(z)$, then in the neighborhood of the origin (the inner region), where $x^2+t^2=O(1)$, $[u_{1,N, 0}(x,t), u_{2,N, 0}(x,t)]$ is approximately a lower $\left(N_{1Q}, N_{2Q}\right)$-th order rogue wave $[u_{1,N_{1Q}, N_{2Q}}(x,t), u_{2,N_{1Q}, N_{2Q}}(x,t)]$ as given in Theorem~3, where $\left(N_{1Q}, N_{2Q}\right)$ are provided in Theorem~1. Internal parameters $(\hat{a}_{1, 1}, \hat{a}_{2,1}, \hat{a}_{4, 1}, \hat{a}_{5,1}, \dots, \hat{a}_{3N_{1Q}-1,\hspace{0.05cm} 1})$ and $(\hat{a}_{1, 2}, \hat{a}_{2,2}, \hat{a}_{4, 2}, \hat{a}_{5,2}, \dots, \hat{a}_{3N_{2Q}-2, \hspace{0.05cm} 2})$ in this lower-order rogue wave are related to those in the original rogue wave as
\[ \label{ahat1a}
\hat{a}_{j,1}=\hat{a}_{j,2}=a_{j}+(N-N_{1Q}-N_{2Q})s_j,  \quad j=1, 2, 4, 5, \cdots,
\]
where $s_j$ is as defined in Theorem 3 and numerically given in Eq.~(\ref{svalueMan}). The error of this lower-order rogue wave approximation is $O(|a_{m}|^{-1})$. Expressed mathematically, when $|a_{m}|\gg 1$ and $x^2+t^2=O(1)$,
\[ \label{uNManaQ2}
\begin{array}{l}
u_{1,N, 0}(x,t; a_{2}, a_{4}, a_{5}, \cdots) = u_{1,N_{1Q},N_{2Q}}(x,t; \hat{a}_{j,1},  \hat{a}_{j,2}, j=1, 2, 4, 5, \dots) + O\left(|a_{m}|^{-1}\right), \\
u_{2,N,0}(x,t; a_{2}, a_{4}, a_{5}, \cdots) = u_{2,N_{1Q},N_{2Q}}(x,t; \hat{a}_{j,1}, \hat{a}_{j,2}, j=1, 2, 4, 5, \dots)+ O\left(|a_{m}|^{-1}\right).
\end{array}
\]
If zero is not a root of $Q_{N}^{[m]}(z)$, then in the inner region, the solution
$[u_{1,N,0}(x, t), u_{2,N,0}(x, t)]$ approaches the constant background $[\rho_{1} e^{{\rm{i}} (k_{1}x + \omega_{1} t)}, \rho_{1}  e^{{\rm{i}} (k_{2}x + \omega_{2} t)}]$ when $|a_m|\gg 1$.
\end{enumerate}
\end{quote}

\begin{quote}
\textbf{Theorem 6} \hspace{0.06cm} For R-type $N$-th order rogue waves $[u_{1,0,N}(x,t), u_{2,0,N}(x,t)]$ in the Manakov system, suppose $|a_m|\gg 1$ and all other internal parameters are $O(1)$. In addition, suppose all nonzero roots of $R_{N}^{[m]}(z)$ are simple. Then, the following asymptotics for these rogue waves holds.

\begin{enumerate}
\item
In the outer region, where $\sqrt{x^2+t^2}=O\left(|a_{m}|^{1/m}\right)$, this rogue wave asymptotically separates into $M_{R}$ fundamental rogue waves, where $M_R$ is given in Eq.~(\ref{MQMR}). These fundamental rogue waves are $[\hat{u}_{1}(x-\hat{x}_{0}, t-\hat{t}_{0})\hspace{0.06cm} e^{{\rm{i}} (k_1x+\omega_{1} t)}, \hspace{0.07cm} \hat{u}_{2}(x-\hat{x}_{0}, t-\hat{t}_{0})\hspace{0.06cm} e^{{\rm{i}} (k_{2}x + \omega_{2} t)}]$, where functions $[\hat{u}_{1}(x,t), \hat{u}_{2}(x,t)]$ are as given in Eqs.~(\ref{u11Mana})-(\ref{u21Mana}), positions $(\hat{x}_{0}, \hat{t}_{0})$ of these fundamental rogue waves are given by
\begin{eqnarray}
&& \hat{x}_{0}= \frac{1}{\Re(p_0)}\Re\left[ \frac{p_{0}^*}{p_{1}}\left(z_{0}a_m^{1/m}-\Delta_{R}\right) \right],  \label{x0ManaR} \\
&& \hat{t}_{0}=\frac{1}{2 \Re(p_0)} \Im\left[ \frac{1}{p_{1}}\left(z_{0}a_m^{1/m}- \Delta_{R}\right)\right],  \label{t0ManaR}
\end{eqnarray}
$z_0$ is any of the $M_R$ nonzero simple roots of $R_{N}^{[m]}(z)$, and $\Delta_{R}$ is a $z_0$-dependent $O(1)$ quantity given by Eq.~(\ref{DeltaRform}) in later text.
The error of this fundamental rogue wave approximation is $O(|a_{m}|^{-1/m})$. Expressed mathematically, when $|a_{m}|\gg 1$ and $(x-\hat{x}_{0})^2+(t-\hat{t}_{0})^2=O(1)$, we have the following solution asymptotics
\[ \label{uNManaR1}
\begin{array}{l}
u_{1,0,N}(x,t) = \hat{u}_{1}(x-\hat{x}_{0},t-\hat{t}_{0}) \hspace{0.06cm} e^{{\rm{i}} (k_1x+\omega_{1} t)} + O\left(|a_{m}|^{-1/m}\right),    \\
u_{2,0,N}(x,t) = \hat{u}_{2}(x-\hat{x}_{0},t-\hat{t}_{0}) \hspace{0.06cm} e^{{\rm{i}} (k_2x+\omega_{2} t)} + O\left(|a_{m}|^{-1/m}\right).
\end{array}
\]
\item
If zero is a root of the Okamoto-hierarchy polynomial $R_{N}^{[m]}(z)$, then in the inner region, where $x^2+t^2=O(1)$, $[u_{1,0,N}(x,t), u_{2,0,N}(x,t)]$ is approximately a lower $\left(N_{1R}, N_{2R}\right)$-th order rogue wave $[u_{1,N_{1R},N_{2R}}(x,t),$ $u_{2,N_{1R}, N_{2R}}(x,t)]$ as given in Theorem~3, where $\left(N_{1R}, N_{2R}\right)$ are provided in Theorem~2. Internal parameters $(\hat{a}_{1, 1}, \hat{a}_{2,1}, \hat{a}_{4, 1}, \hat{a}_{5,1}, \dots, \hat{a}_{3N_{1R}-1,\hspace{0.05cm} 1})$ and $(\hat{a}_{1, 2}, \hat{a}_{2,2}, \hat{a}_{4, 2}, \hat{a}_{5,2}, \dots, \hat{a}_{3N_{2R}-2, \hspace{0.05cm} 2})$
in this lower-order rogue wave are related to those in the original rogue wave as
\[
\hat{a}_{j,1}=\hat{a}_{j,2}=a_{j}+(N-N_{1R}-N_{2R})s_j, \quad j=1, 2, 4, 5, \cdots,
\]
where $s_j$ as is defined in Theorem 3 and numerically given in Eq.~(\ref{svalueMan}). The error of this lower-order rogue wave approximation is $O(|a_{m}|^{-1})$. Expressed mathematically, when $|a_{m}|\gg 1$ and $x^2+t^2=O(1)$,
\[ \label{uNManaR2}
\begin{array}{l}
u_{1,0,N}(x,t; a_{2}, a_{4}, a_{5}, \cdots) = u_{1,N_{1R},N_{2R}}(x,t; \hat{a}_{j,1},  \hat{a}_{j,2}, j=1, 2, 4, 5, \dots) + O\left(|a_{m}|^{-1}\right), \\
u_{2,0,N}(x,t; a_{2}, a_{4}, a_{5}, \cdots) = u_{2,N_{1R},N_{2R}}(x,t; \hat{a}_{j,1},  \hat{a}_{j,2}, j=1, 2, 4, 5, \dots)+ O\left(|a_{m}|^{-1}\right).
\end{array}
\]
If zero is not a root of $R_{N}^{[m]}(z)$, then in the inner region, the solution
$[u_{1,0,N}(x, t), u_{2,0,N}(x, t)]$ approaches the constant background $[\rho_{1} e^{{\rm{i}} (k_{1}x + \omega_{1} t)}, \rho_{1}  e^{{\rm{i}} (k_{2}x + \omega_{2} t)}]$ when $|a_m|\gg~1$.
\end{enumerate}
\end{quote}

\vspace{0.1cm}
Theorems 5 and 6 show that, when the internal parameter $|a_m|$ is large, then in the outer region, patterns of Q- and R-type Manakov rogue waves comprise fundamental rogue waves, whose positions are determined by root structures of $Q_{N}^{[m]}(z)$ and $R_{N}^{[m]}(z)$ polynomials through formulae (\ref{x0ManaQ})-(\ref{t0ManaQ}) and (\ref{x0ManaR})-(\ref{t0ManaR}). To the leading order of these positions, i.e., to $O\left(|a_{m}|^{1/m}\right)$, rogue patterns are linear transformations of the underlying root structures. However, the next-order corrections, of $O(1)$, to these leading-order terms, induced by $\Delta_Q$ and $\Delta_R$ in Eqs.~(\ref{x0ManaQ})-(\ref{t0ManaQ}) and (\ref{x0ManaR})-(\ref{t0ManaR}), depend on the root $z_0$ in a nonlinear way (see Eqs.~(\ref{DeltaQform}) and (\ref{DeltaRform}) in later text). These next-order nonlinear corrections will introduce deformations to rogue patterns and make them look different from linear transformations of root structures, as we will see graphically in the next section. This behavior contrasts rogue patterns reported in \cite{NLSRWs2021,Yanguniversal} for some other types of rogue waves, where those patterns are just linear transformations of root structures of the Yablonskii-Vorob'ev polynomial hierarchy, even after next-order position corrections are included. We do note, though, that these nonlinear deformations of rogue patterns in the present case are subdominant compared to the leading-order term, and will become less significant as $|a_m|$ gets larger. In other words, as $|a_m|$ increases, rogue patterns for Q- and R-type Manakov rogue waves will look more and more like the linear transformation of root structures of $Q_{N}^{[m]}(z)$ and $R_{N}^{[m]}(z)$.

Theorems 5 and 6 also show that, when the internal parameter $|a_m|$ is large, then in the inner region, the original rogue wave reduces to a lower-order rogue wave, or to the constant background, depending on whether zero is a root of $Q_{N}^{[m]}(z)$ or $R_{N}^{[m]}(z)$. If zero is a root, then its multiplicity will determine the order of this reduced rogue wave.

A small note we would like to add is regarding the fundamental rogue wave (\ref{u11Mana})-(\ref{u21Mana}) we predicted in the outer regions of Theorems 5 and 6. If we choose the background wavenumbers as $k_2=-k_1$, which is always possible through a Galilean transformation, then $p_0$ would be real, see Eq.~(\ref{p0Mana}). In this case, we can show that this fundamental rogue wave would admit the symmetry of $\hat{u}_{2}(x,t)=\hat{u}_{1}(-x,t)$, i.e., $\hat{u}_{2}(x,t)$ would be a mirror image of $\hat{u}_{1}(x,t)$ around the $t$-axis in the $(x, t)$ plane. This symmetry is clearly visible in the graphs we will present in Sec. \ref{seccompareMan} later.

\subsection{Rogue-pattern results in the three-wave interaction system}
Our results on Q-type and R-type rogue patterns in the three-wave interaction system under a large internal parameter are summarized in the following two theorems.

\begin{quote}
\textbf{Theorem 7} \hspace{0.06cm} For Q-type $N$-th order rogue waves $[u_{1,N, 0}(x,t), u_{2,N,0}(x,t), u_{3,N,0}(x, t)]$ in the three-wave resonant interaction system, suppose $|a_m|\gg 1$ and all other internal parameters are $O(1)$. In addition, suppose all nonzero roots of $Q_{N}^{[m]}(z)$ are simple. Then, the following asymptotics for these rogue waves holds.

\begin{enumerate}
\item
In the outer region, where $\sqrt{x^2+t^2}=O\left(|a_{m}|^{1/m}\right)$, this rogue wave asymptotically separates into $M_{Q}$ fundamental rogue waves, where $M_Q$ is given in Eq.~(\ref{MQMR}). These fundamental rogue waves are $[\hat{u}_{1}(x-\hat{x}_{0}, t-\hat{t}_{0})\hspace{0.06cm} e^{{\rm{i}} (k_1x+\omega_{1} t)}, \hspace{0.07cm} \hat{u}_{2}(x-\hat{x}_{0}, t-\hat{t}_{0})\hspace{0.06cm} e^{{\rm{i}} (k_{2}x + \omega_{2} t)}, \hat{u}_{3}(x-\hat{x}_{0}, t-\hat{t}_{0})\hspace{0.06cm} e^{-{\rm{i}} [(k_1+k_{2})x + (\omega_{1}+\omega_{2}) t]}]$, where
\[\label{u13wave}
\hat{u}_{1}(x,t)=\rho_1 \frac{\hat{g}_{1}}{\hat{f}}, \quad
\hat{u}_{2}(x,t)=\rho_2 \frac{\hat{g}_{2}}{\hat{f}}, \quad
\hat{u}_{3}(x,t)= {\rm{i}}\hspace{0.04cm}\rho_{3} \frac{\hat{g}_{3}}{\hat{f}},
\]
$(\hat{f}, \hat{g}_{1}, \hat{g}_{2}, \hat{g}_{3})$ are obtained from $(f, g_1, g_2, g_3)$ in Eq.~(\ref{SchpolysolufN2}) of Theorem 4 by setting $N_1=0, N_2=1$ and $a_{2,1}=0$, or more explicitly,
\begin{eqnarray*}
&& \hat{f} = \left| \left(\alpha_{1}-\beta_{1}\right) x+(c_{1}\beta_{1}-c_{2}\alpha_{1})t \right|^2+\zeta_{0},  \label{f10} \\
&& \hat{g}_{1}= \left[\left(\alpha_{1}-\beta_{1}\right) x+(c_{1}\beta_{1}-c_{2}\alpha_{1})t + \theta_{1}\right] \left[\left(\alpha_{1}^*-\beta_{1}^*\right) x+(c_{1}\beta_{1}^*-c_{2}\alpha_{1}^*)t - \theta_{1}^* \right]+\zeta_{0},\\
&& \hat{g}_{2}=  \left[\left(\alpha_{1}-\beta_{1}\right) x+(c_{1}\beta_{1}-c_{2}\alpha_{1})t - \lambda_{1}\right] \left[\left(\alpha_{1}^*-\beta_{1}^*\right) x+(c_{1}\beta_{1}^*-c_{2}\alpha_{1}^*)t + \lambda_{1}^* \right]+\zeta_{0}, \\
&& \hat{g}_{3}=  \left[\left(\alpha_{1}-\beta_{1}\right) x+(c_{1}\beta_{1}-c_{2}\alpha_{1})t - \theta_{1}+\lambda_{1}\right] \left[\left(\alpha_{1}^*-\beta_{1}^*\right) x+(c_{1}\beta_{1}^*-c_{2}\alpha_{1}^*)t + \theta_{1}^*-\lambda_{1}^* \right]+\zeta_{0},  \label{g310} \\
&& \alpha_1=-\frac{p_1 \gamma_1}{p_{0}^2  (c_{1}-c_{2})},\ \beta_{1}=-\frac{p_1 \gamma_2}{(p_{0}-{\rm{i}})^2  (c_{1}-c_{2})}, \ \theta_{1}=\frac{p_{1}}{p_{0}-{\rm{i}}}, \ \lambda_{1}=\frac{p_{1}}{p_{0}}, \ \zeta_{0} =\frac{|p_{1}|^2 }{(p_{0}+p_{0}^*)^2}, \label{SpecialNum1}
\end{eqnarray*}
positions $(\hat{x}_{0}, \hat{t}_{0})$ of these fundamental rogue waves are
\begin{eqnarray} \label{x0t03waveQ}
\hat{x}_{0}=\frac{\Im  \left[ \frac{z_{0} a_{m}^{1/m} -\hat{\Delta}_{Q}}{c_{1} \beta_{1}- c_{2}\alpha_{1}} \right]}{\Im  \left[ \frac{\alpha_{1}-\beta_{1}}{c_{1} \beta_{1}- c_{2}\alpha_{1}} \right]}, \quad
\hat{t}_{0}=\frac{\Im  \left[ \frac{z_{0}  a_{m}^{1/m} -\hat{\Delta}_{Q}}{\alpha_{1}-\beta_{1}} \right]}{\Im  \left[ \frac{c_{1} \beta_{1}- c_{2}\alpha_{1}}{\alpha_{1}-\beta_{1}} \right]},
\end{eqnarray}
$z_0$ is any of the $M_Q$ nonzero simple roots of $Q_{N}^{[m]}(z)$, and $\hat{\Delta}_{Q}$ is a $z_0$-dependent $O(1)$ quantity given by Eq.~(\ref{DeltaQhatform}) in later text.
The error of this fundamental rogue wave approximation is $O(|a_{m}|^{-1/m})$. Expressed mathematically, when $|a_{m}|\gg 1$ and $(x-\hat{x}_{0})^2+(t-\hat{t}_{0})^2=O(1)$, we have the following solution asymptotics
\begin{equation}\label{uN3waveQ1}
\begin{array}{l}
u_{1,N, 0}(x,t) = \hat{u}_{1}(x-\hat{x}_{0},t-\hat{t}_{0}) \hspace{0.06cm} e^{{\rm{i}} (k_1x+\omega_{1} t)} + O\left(|a_{m}|^{-1/m}\right),   \\
u_{2,N, 0}(x,t) = \hat{u}_{2}(x-\hat{x}_{0},t-\hat{t}_{0}) \hspace{0.06cm} e^{{\rm{i}} (k_2x+\omega_{2} t)} + O\left(|a_{m}|^{-1/m}\right),    \\
u_{3,N, 0}(x,t) = \hat{u}_{3}(x-\hat{x}_{0},t-\hat{t}_{0}) \hspace{0.06cm} e^{-{\rm{i}} [(k_1+k_{2})x + (\omega_{1}+\omega_{2}) t]}+ O\left(|a_{m}|^{-1/m}\right).
\end{array}
\end{equation}
\item
If zero is a root of the Okamoto-hierarchy polynomial $Q_{N}^{[m]}(z)$, then
in the inner region, where $x^2+t^2=O(1)$, $[u_{1,N,0}(x,t), u_{2,N,0}(x,t), u_{3,N,0}(x,t)]$ is approximately a lower $\left(N_{1Q}, N_{2Q}\right)$-th order rogue wave $[u_{1,N_{1Q}, N_{2Q}}(x,t), u_{2,N_{1Q}, N_{2Q}}(x,t), u_{3,N_{1Q}, N_{2Q}}(x,t)]$ as given in Theorem~4, where $\left(N_{1Q}, N_{2Q}\right)$ are provided in Theorem~1.
Internal parameters $(\hat{a}_{1, 1}, \hat{a}_{2,1}, \hat{a}_{4, 1}, \hat{a}_{5,1}, \dots, \hat{a}_{3N_{1Q}-1,\hspace{0.05cm} 1})$ and $(\hat{a}_{1, 2}, \hat{a}_{2,2}, \hat{a}_{4, 2}, \hat{a}_{5,2}, \dots,$ $\hat{a}_{3N_{2Q}-2, \hspace{0.05cm} 2})$ in this lower-order rogue wave are related to those in the original rogue wave as
\[
\hat{a}_{j,1}=\hat{a}_{j,2}=a_{j}+(N-N_{1Q}-N_{2Q})s_j, \quad j=1, 2, 4, 5, \cdots,
\]
where $s_j$ is as defined in Theorem 4 and numerically given in Eq.~(\ref{svalue3wave}). The error of this lower-order rogue wave approximation is $O(|a_{m}|^{-1})$. Expressed mathematically, when $|a_{m}|\gg 1$ and $x^2+t^2=O(1)$,
\begin{equation}\label{uN3waveQ2}
\begin{array}{l}
u_{1,N, 0}(x,t; a_{2}, a_{4}, a_{5}, \cdots) = u_{1,N_{1Q},N_{2Q}}(x,t; \hat{a}_{j,1},  \hat{a}_{j,2}, j=1, 2, 4, 5, \dots) + O\left(|a_{m}|^{-1}\right), \\
u_{2,N,0}(x,t; a_{2}, a_{4}, a_{5}, \cdots) = u_{2,N_{1Q},N_{2Q}}(x,t; \hat{a}_{j,1},  \hat{a}_{j,2}, j=1, 2, 4, 5, \dots) + O\left(|a_{m}|^{-1}\right), \\
u_{3,N,0}(x,t; a_{2}, a_{4}, a_{5}, \cdots) = u_{3,N_{1Q},N_{2Q}}(x,t; \hat{a}_{j,1},  \hat{a}_{j,2}, j=1, 2, 4, 5, \dots) + O\left(|a_{m}|^{-1}\right).
\end{array}
\end{equation}
If zero is not a root of $Q_{N}^{[m]}(z)$, then in the inner region, the solution
$[u_{1,N,0}(x, t), u_{2,N,0}(x, t), u_{3,N,0}(x, t)]$ approaches the constant background
$[\rho_1 e^{{\rm{i}} (k_1x+\omega_{1} t)}, \rho_2 e^{{\rm{i}} (k_{2}x + \omega_{2} t)}, {\rm{i}} \hspace{0.03cm} \rho_3 e^{-{\rm{i}} [(k_1+k_{2})x + (\omega_{1}+\omega_{2}) t]}]$ when $|a_m|\gg 1$.
\end{enumerate}
\end{quote}

\begin{quote}
\textbf{Theorem 8} \hspace{0.06cm} For R-type $N$-th order rogue waves $[u_{1,0, N}(x,t), u_{2,0,N}(x,t), u_{3,0,N}(x, t)]$ in the three-wave resonant interaction system, suppose $|a_m|\gg 1$ and all other internal parameters are $O(1)$. In addition, suppose all nonzero roots of $R_{N}^{[m]}(z)$ are simple. Then, the following asymptotics for these rogue waves holds.

\begin{enumerate}
\item
In the outer region, where $\sqrt{x^2+t^2}=O\left(|a_{m}|^{1/m}\right)$, this rogue wave asymptotically separates into $M_{R}$ fundamental rogue waves, where $M_R$ is given in Eq.~(\ref{MQMR}). These fundamental rogue waves are $[\hat{u}_{1}(x-\hat{x}_{0}, t-\hat{t}_{0})\hspace{0.06cm} e^{{\rm{i}} (k_1x+\omega_{1} t)}, \hspace{0.07cm} \hat{u}_{2}(x-\hat{x}_{0}, t-\hat{t}_{0})\hspace{0.06cm} e^{{\rm{i}} (k_{2}x + \omega_{2} t)}, \hat{u}_{3}(x-\hat{x}_{0}, t-\hat{t}_{0})\hspace{0.06cm} e^{-{\rm{i}} [(k_1+k_{2})x + (\omega_{1}+\omega_{2}) t]}]$, where functions
$[\hat{u}_{1}(x,t), \hat{u}_{2}(x,t), \hat{u}_{3}(x,t)]$ are as given in Eq.~(\ref{u13wave}), positions $(\hat{x}_{0}, \hat{t}_{0})$ of these fundamental rogue waves are
\begin{eqnarray} \label{x0t03waveR}
\hat{x}_{0}=\frac{\Im  \left[ \frac{z_{0} a_{m}^{1/m} -\hat{\Delta}_{R}}{c_{1} \beta_{1}- c_{2}\alpha_{1}} \right]}{\Im  \left[ \frac{\alpha_{1}-\beta_{1}}{c_{1} \beta_{1}- c_{2}\alpha_{1}} \right]}, \quad
\hat{t}_{0}=\frac{\Im  \left[ \frac{z_{0}  a_{m}^{1/m} -\hat{\Delta}_{R}}{\alpha_{1}-\beta_{1}} \right]}{\Im  \left[ \frac{c_{1} \beta_{1}- c_{2}\alpha_{1}}{\alpha_{1}-\beta_{1}} \right]},
\end{eqnarray}
$z_0$ is any of the $M_R$ nonzero simple roots of $R_{N}^{[m]}(z)$, and $\hat{\Delta}_{R}$ is a $z_0$-dependent $O(1)$ quantity given by Eq.~(\ref{DeltaRhatform}) in the later text.
The error of this fundamental rogue wave approximation is $O(|a_{m}|^{-1/m})$. Expressed mathematically, when $|a_{m}|\gg 1$ and $(x-\hat{x}_{0})^2+(t-\hat{t}_{0})^2=O(1)$, we have the following solution asymptotics
\begin{equation}\label{uN3waveR1}
\begin{array}{l}
u_{1,0, N}(x,t) = \hat{u}_{1}(x-\hat{x}_{0},t-\hat{t}_{0}) \hspace{0.06cm} e^{{\rm{i}} (k_1x+\omega_{1} t)} + O\left(|a_{m}|^{-1/m}\right),   \\
u_{2,0, N}(x,t) = \hat{u}_{2}(x-\hat{x}_{0},t-\hat{t}_{0}) \hspace{0.06cm} e^{{\rm{i}} (k_2x+\omega_{2} t)} + O\left(|a_{m}|^{-1/m}\right),    \\
u_{3,0, N}(x,t) = \hat{u}_{3}(x-\hat{x}_{0},t-\hat{t}_{0}) \hspace{0.06cm} e^{-{\rm{i}} [(k_1+k_{2})x + (\omega_{1}+\omega_{2}) t]}+ O\left(|a_{m}|^{-1/m}\right).
\end{array}
\end{equation}
\item
If zero is a root of the Okamoto-hierarchy polynomial $R_{N}^{[m]}(z)$, then
in the inner region, where $x^2+t^2=O(1)$, $[u_{1,0,N}(x,t), u_{2,0,N}(x,t), u_{3,0,N}(x,t)]$ is approximately a lower $\left(N_{1R}, N_{2R}\right)$-th order rogue wave $[u_{1,N_{1R}, N_{2R}}(x,t), u_{2,N_{1R}, N_{2R}}(x,t), u_{3,N_{1R}, N_{2R}}(x,t)]$ as given in Theorem~4, where $\left(N_{1R}, N_{2R}\right)$ are provided in Theorem~2.
Internal parameters $(\hat{a}_{1, 1}, \hat{a}_{2,1}, \hat{a}_{4, 1}, \hat{a}_{5,1}, \dots, \hat{a}_{3N_{1R}-1,\hspace{0.05cm} 1})$ and $(\hat{a}_{1, 2}, \hat{a}_{2,2}, \hat{a}_{4, 2}, \hat{a}_{5,2}, \dots,$ $\hat{a}_{3N_{2R}-2, \hspace{0.05cm} 2})$
in this lower-order rogue wave are related to those in the original rogue wave as
\[
\hat{a}_{j,1}=\hat{a}_{j,2}=a_{j}+(N-N_{1R}-N_{2R})s_j, \quad j=1, 2, 4, 5, \cdots,
\]
where $s_j$ is as defined in Theorem 4 and numerically given in Eq.~(\ref{svalue3wave}).
The error of this lower-order rogue wave approximation is $O(|a_{m}|^{-1})$. Expressed mathematically, when $|a_{m}|\gg 1$ and $x^2+t^2=O(1)$,
\begin{equation}\label{uN3waveR2}
\begin{array}{l}
u_{1,0, N}(x,t; a_{2}, a_{4}, a_{5}, \cdots) = u_{1,N_{1R},N_{2R}}(x,t; \hat{a}_{j,1},  \hat{a}_{j,2}, j=1, 2, 4, 5, \dots) + O\left(|a_{m}|^{-1}\right), \\
u_{2,0,N}(x,t; a_{2}, a_{4}, a_{5}, \cdots) = u_{2,N_{1R},N_{2R}}(x,t; \hat{a}_{j,1},  \hat{a}_{j,2}, j=1, 2, 4, 5, \dots) + O\left(|a_{m}|^{-1}\right), \\
u_{3,0,N}(x,t; a_{2}, a_{4}, a_{5}, \cdots) = u_{3,N_{1R},N_{2R}}(x,t; \hat{a}_{j,1},  \hat{a}_{j,2}, j=1, 2, 4, 5, \dots) + O\left(|a_{m}|^{-1}\right).
\end{array}
\end{equation}
If zero is not a root of $R_{N}^{[m]}(z)$, then in the inner region, the solution
$[u_{1,0,N}(x, t), u_{2,0,N}(x, t), u_{3,0,N}(x, t)]$ approaches the constant background
$[\rho_1 e^{{\rm{i}} (k_1x+\omega_{1} t)}, \rho_2 e^{{\rm{i}} (k_{2}x + \omega_{2} t)}, {\rm{i}}
\hspace{0.03cm} \rho_3 e^{-{\rm{i}} [(k_1+k_{2})x + (\omega_{1}+\omega_{2}) t]}]$ when $|a_m|\gg 1$.
\end{enumerate}
\end{quote}

Similar to the Manakov case, patterns of Q- and R-type rogue waves in the three wave interaction system are linear transformations of root structures of $Q_{N}^{[m]}(z)$ and $R_{N}^{[m]}(z)$ to the leading order, but are nonlinear transformations of those root structures when the next-order  position corrections are included.

Proofs of these four theorems will be presented in Sec. \ref{sec:Manaproof}.

\section{Comparison between analytical predictions and true rogue solutions}
In this section, we compare our analytical predictions of rogue patterns in Theorems 5-8 to true rogue solutions in the Manakov and three-wave-interaction systems.

\subsection{Comparison in the Manakov system}  \label{seccompareMan}
For the Manakov system, we choose background wavenumbers $k_1=-k_2=1/\sqrt{12}$. Then background amplitudes are obtained from conditions (\ref{CubicCondition}) as $\rho_1=\rho_2=\sqrt{2/3}$, and background wave frequencies can be obtained from equations~(\ref{manaomega}).

\subsubsection{Q-type}
\begin{figure}[htb]
\begin{center}
\includegraphics[scale=0.60, bb=60 000 385 450]{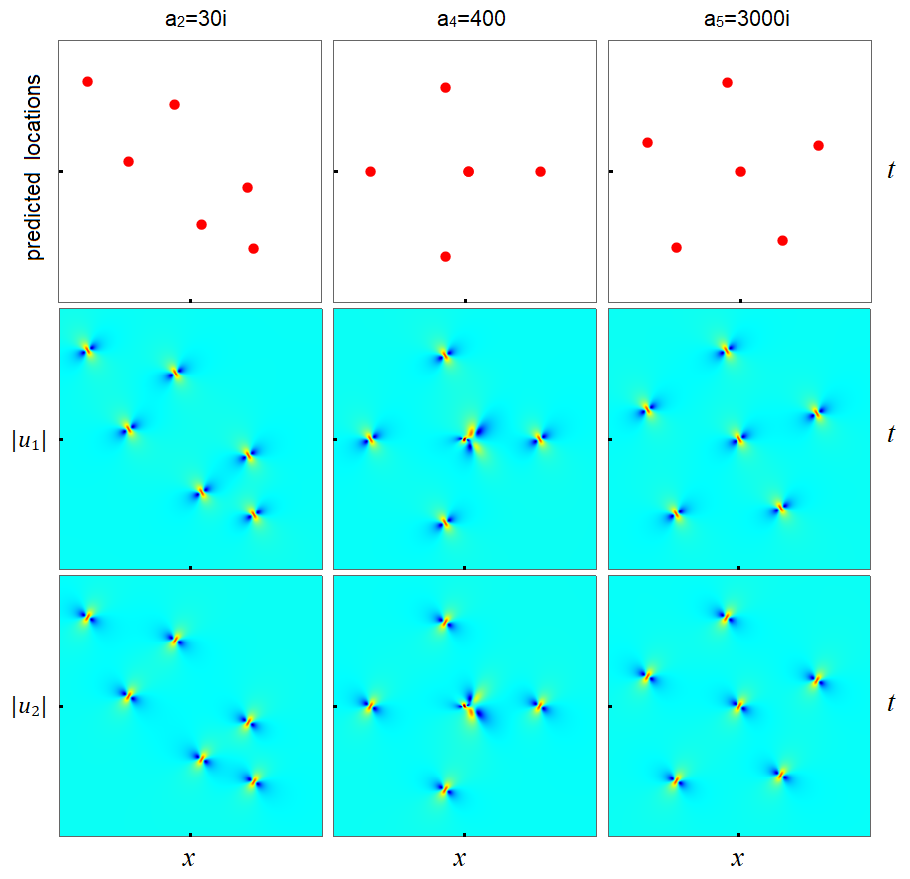}
\caption{Predicted Q-type second-order Manakov rogue waves from Theorem 5. Each column is for a rogue wave with a single large parameter $a_m$, whose value is indicated on top, and all other internal parameters are set as zero. Top row: predicted $(\hat{x}_0, \hat{t}_0)$ locations by formulae (\ref{x0ManaQ})-(\ref{t0ManaQ}) applied to all roots of $Q_{2}^{[m]}(z)$. Middle row: predicted $|u_1(x,t)|$. Bottom row: predicted $|u_2(x, t)|$. In all panels, the $(x, t)$ internals are $ -40 \le x, t \le 40$.   \label{f:ManQpre}}
\end{center}
\end{figure}

First, we consider Q-type Manakov rogue waves. Specifically, we take $N=2$; thus these are second-order waves with three internal parameters $(a_2, a_4, a_5)$. We set one of these parameters large and the other parameters zero. Then, when that large parameter is chosen as one of
\[ \label{para_ManaQ}
a_{2}=30{\rm{i}}, \quad a_{4}=400, \quad a_{5}=3000{\rm{i}},
\]
the three predicted rogue waves from Theorem 5 are displayed in the three columns of Fig.~\ref{f:ManQpre}, respectively. The top row of this figure shows the predicted $(\hat{x}_0, \hat{t}_0)$ locations by formulae (\ref{x0ManaQ})-(\ref{t0ManaQ}) applied to all roots of $Q_{2}^{[m]}(z)$. In these formulae, $p_0=1/2$ from Eq.~(\ref{p0Mana}), $(p_1, p_2)=\left(12^{-1/3}, 144^{-1/3} \right)$ from Eq.~(\ref{defpk}), and $\Delta_Q$ is calculated from Eq.~(\ref{DeltaQform}). Note that these $(\hat{x}_0, \hat{t}_0)$ predictions contain not only the dominant $O(a_m^{1/m})$ contribution, but also the subdominant $O(1)$ contribution.

According to Theorem 5, at each of the $(\hat{x}_{0}, \hat{t}_{0})$ locations obtained from formulae (\ref{x0ManaQ})-(\ref{t0ManaQ}) for nonzero roots of $Q_{2}^{[m]}(z)$, a fundamental Manakov rogue wave $[\hat{u}_{1}(x-\hat{x}_{0}, t-\hat{t}_{0})\hspace{0.06cm} e^{{\rm{i}} (k_1x+\omega_{1} t)}, \hspace{0.07cm} \hat{u}_{2}(x-\hat{x}_{0}, t-\hat{t}_{0})\hspace{0.06cm} e^{{\rm{i}} (k_{2}x + \omega_{2} t)}]$ is predicted, where $[\hat{u}_{1}(x,t), \hat{u}_{2}(x,t)]$ are as given in Eqs.~(\ref{u11Mana})-(\ref{u21Mana}). The amplitude fields $|\hat{u}_{1}|$ and $|\hat{u}_{2}|$ of these fundamental rogue waves are plotted in the middle and bottom rows of Fig.~\ref{f:ManQpre}, respectively.

Theorem 5 also predicts that, if zero is a root of $Q_{2}^{[m]}(z)$, as is the case for $m=4$ and 5, then in the inner region, i.e., the region near the $(\hat{x}_{0}, \hat{t}_{0})$ location from formulae (\ref{x0ManaQ})-(\ref{t0ManaQ}) for $z_0=0$, a lower $(N_{1Q}, N_{2Q})$-th order rogue wave would appear. These $(N_{1Q}, N_{2Q})$ values are calculated from Theorem 1 as
\[ \label{N12Q}
(N_{1Q}, N_{2Q})=(0,0), \hspace{0.1cm} (1,1), \hspace{0.1cm} (0,1),
\]
for the three solutions in Fig.~\ref{f:ManQpre}, respectively. The first set of $(0,0)$ indicates that zero is not a root of $Q_{2}^{[2]}(z)$, hence no lower-order rogue wave in the inner region. The third set of $(0,1)$ indicates that the lower-order rogue wave in the inner region is a fundamental rogue wave, while the second set of $(1,1)$ indicates that the rogue wave in the inner region is a non-fundamental rogue wave. Internal parameters in these predicted lower $(N_{1Q},N_{2Q})$-th order rogue waves are all zero, due to our choices of internal parameters in the original rogue waves and
the $s_j$ values shown in Eq.~(\ref{svalueMan}). Plotting these $(N_{1Q},N_{2Q})$-th order rogue waves, we get the predicted center-region solutions in the middle and bottom rows of Fig.~\ref{f:ManQpre}.

Looking at these predicted rogue solutions in Fig.~\ref{f:ManQpre}, we see that the large-$a_2$  solution exhibits a skewed double-triangle, reminiscent of the double-triangle root structure of $Q_{2}^{[2]}(z)$ in Fig.~1. The large-$a_4$ solution exhibits a square, reminiscent of the square-shaped root structure of $Q_{2}^{[4]}(z)$ in Fig.~1. The large-$a_5$ solution exhibits a pentagon, reminiscent of the pentagon-shaped root structure of $Q_{2}^{[5]}(z)$ in Fig.~1. This pentagon-shaped rogue pattern has been seen in the NLS and other equations before \cite{KAAN2011,GLML2012,OhtaJY2012,Yanguniversal}, but the double-triangle and square patterns are new.

\begin{figure}[htb]
\begin{center}
\includegraphics[scale=0.60, bb=170 000 285 300]{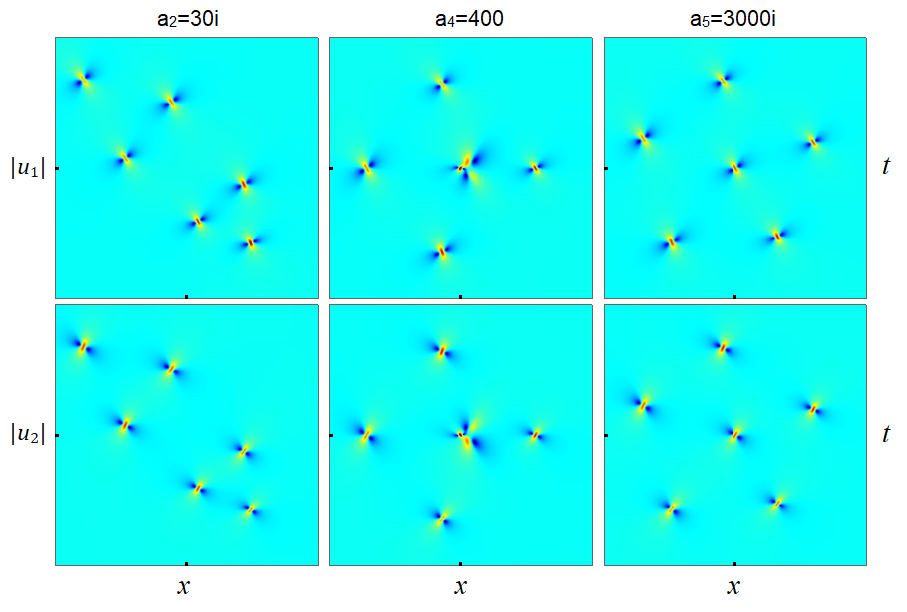}
\caption{True Q-type second-order Manakov rogue waves for the same parameters and $(x, t)$ intervals as in Fig.~\ref{f:ManQpre}.  \label{f:ManQtr} }
\end{center}
\end{figure}

Now, we compare these predictions to true solutions. The corresponding true solutions are plotted directly from Theorem 3 and displayed in Fig.~\ref{f:ManQtr}. Comparing these true solution graphs with the predicted ones in Fig.~\ref{f:ManQpre}, they clearly match each other very well.

\begin{figure}[htb]
\begin{center}
\includegraphics[scale=0.5, bb=420 000 385 280]{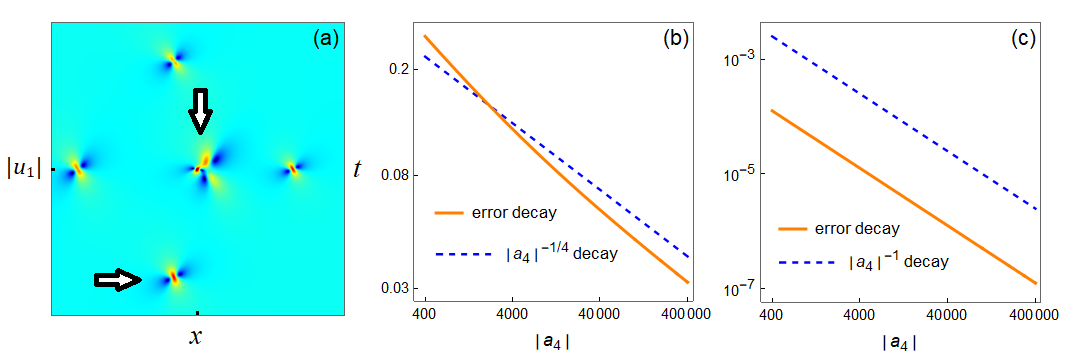}
\caption{Decay of errors in our predictions of Theorem 5 for the outer and inner regions of the Q-type second-order Manakov rogue wave with various large real values of $a_4$, while the other internal parameters are set as zero. (a) $|u_1(x,t)|$ of the true rogue wave with $a_4=400$. (b) Decay of error versus $a_4$ for the outer fundamental rogue wave marked by the lower arrow in panel (a), together with the $|a_4|^{-1/4}$ decay for comparison. (c) Decay of error versus $a_4$ at $x=t=0$ of the inner region marked by the upper arrow in panel (a), together with the $|a_4|^{-1}$ decay for comparison. \label{f:errordecay} }
\end{center}
\end{figure}

To quantitatively compare our prediction with the true solution and verify Theorem 5's error decay rates with the large parameter $a_m$, we choose $a_4$ to be the large parameter, corresponding to the second-column solution in Figs.~\ref{f:ManQpre} and \ref{f:ManQtr}. For simplicity, we choose all $a_4$ to be real. As before, the other two internal parameters $(a_2, a_5)$ in the rogue wave will be set as zero. We will vary this $a_4$ value, from $400$ to $400000$, and for each value, we measure the errors of our prediction in the outer and inner regions and then plot these errors versus $a_4$. In the outer region, this error is defined as the distance in the $(x, t)$ plane between the predicted and true positions of the fundamental rogue wave marked by the lower arrow in panel (a) of Fig.~\ref{f:errordecay}. In the inner region, marked by the upper arrow in panel (a), the error is defined as the magnitude of the difference between the predicted and true solution values at the origin $x=t=0$. These error curves are plotted in panels (b) and (c), for the outer and inner regions, respectively. For comparison, decay rates of $|a_4|^{-1/4}$ and $|a_4|^{-1}$ are also plotted in the corresponding panels. These error curves clearly show that, the error decay rate is $|a_4|^{-1/4}$ in the outer region and $|a_4|^{-1}$ in the inner region, which fully agree with our theoretical predictions in Theorem 5.

\subsubsection{R-type}
\begin{figure}[htb]
\begin{center}
\includegraphics[scale=0.6, bb=210 000 385 450]{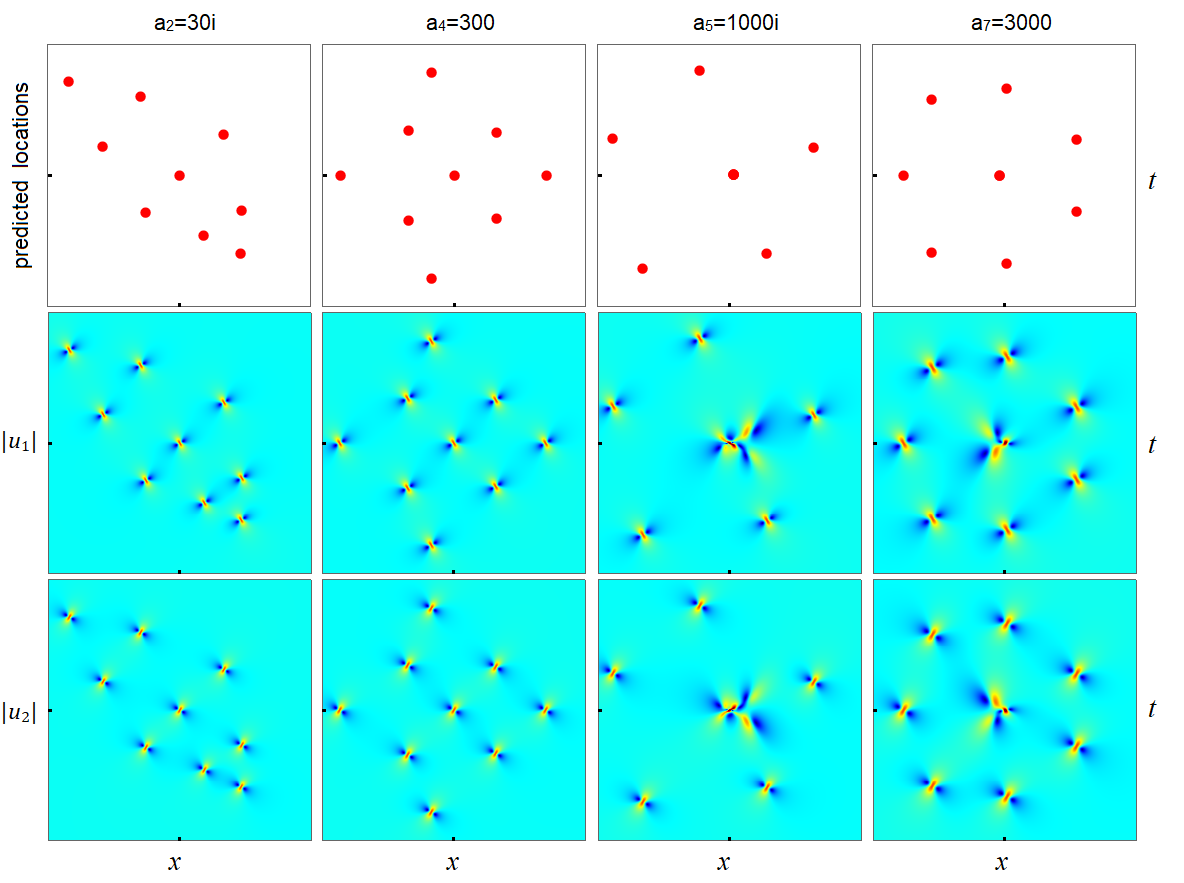}
\caption{Predicted R-type third-order Manakov rogue waves from Theorem 6. Each column is for a rogue wave with a single large parameter $a_m$, whose value is indicated on top, and all other internal parameters are set as zero. Top row: predicted $(\hat{x}_0, \hat{t}_0)$ locations by formulae (\ref{x0ManaR})-(\ref{t0ManaR}) applied to all roots of $R_{3}^{[m]}(z)$. Middle row: predicted $|u_1(x,t)|$. Bottom row: predicted $|u_2(x, t)|$.  The $(x, t)$ internals in the four columns are $-46 \le x, t \le 46$, $ -41 \le x, t \le 41 $, $-35 \le x, t \le 35$, and $-28 \le x, t \le 28$, respectively. \label{f:ManRpre}}
\end{center}
\end{figure}

Next, we compare R-type rogue waves in the Manakov system. Here, we set $N=3$. Thus, these are third-order waves with internal parameters $(a_2, a_4, a_5, a_7)$. We choose one of these parameters large and the other parameters zero. Then, when that large parameter is chosen as one of
\[ \label{para_ManaR}
a_{2}=30{\rm{i}}, \quad a_{4}=300, \quad a_{5}=1000{\rm{i}}, \quad a_{7}=3000,
\]
the four predicted rogue waves from Theorem 6 are displayed in the four columns of Fig.~\ref{f:ManRpre}, respectively. The top row of this figure shows the predicted $(\hat{x}_0, \hat{t}_0)$ locations by formulae (\ref{x0ManaR})-(\ref{t0ManaR}) applied to all roots of $R_{3}^{[m]}(z)$. At each of the $(\hat{x}_0, \hat{t}_0)$ locations resulting from nonzero roots of $R_{3}^{[m]}(z)$, Theorem 6 predicts a fundamental Manakov rogue wave, whose amplitude fields $|\hat{u}_{1}|$ and $|\hat{u}_{2}|$ are plotted in the middle and bottom rows of Fig.~\ref{f:ManRpre}, respectively. Our prediction for the center regions in these rows is based on Eq.~(\ref{uNManaR2}) of Theorem 6. In this prediction, the $(N_{1R}, N_{2R})$ values for these four rogue solutions are obtained from Theorem 2 as
\[ \label{N12R}
(N_{1R}, N_{2R})=(0,1), \hspace{0.1cm} (0,1), \hspace{0.1cm} (1,2), \hspace{0.1cm} (1,0),
\]
respectively. These values show that the center region of the first two rogue solutions hosts a fundamental rogue wave, while that region in the last two rogue solutions hosts a non-fundamental rogue wave. Internal parameters in these predicted lower $(N_{1R},N_{2R})$-th order rogue waves of the center region are all zero, due to our choices of internal parameters in the original rogue waves as well as the $s_j$ values shown in Eq.~(\ref{svalueMan}). Plotting these $(N_{1R},N_{2R})$-th order rogue waves from Theorem 3, we get the predicted center-region solutions in the middle and bottom rows of Fig.~\ref{f:ManRpre}.

These predicted rogue solutions in Fig.~\ref{f:ManRpre} exhibit various patterns, such as a skewed and deformed rhombus (first column), a deformed square (second column), a deformed pentagon (third column), and a heptagon (last column). Of these patterns, rhombus-shaped and square-shaped ones are new.

\begin{figure}[htb]
\begin{center}
\includegraphics[scale=0.6, bb=310 000 285 300]{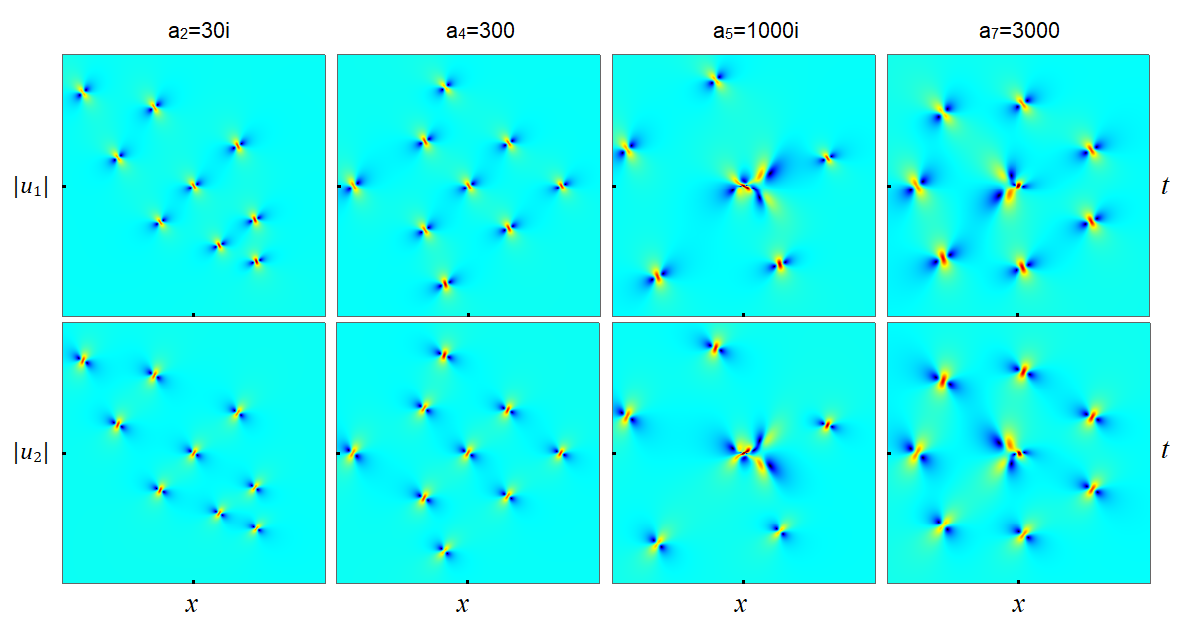}
\caption{True R-type third-order Manakov rogue waves for the same parameters and $(x, t)$ intervals as in Fig.~\ref{f:ManRpre}. \label{f:ManRtr} }
\end{center}
\end{figure}

Now, we compare these predictions to true solutions. The corresponding true solutions are plotted directly from Theorem 3 and displayed in Fig.~\ref{f:ManRtr}. These true solutions clearly match the predicted ones in Fig.~\ref{f:ManRpre} very well.

In addition to this visual agreement, we have also performed error analysis for these R-type waves, similar to what we have done for Q-type waves in Fig.~\ref{f:errordecay}. This error analysis confirmed the error decay rates we predicted in Theorem 6 for the outer and inner regions. Details are omitted for brevity.

\subsubsection{Higher-order Manakov rogue patterns}
Rogue patterns we have seen in Figs.~\ref{f:ManQpre}-\ref{f:ManQtr} for $N=2$ and Figs.~\ref{f:ManRpre}-\ref{f:ManRtr} for $N=3$ are relatively simple. Using Theorems 5-6 and root structures of $Q_{N}^{[m]}(z)$ and $R_{N}^{[m]}(z)$ polynomials in Figs.~\ref{f:rootsQN}-\ref{f:rootsRN}, we can predict more complex Manakov rogue patterns by increasing the order $N$.

\begin{figure}[htb]
\begin{center}
\includegraphics[scale=0.6, bb=210 000 385 320]{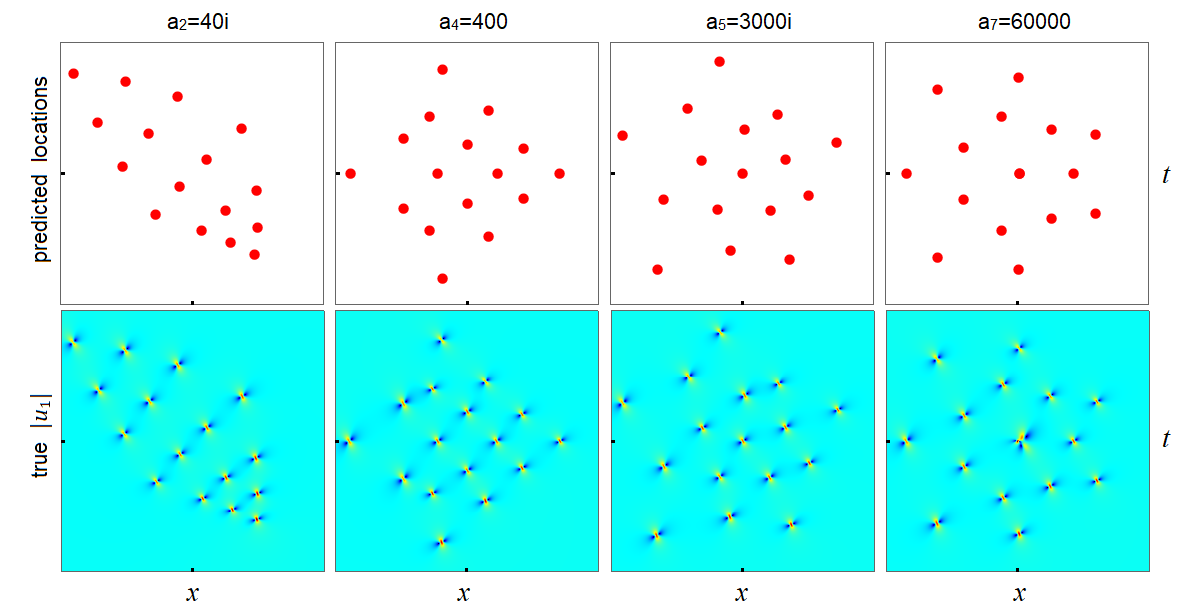}
\caption{Comparison between predicted rogue patterns and true solutions for R-type fourth-order Manakov rogue waves with a single large parameter $a_m$, as shown on the top of each column, and other internal parameters are set as zero. Upper row: predicted $(\hat{x}_0, \hat{t}_0)$ locations by formulae (\ref{x0ManaR})-(\ref{t0ManaR}) applied to all roots of $R_{4}^{[m]}(z)$. Lower row: true $|u_1(x, t)|$ solutions. In all panels, $-64 \le x, t\le 64$.  \label{f:RN45}}
\end{center}
\end{figure}

As an example, we consider fourth-order R-type Manakov rogue waves. We take a single large parameter $a_m$ as one of
\[ \label{para_Mana2}
a_{2}=40 \textrm{i}, \quad a_{4}=400, \quad a_{5}=3000 \textrm{i}, \quad a_{7}=60000,
\]
and the other internal parameters are set as zero. Then, our predictions of rogue wave locations $(\hat{x}_0, \hat{t}_0)$ from formulae (\ref{x0ManaR})-(\ref{t0ManaR}) of Theorem~6 for roots of the $R_{4}^{[m]}(z)$ polynomials are plotted in the upper row of Fig.~\ref{f:RN45}. According to Theorem~6, each $(\hat{x}_0, \hat{t}_0)$ location away from the pattern center hosts a fundamental rogue wave. The $(\hat{x}_0, \hat{t}_0)$ location near the pattern center, generated by the zero root $z_0=0$ of $R_{4}^{[m]}(z)$ and appearing in the large-$a_5$ and $a_7$ panels only, signals a rogue wave of lower order $(N_{1R}, N_{2R})$, whose values are $(0,1)$ and $(2,2)$ for the large-$a_5$ and $a_7$ cases, respectively. This means that at the pattern center of large-$a_5$ and $a_7$ panels, a fundamental rogue wave and a non-fundamental $(2,2)$-th order rogue wave are predicted respectively.

To verify these predictions, true solutions are plotted in the lower row of Fig.~\ref{f:RN45}. The agreement between predicted and true solutions is obvious.

Predictions and confirmations for higher-order Q-type Manakov rogue waves can also be obtained. In this case, a little caution is warranted. As one can see from Fig.~1, a distinctive feature of some $Q_{N}^{[m]}(z)$ root structures is that some nonzero roots are extremely close to each other, see the $m=4$ column with $N=3$ and 4. When that happens, in order for our asymptotic theory to hold, the $|a_m|$ value would have to be chosen very large, so that the $(\hat{x}_0, \hat{t}_0)$ locations from formulae (\ref{x0ManaR})-(\ref{t0ManaR}) for those extremely close roots can be well separated in the $(x, t)$ plane in order for them to host a fundamental rogue wave each.

\subsubsection{Effect of parameter size on rogue shapes}  \label{sec_paraeffect}
From the above comparisons, we have established that Manakov rogue patterns can be accurately predicted by root structures of Okamoto-hierarchy polynomials through mappings (\ref{x0ManaQ})-(\ref{t0ManaQ}) and (\ref{x0ManaR})-(\ref{t0ManaR}). The reader may have noticed that, rogue shapes in the above figures are often twisted and less orderly, even though their corresponding root structures of Okamoto-hierarchy polynomials are. For example, in the R-type third-order rogue wave of Figs.~\ref{f:ManRpre}-\ref{f:ManRtr} with large $a_4$, the upper-left and lower-left sides of rogue patterns are strongly bent in, resulting in an irregular square, but the corresponding root structure of $R_{3}^{[4]}(z)$ in Fig.~2 is a regular square.

The reason for this irregularity in Manakov rogue patterns is apparently due to the next-order correction term in mappings (\ref{x0ManaQ})-(\ref{t0ManaQ}) and (\ref{x0ManaR})-(\ref{t0ManaR}) from the root structure of Okamoto-hierarchy polynomials to rogue positions in the $(x, t)$ plane. While the leading term in those formulae is a linear mapping, the next-order correction term is a nonlinear mapping in view of formulae (\ref{DeltaQform}) and (\ref{DeltaRform}). This nonlinear part of the mappings causes deformations in rogue shapes and makes them irregular even if the underlying root structures are.

\begin{figure}[htb]
\begin{center}
\includegraphics[scale=0.60, bb=170 000 285 310]{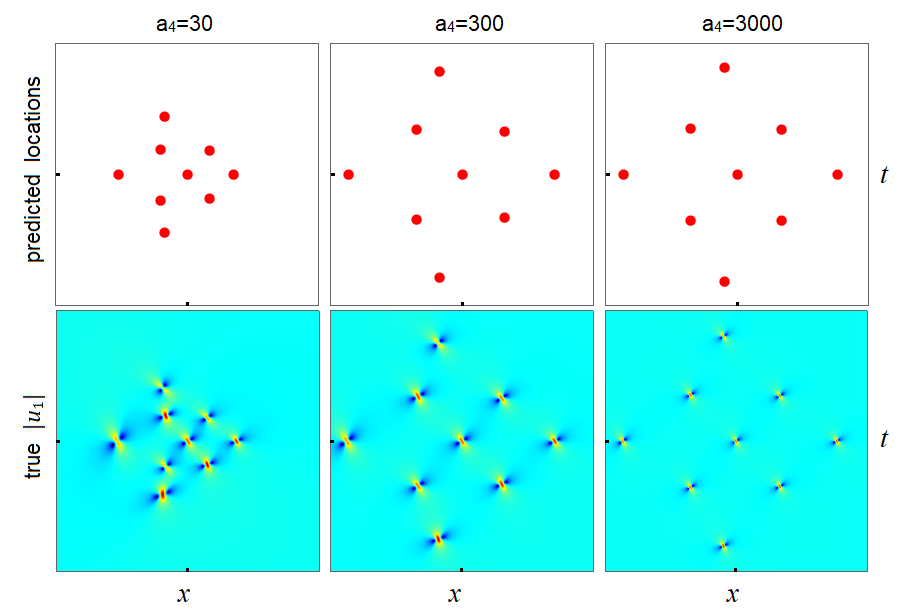}
\caption{Effect of parameter size $a_4$ on R-type third-order Manakov rogue shapes (all other internal parameters are set as zero). Upper row: predicted $(\hat{x}_0, \hat{t}_0)$ locations by formulae (\ref{x0ManaR})-(\ref{t0ManaR}) applied to all roots of $R_{3}^{[4]}(z)$. Lower row: true $|u_1(x, t)|$. The $(x, t)$ intervals in the three columns are $-41 \le x, t \le 41$, $ -41 \le x, t \le 41$, and $ -70 \le x, t \le 70$, respectively.   \label{f:ManQam} }
\end{center}
\end{figure}

It is important to recognize that, this next-order correction term is subdominant, and its effect will get weaker when $|a_m|$ gets larger. Thus, if we increase $|a_m|$, this irregularity in rogue shape would diminish, and the rogue pattern would approach a linearly transformed root structure of Okamoto-hierarchy polynomials, which would be regular if the underlying root structure is. To confirm this prediction, we take that R-type third-order rogue wave of Figs.~\ref{f:ManRpre}-\ref{f:ManRtr} with large $a_4$, and vary its $a_4$ value, with other internal parameters still set as zero. For three $a_4$ values of 30, 300 and 3000, predicted rogue locations $(\hat{x}_0, \hat{t}_0)$ from formulae (\ref{x0ManaR})-(\ref{t0ManaR}) of Theorem~6 are plotted in the upper row of Fig.~\ref{f:ManQam}, and true solutions (only the $|u_1|$ part) are plotted in the lower row. We see that when $a_4=30$,
both the predicted and true solutions are highly irregular, almost random-like. But as $a_4$ increases to 300, this irregularity is significantly reduced and is visible only at the upper-left and lower-left sides of the figure. When $a_4$ further increases to 3000, this irregularity is almost completely gone, and the rogue shape closely resembles the root structure of $R_{3}^{[4]}(z)$.

\subsection{Comparison in the three-wave system}
Now, we consider the three-wave system (\ref{3WRIModel}). In this system, we choose velocity values as $(c_1, c_2, c_3)=(1, 9/20, 0)$, and the first wave's background amplitude $\rho_1=1$. Then, the other two waves' background amplitudes can be derived from Eq.~(\ref{CubicRestrict}) as $\rho_2=2 \sqrt{5}/3$ and $\rho_3=\sqrt{11}/3$ (we have taken the plus signs). Wave numbers and frequencies of the three background waves can be determined from Eq.~(\ref{Pararlation}).

\subsubsection{Q-type}
\begin{figure}[htb]
\begin{center}
\includegraphics[scale=0.6, bb=150 000 285 590]{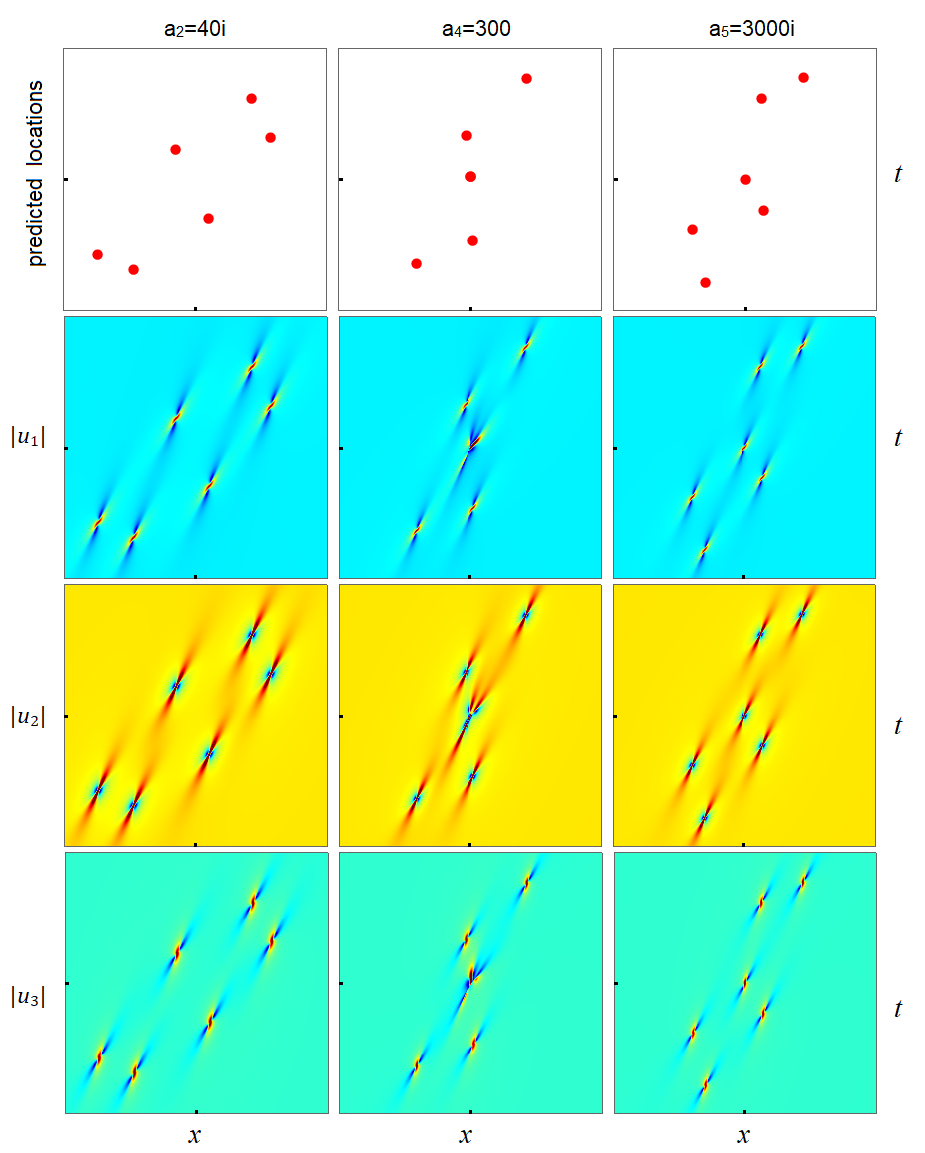}
\caption{Predicted Q-type second-order rogue waves from Theorem~7 in the three-wave system. Each column shows a predicted rogue wave with a single large parameter $a_m$, whose value is indicated on top, and all other internal parameters are set as zero. First row: predicted $(\hat{x}_0, \hat{t}_0)$ locations from formulae (\ref{x0t03waveQ}) applied to all roots of $Q_{2}^{[m]}(z)$. Second row: predicted $|u_1(x,t)|$. Third row: predicted $|u_2(x, t)|$. Last row: predicted $|u_3(x, t)|$. The $(x, t)$ internals in the three columns are $ -19 \le x, t \le 19$, $ -23 \le x, t \le 23$, and $ -25 \le x, t \le 25$, respectively. \label{f:3waveQpre}  }
\end{center}
\end{figure}

We first compare Q-type rogue waves of the three-wave system and set $N=2$. Regarding their three internal parameters $(a_{2}, a_{4}, a_{5})$, we choose one of them large and the other two zero. Then, when that large parameter is taken as one of
\[ \label{para_3WQ}
a_{2}=40{\rm{i}}, \quad a_{4}=300, \quad a_{5}=3000{\rm{i}},
\]
the three predicted rogue waves from Theorem 7 are displayed in the three columns of Fig.~\ref{f:3waveQpre}, respectively. The first row of this figure shows the predicted $(\hat{x}_0, \hat{t}_0)$ locations from formulae (\ref{x0t03waveQ}) applied to all roots of $Q_{2}^{[m]}(z)$. In these formulae, $(\alpha_1, \beta_1)\approx (-1.2632+1.8990{\rm{i}}, 0.4558+0.9195{\rm{i}})$ from the expansions (\ref{schucoefalpha2})-(\ref{schucoefbeta2}), and $\hat{\Delta}_Q$ is calculated from Eq.~(\ref{DeltaQhatform}). At each of the $(\hat{x}_{0}, \hat{t}_{0})$ locations resulting from nonzero roots of $Q_{2}^{[m]}(z)$, a fundamental rogue wave of the three-wave system is predicted, whose amplitude fields $(|\hat{u}_1|, |\hat{u}_2|, |\hat{u}_3|)$ are plotted in the second to fourth rows of Fig.~\ref{f:3waveQpre}, respectively. Our prediction for the center regions of these rows is based on Eq.~(\ref{uN3waveQ2}) of Theorem 7. In this prediction, the $(N_{1Q}, N_{2Q})$ values for these three rogue waves are the same as those given in Eq.~(\ref{N12Q}) earlier. Internal parameters in these predicted lower $(N_{1Q},N_{2Q})$-th order rogue waves in the center region are all zero, due to our choices of internal parameters in the original rogue waves and
the $s_j$ values shown in Eq.~(\ref{svalue3wave}). Plotting these $(N_{1Q},N_{2Q})$-th order rogue waves from Theorem 4, we get the predicted center-region solutions in the second to fourth rows of Fig.~\ref{f:3waveQpre}.

It is easy to see that these predicted rogue patterns in Fig.~\ref{f:3waveQpre}, although being produced from the root structures of $Q_{2}^{[m]}(z)$ polynomials in Fig.~1, look totally different from those root structures. The reason is the nonlinear mapping of the next-order correction term in formulae (\ref{x0t03waveQ}), which induces strong deformations to the linearly mapped result from the leading-order term in (\ref{x0t03waveQ}). These deformations, under our current velocity choices of $(c_1, c_2, c_3)$, are much stronger than in the previous Manakov case, at comparable $a_m$ values. As we have explained in Sec.~\ref{sec_paraeffect} earlier, if we significantly increase the $|a_m|$ values, these deformations will become weaker, and rogue patterns will approach linearly transformed root structures of Okamoto-hierarchy polynomials and will thus be more recognizable

\begin{figure}[htb]
\begin{center}
\includegraphics[scale=0.6, bb=150 000 285 440]{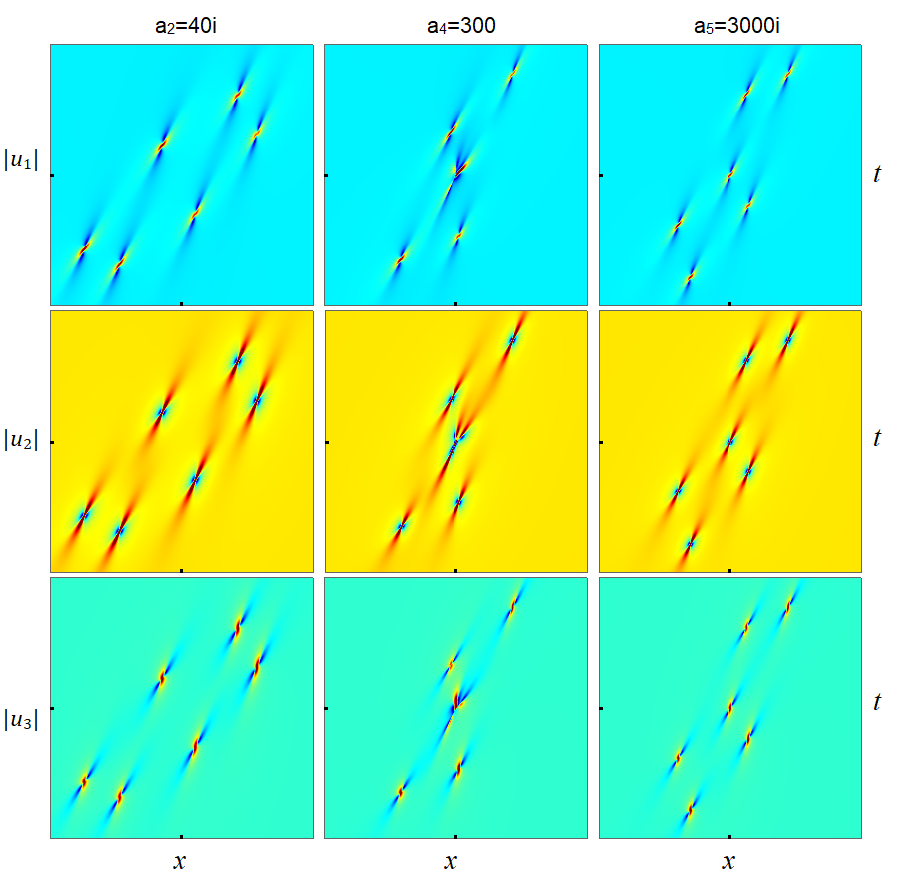}
\caption{True Q-type second-order rogue waves of the three-wave system for the same parameters and $(x, t)$ intervals as in Fig.~\ref{f:3waveQpre}. \label{f:3waveQtr}}
\end{center}
\end{figure}

To compare these predictions to true solutions, we plot in Fig.~\ref{f:3waveQtr} the corresponding true solutions from Theorem 4. It is easy to see that the agreement is excellent, confirming the validity of Theorem 7. This agreement also indicates that, predictions from our Theorem 7 are highly accurate, even when rogue patterns are strongly deformed from Okamoto root structures.

\subsubsection{R-type}
\begin{figure}[htb]
\begin{center}
\includegraphics[scale=0.6, bb=300 000 285 420]{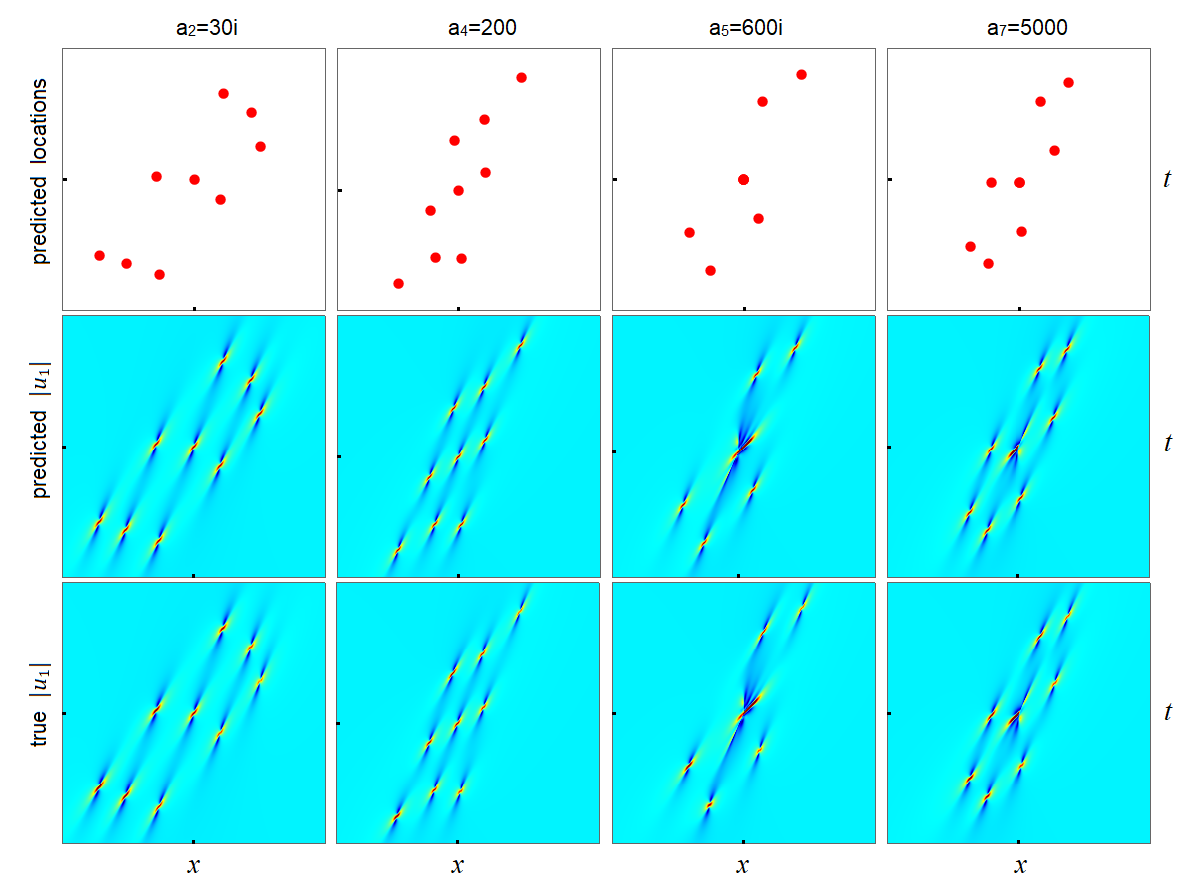}
\caption{Comparison between predicted and true R-type third-order rogue waves of the three-wave system. Each column is for a rogue wave with a single large parameter $a_m$, whose value is indicated on top, and all other internal parameters are set as zero. Top row: predicted $(\hat{x}_0, \hat{t}_0)$ locations by formulae (\ref{x0t03waveR}) applied to all roots of $R_{3}^{[m]}(z)$. Middle row: predicted $|u_1(x,t)|$. Third row: true $|u_1(x,t)|$. The $(x, t)$ internals in the four columns are $-21 \le x, t \le 21 $, $ -23 \le x, t \le 27 $, $-24 \le x, t \le 24$, and $-23 \le x, t \le 23$, respectively.  \label{f:3waveR}}
\end{center}
\end{figure}

Next, we consider R-type rogue waves, and set $N=3$. Regarding their internal parameters $(a_{2}, a_{4}, a_{5}, a_{7})$, we choose one of them large, and the others zero. Then, when that large parameter $a_m$ is taken as one of
\[ \label{para_3WR}
a_{2}=30{\rm{i}}, \quad a_{4}=200, \quad a_{5}=600{\rm{i}}, \quad a_{7}=5000,
\]
predicted rogue waves from Theorem 8 are displayed in the first two rows of Fig.~\ref{f:3waveR}.
The first row of this figure shows the predicted $(\hat{x}_0, \hat{t}_0)$ locations by formulae (\ref{x0t03waveR}) applied to all roots of $R_{3}^{[m]}(z)$. The second row shows the predicted amplitude fields $|u_1|$ (the other two fields $|u_2|$ and $|u_3|$ are not shown for brevity). These amplitude fields in the outer region are predicted by the fundamental rogue wave in Theorem 8, and these fields in the inner region are predicted by the lower $(N_{1R},N_{2R})$-th order rogue wave with all-zero internal parameters, and their $(N_{1R},N_{2R})$ values are as given in Eq.~(\ref{N12R}).

As in the earlier Q-case, predicted rogue patterns in Fig.~\ref{f:3waveR} also look very different from the underlying root structures of $R_{3}^{[m]}(z)$ polynomials in Fig.~2.

In the bottom row of this same figure, the corresponding true solutions $|u_1|$ from Theorem 4 are plotted. Again, perfect agreement is seen between our prediction and the true solution, confirming the predictive power of our Theorem~8.

\section{Proofs of Theorems 5 to 8} \label{sec:Manaproof}

In this section, we prove Theorems 5-8 on rogue patterns in the Manakov and three-wave systems. Our proof is based on an asymptotic analysis of the two systems' rogue wave solutions, or equivalently, the determinant $\sigma_{n,k}$ in Eqs.~(\ref{sigma1})-(\ref{sigma2}), in the large $|a_{m}|$ limit.

\textbf{Proof of Theorem 5 for the outer region.} \hspace{0.05cm}
Suppose $|a_{m}|$ is large and the other parameters $O(1)$ in the Q-type Manakov rogue solution. Then, when $(x, t)$ is in the outer region of $\sqrt{x^2+t^2}=O\left(|a_{m}|^{1/m}\right)$, we have
\begin{eqnarray}\label{Skxnasym}
&& S_{j}(\textbf{\emph{x}}^{+}(n,k) +\nu \textbf{\emph{s}}) =
S_{j}(x_{1}^{+}, x_{2}^{+}, \nu s_{3}, x_{4}^{+}, x_{5}^{+}, \nu s_{6}, \cdots,x_{m}^{+}+\nu s_m,\cdots) \sim S_j(\textbf{v}),
\end{eqnarray}
where
\[  \label{vdef}
\textbf{v}=( p_1 x+ 2 p_0 p_1\textrm{i}t, \hspace{0.05cm} 0, \cdots, 0, a_{m}, 0, \cdots).
\]
Here, the fact of $s_1=s_2=s_4=s_5=0$ from Eq.~(\ref{svalueMan}) has been used.

From the definition (\ref{Elemgenefunc}) of Schur polynomials and variable scalings, it is easy to see that
\begin{equation} \label{Skorder}
S_j(\textbf{v})=a_m^{j/m}p_{j}^{[m]}(z),
\end{equation}
where
\begin{equation} \label{Az}
z=a_m^{-1/m}(p_1 x+ 2 p_0 p_1\textrm{i}t).
\end{equation}
Using these formulae, we find that
\begin{equation} \label{Splusasym}
\det \left[S_{3i-j}(\textbf{\emph{x}}^{+}(n,k) + \nu_j \textbf{\emph{s}}) \right]\sim
c_N^{-1}a_m^{N(N+1)/m}Q_{N}^{[m]}(z).
\end{equation}
Similarly,
\begin{equation} \label{Sminusasym}
\det \left[S_{3i-j}(\textbf{\emph{x}}^{-}(n,k) + \nu_j \textbf{\emph{s}}^*) \right]\sim
c_N^{-1}\left(a_m^*\right)^{N(N+1)/m}Q_{N}^{[m]}(z^*).
\end{equation}
Here, $S_j\equiv 0$ when $j<0$.

To proceed further, we use determinant identities and the Laplace expansion to rewrite $\sigma_{n,k}^{(Q)}$ in Eq. (\ref{sigma1}) as \cite{OhtaJY2012}
\begin{eqnarray} \label{sigmanLap}
&& \hspace{-2.0cm} \sigma_{n,k}^{(Q)}=\sum_{0\leq\nu_{1} < \nu_{2} < \cdots < \nu_{N}\leq 3N-1}
\det_{1 \leq i, j\leq N} \left[(h_0)^{\nu_j} S_{3i-1-\nu_j}(\textbf{\emph{x}}^{+}(n,k) +\nu_j \textbf{\emph{s}}) \right]  \times \det_{1 \leq i, j\leq N}\left[ (h_0^*)^{\nu_j} S_{3i-1-\nu_j}(\textbf{\emph{x}}^{-}(n,k) + \nu_j \textbf{\emph{s}}^*)\right],
\end{eqnarray}
where $h_0=p_1/(p_0+p_0^*)$. Since the highest order term of $a_{m}$ in this $\sigma_{n,k}^{(Q)}$ comes from the index choices of $\nu_{j}=j-1$, then
\begin{equation} \label{sigmanmax}
\sigma_{n,k}^{(Q)} \sim |\alpha|^2 \hspace{0.05cm} |a_{m}|^{2N(N+1)/m} \left|Q_{N}^{[m]}(z)\right|^2,
\end{equation}
where $\alpha=(h_0)^{N(N-1)/2}c_N^{-1}$.
Since this leading-order asymptotics of $\sigma_{n,k}^{(Q)}$ is independent of $n$ and $k$, it shows that, for $|a_{m}|\gg 1$, $\sigma_{1,0}/\sigma_{0,0}\sim 1$ and $\sigma_{0,1}/\sigma_{0,0}\sim 1$. Thus, the solution $[u_{1,N, 0}(x,t), u_{2,N, 0}(x,t)]$ is on the constant-amplitude continuous-wave
background $[\rho_1 e^{{\rm{i}} (k_1x+\omega_{1} t)}, \rho_1 e^{{\rm{i}} (k_2x+\omega_{2} t)}]$, except at or near $(x, t)$ locations $\left(\tilde{x}_{0}, \tilde{t}_{0}\right)$ where
\[  \label{z0def}
z_0=a_m^{-1/m}(p_1 \tilde{x}_{0}+2\textrm{i}p_0 p_1\tilde{t}_{0})
\]
is a root of the polynomial $Q_{N}^{[m]}(z)$, and such $\left(\tilde{x}_{0}, \tilde{t}_{0}\right)$ locations are the leading-order terms of $(\hat{x}_0, \hat{t}_0)$ in Eqs.~(\ref{x0ManaQ})-(\ref{t0ManaQ}) of Theorem 5. Due to the requirement of $\sqrt{x^2+t^2}=O\left(|a_{m}|^{1/m}\right)$, $z_0$ should not be zero.

Next, we show that when $(x, t)$ is in the neighborhood of each of the $\left(\tilde{x}_{0}, \tilde{t}_{0}\right)$ locations given by Eq.~(\ref{z0def}), the Q-type Manakov rogue wave $[u_{1,N, 0}(x,t), u_{2,N, 0}(x,t)]$ approaches a fundamental Manakov rogue wave that is located within $O(1)$ distance from $\left(\tilde{x}_{0}, \tilde{t}_{0}\right)$. In order to derive this more refined asymptotics, we need to calculate terms in Eq.~(\ref{sigmanLap}) whose order is lower than $|a_{m}|^{2N(N+1)/m}$, since that highest order term (\ref{sigmanmax}) vanishes at $\left(\tilde{x}_{0}, \tilde{t}_{0}\right)$.

First, we denote
\[
\hat{x}_2^+(x,t) = p_2 x+(2p_0p_2+p_1^2)(\textrm{i}t),
\]
which are the dominant terms of $x_2^+(x,t)$ in Eq.~(\ref{defxrp}) with the index `$I$' removed, when $(x, t)$ is in the outer region. Then, for $(x, t)$ in the neighborhood of $\left(\tilde{x}_{0}, \tilde{t}_{0}\right)$, we have a more refined asymptotics for $S_j(\textbf{\emph{x}}^{+}(n,k)+\nu \textbf{\emph{s}})$ as
\begin{eqnarray}\label{Skasym2}
&& S_{j}(\textbf{\emph{x}}^{+}(n,k) +\nu \textbf{\emph{s}}) =
S_{j}(x_{1}^{+}, x_{2}^{+}, \nu s_{3}, x_{4}^{+}, x_{5}^{+}, \nu s_{6}, \cdots,x_{m}^{+}+\nu s_m,\cdots) \nonumber \\
&&=\left[ S_j(\hat{\textbf{v}})+ \hat{x}_2^+(\tilde{x}_0,\tilde{t}_0) S_{j-2}(\hat{\textbf{v}})\right] \left[1+O\left(a_{m}^{-2/m}\right)\right], \quad \quad |a_{m}| \gg 1,
\end{eqnarray}
where
\[\label{vhatdef}
\hat{\textbf{v}}=\left(x_1^+, 0, \cdots, 0, a_{m}, 0, \cdots\right)=\left(p_1 x+ 2 p_0 p_1\textrm{i}t+n \theta_1+ k \lambda_1, 0, \cdots, 0, a_{m}, 0, \cdots\right).
\]
Here, the normalization of $a_1=0$ in $x_1^+$ has been used. Polynomials $S_j(\hat{\textbf{v}})$ are related to $p_{j}^{[m]}(z)$ in Eq.~(\ref{pkmz}) as
\begin{equation} \label{Skorder2}
S_j(\hat{\textbf{v}})=a_m^{j/m}p_{j}^{[m]}(\hat{z}),
\end{equation}
where $\hat{z}=a_m^{-1/m}(p_1 x+ 2 p_0 p_1\textrm{i}t+n \theta_1+ k \lambda_1)$.

Now, we derive leading order terms of $a_{m}$ in Eq. (\ref{sigmanLap}) when $(x, t)$ is in the $O(1)$ neighborhood of $\left(\tilde{x}_{0}, \tilde{t}_{0}\right)$. These leading order terms come from two index choices, the first being $\nu=(0, 1, \cdots, N-1)$, and the second being $\nu=(0, 1, \cdots, N-2, N)$.

(i) With the first index choice, in view of Eqs. (\ref{Skasym2})-(\ref{Skorder2}), dominant contributions to the first determinant involving $\textbf{\emph{x}}^{+}(n,k)$ in Eq.~(\ref{sigmanLap}) contain two parts. One part, coming from the $S_j(\hat{\textbf{v}})$ term in Eq.~(\ref{Skasym2}) for each element of that determinant, is
\[ \label{Phinnxtnew}
\alpha \hspace{0.06cm} a_{m}^{N(N+1)/m}Q_{N}^{[m]}(\hat{z}) \left[1+O\left(a_{m}^{-2/m}\right)\right],
\]
where $\alpha$ is given below Eq.~(\ref{sigmanmax}). Expanding $Q_{N}^{[m]}(\hat{z})$ around $\hat{z}=z_0$, where $z_0$ is given in Eq.~(\ref{z0def}), and recalling $Q_{N}^{[m]}(z_0)=0$, we have
\[
Q_{N}^{[m]}(\hat{z})=a_m^{-1/m}\left[p_1(x-\tilde{x}_0)+ 2\textrm{i} p_0 p_1(t-\tilde{t}_0)+n \theta_1+ k \lambda_1\right]\left[Q_{N}^{[m]}\right]'(z_0)
\left[1+O\left(a_m^{-1/m}\right)\right].
\]
Inserting this equation into (\ref{Phinnxtnew}), this part of the contribution becomes
\[ \label{contribution1}
\alpha \hspace{0.06cm} a_{m}^{[N(N+1)-1]/m}
\left[p_1(x-\tilde{x}_0)+ 2 \textrm{i} p_0 p_1(t-\tilde{t}_0)+n \theta_1+ k \lambda_1\right] \hspace{0.06cm} \left[Q_{N}^{[m]}\right]'(z_0)\left[1+O\left(a_{m}^{-1/m}\right)\right].
\]
The other part of the contribution to the first determinant in Eq.~(\ref{sigmanLap}) comes from the $S_{j-2}(\hat{\textbf{v}})$ term of (\ref{Skasym2}) for each single column of that determinant and the $S_j(\hat{\textbf{v}})$ term for the rest of the columns. This part of the contribution gives
\[
\hat{x}_2^+(\tilde{x}_0,\tilde{t}_0) \sum_{j=1}^{N} \det_{1 \leq i \leq N} \left[S_{3i-1}(\hat{\textbf{v}}),  \cdots, h_0^{j-1}S_{3i-j-2}(\hat{\textbf{v}}), \cdots, h_0^{N-1}S_{3i-N}(\hat{\textbf{v}})\right]\left[1+O\left(a_{m}^{-1/m}\right)\right].
\]
Replacing $S_j(\hat{\textbf{v}})$ by its leading-order term $S_j(\textbf{v})$, utilizing the relation (\ref{Skorder}), and further replacing $z$ by its leading-order term $z_0$, the above contribution reduces to
\[
\alpha \hspace{0.03cm} a_{m}^{[N(N+1)-2]/m} \hat{x}_2^+(\tilde{x}_0,\tilde{t}_0) \hspace{0.03cm} \sum_{j=1}^{N} \det_{1 \leq i \leq N} \left[p^{[m]}_{3i-1}(z_0),  \cdots, p^{[m]}_{3i-j-2}(z_0), \cdots, p^{[m]}_{3i-N}(z_0)\right]\left[1+O\left(a_{m}^{-1/m}\right)\right].
\]
Since $\hat{x}_2^+(\tilde{x}_0,\tilde{t}_0)=O(a_m^{1/m})$, this contribution is of the same order as that in Eq.~(\ref{contribution1}). Combining these two contributions, the first determinant involving $\textbf{\emph{x}}^{+}(n,k)$ in Eq.~(\ref{sigmanLap}) is found to be
\[ \label{tildext0}
\alpha \hspace{0.06cm} a_{m}^{[N(N+1)-1]/m}\left[p_1(x-\tilde{x}_0)+ 2 p_0 p_1\textrm{i}(t-\tilde{t}_0)+n \theta_1+ k \lambda_1+\Delta_Q \right]\left[Q_{N}^{[m]}\right]'(z_0)
\left[1+O\left(a_m^{-1/m}\right)\right],
\]
where
\[
\Delta_{Q}=\frac{\hat{x}_2^+(\tilde{x}_0,\tilde{t}_0)}{a_m^{1/m}}\frac{
\sum_{j=1}^{N} \det_{1 \leq i \leq N} \left[p^{[m]}_{3i-1}(z_0),  \cdots, p^{[m]}_{3i-j-2}(z_0), \cdots, p^{[m]}_{3i-N}(z_0)\right]}{\left[Q_{N}^{[m]}\right]'(z_0)}.
\]
Denoting $\varphi_m\equiv \mbox{arg}\left(a_m^{1/m}\right)$, then using the $(\tilde{x}_0,\tilde{t}_0)$ expressions as obtained from Eq.~(\ref{z0def}), we find that
\[
\frac{\hat{x}_2^+(\tilde{x}_0,\tilde{t}_0)}{a_m^{1/m}}=\frac{p_2}{p_1}z_0
+\frac{{\rm{i}}p_1^2}{2\hspace{0.04cm} \Re(p_0)}\frac{\Im\left(z_0 \hspace{0.03cm} e^{i\varphi_m}/p_1\right)}
{e^{i\varphi_m}}.
\]
Thus,
\[ \label{DeltaQform}
\Delta_{Q}=\left(\frac{p_2}{p_1}z_0
+\frac{{\rm{i}}p_1^2}{2\hspace{0.04cm} \Re(p_0)}\frac{\Im\left(z_0 \hspace{0.03cm} e^{i\varphi_m}/p_1\right)}{e^{i\varphi_m}}\right)
\frac{\sum_{j=1}^{N} \det_{1 \leq i \leq N} \left[p^{[m]}_{3i-1}(z_0),  \cdots, p^{[m]}_{3i-j-2}(z_0), \cdots, p^{[m]}_{3i-N}(z_0)\right]}{\left[Q_{N}^{[m]}\right]'(z_0)}.
\]
Here, the determinant inside the summation of the above formula is the determinant of $Q_{N}^{[m]}(z_0)$, i.e., $\det_{1 \leq i, j \leq N} \left[p^{[m]}_{3i-j}(z_0)\right]$, except that indices of its $j$-th column are reduced by two.
This $\Delta_Q$ is an $O(1)$ quantity that is dependent on $m, N, z_0, \varphi_m, p_0, p_1$ and $p_2$. Absorbing this $\Delta_Q$ term into $(\tilde{x}_0, \tilde{t}_0)$ in Eq.~(\ref{tildext0}), we find that the contribution to the first determinant in Eq.~(\ref{sigmanLap}) under the first index choice of $\nu=(0, 1, \cdots, N-1)$ is
\[ \label{tildext02}
\alpha \hspace{0.06cm} a_{m}^{[N(N+1)-1]/m}\left[p_1(x-\hat{x}_0)+ 2 p_0 p_1\textrm{i}(t-\hat{t}_0)+n \theta_1+ k \lambda_1\right]\left[Q_{N}^{[m]}\right]'(z_0)
\left[1+O\left(a_m^{-1/m}\right)\right],
\]
where $(\hat{x}_0, \hat{t}_0)$ are given in Eqs.~(\ref{x0ManaQ})-(\ref{t0ManaQ}) of Theorem 5.

Similarly, the second determinant involving $\textbf{\emph{x}}^{-}(n,k)$ in Eq.~(\ref{sigmanLap}) under the first index choice of $\nu=(0, 1, \cdots, N-1)$ contributes the term
\[
\alpha^* \hspace{0.06cm} (a_{m}^*)^{[N(N+1)-1]/m}\left[p_1^*(x-\hat{x}_0)- 2 p_0^* p_1^*\textrm{i}(t-\hat{t}_0)-n \theta_1^*-k \lambda_1^* \right]\left[Q_{N}^{[m]}\right]'(z_0^*)
\left[1+O\left(a_m^{-1/m}\right)\right].
\]

(ii) Under the second index choice of $\nu=(0, 1, \cdots, N-2, N)$ in Eq.~(\ref{sigmanLap}), the leading-order contribution to the first determinant involving $\textbf{\emph{x}}^{+}(n,k)$ can be calculated from the asymptotics (\ref{Skxnasym}) and the relation (\ref{Skorder}) as
\[
h_0^{N(N-1)/2+1}a_m^{[N(N+1)-1]/m}
\det_{1 \leq i \leq N} \left[p^{[m]}_{3i-1}(z_0),  p^{[m]}_{3i-2}(z_0),
\cdots, p^{[m]}_{3i-(N-1)}(z_0),  p^{[m]}_{3i-N-1}(z_0)\right]\left[1+O\left(a_{m}^{-1/m}\right)\right].
\]
Recalling $p^{[m]}_{j-1}(z)=[p^{[m]}_j]'(z)$, the above contribution can be rewritten as
\[ \label{Phinnxtnew2}
h_0 \hspace{0.04cm} \alpha \hspace{0.06cm} a_m^{[N(N+1)-1]/m} \left[Q_{N}^{[m]}\right]'(z_0) \left[1+O\left(a_{m}^{-1/m}\right)\right].
\]
Similarly, the second determinant involving $\textbf{\emph{x}}^{-}(n,k)$ in Eq.~(\ref{sigmanLap}) contributes
\[
h_0^* \hspace{0.03cm} \alpha^* \hspace{0.02cm} (a_{m}^*)^{[N(N+1)-1]/m}\left[Q_{N}^{[m]}\right]'(z_0^*) \left[1+O\left(a_{m}^{-1/m}\right)\right].
\]

Summarizing the above contributions to Eq.~(\ref{sigmanLap}), we find that
\begin{eqnarray}
&& \sigma_{n,k}^{(Q)}(x,t) = |\alpha|^2 \hspace{0.06cm} \left|\left[Q_{N}^{[m]}\right]'(z_0)\right|^2 |a_{m}|^{[N(N+1)-1]/m}\times   \nonumber \\
&& \hspace{0.4cm} \left( \left[ p_1(x-\hat{x}_0)+ 2 \textrm{i} p_0 p_1(t-\hat{t}_0)+n \theta_1+ k \lambda_1\right]
\left[ p_1^*(x-\hat{x}_0) - 2 \textrm{i} p_0^* p_1^*(t-\hat{t}_0)- n \theta_1^* - k \lambda_1^*\right]+ |h_0|^2
\right)\left[1+O\left(a_{m}^{-1/m}\right)\right]. \hspace{1cm}
\end{eqnarray}
Under our assumption of all nonzero roots of $Q_{N}^{[m]}(z)$ being simple, $\left[Q_{N}^{[m]}\right]'(z_0)\ne 0$. Thus, the above leading-order asymptotics for $\sigma_{n,k}^{(Q)}(x,t)$ does not vanish. It is easy to see that this expression of $\sigma_{n,k}^{(Q)}(x,t)$,
combined with Eqs.~(\ref{Schpolysolu1})-(\ref{SchpolysolufN}), gives a fundamental rogue wave
$[\hat{u}_{1}(x-\hat{x}_{0}, t-\hat{t}_{0})\hspace{0.06cm} e^{{\rm{i}} (k_1x+\omega_{1} t)}, \hspace{0.07cm} \hat{u}_{2}(x-\hat{x}_{0}, t-\hat{t}_{0})\hspace{0.06cm} e^{{\rm{i}} (k_{2}x + \omega_{2} t)}]$ as given in Theorem 5, and the error of this fundamental rogue wave prediction is $O\left(a_{m}^{-1/m}\right)$. This completes the proof of Theorem~5 for the outer region.

\vspace{0.1cm}
\textbf{Proof of Theorem 5 for the inner region.} \hspace{0.05cm}
To analyze the large-$a_{m}$ behavior of Q-type Manakov rogue waves in the inner region, where $x^2+t^2=O(1)$, we first rewrite the $\sigma_{n,k}^{(Q)}$ determinant (\ref{sigma1}) into a $4N\times 4N$ determinant \cite{OhtaJY2012}
\[ \label{3Nby3Ndet2}
\sigma_{n,k}^{(Q)}=\left|\begin{array}{cc}
\textbf{O}_{N\times N} & \Phi_{N\times 3N} \\
-\Psi_{3N\times N} & \textbf{I}_{3N\times 3N} \end{array}\right|,
\]
where
\[
\Phi_{i,j}=\left(\frac{p_1}{p_0+p_0^*}\right)^{j-1} S_{3i-j}\left[\textbf{\emph{x}}^{+}(n,k) + (j-1) \textbf{\emph{s}}\right], \quad
\Psi_{i,j}=\left(\frac{p_1^*}{p_0+p_0^*}\right)^{i-1} S_{3j-i}\left[\textbf{\emph{x}}^{-}(n,k) + (i-1) \textbf{\emph{s}}^*\right].
\]
Defining $\textbf{\emph{y}}^{\pm}$ to be the vector $\textbf{\emph{x}}^{\pm}$ without the $a_{m}$ term, i.e., let
\[ \label{defypm}
\textbf{\emph{x}}^{+}=\textbf{\emph{y}}^{+}+(0, \cdots, 0, a_{m}, 0, \cdots), \quad
\textbf{\emph{x}}^{-}=\textbf{\emph{y}}^{-}+(0, \cdots, 0, a_{m}^*, 0, \cdots),
\]
it is easy to see using the definition (\ref{Elemgenefunc}) of Schur polynomials that the
Schur polynomials of $\textbf{\emph{x}}^{\pm}$ are related to those of $\textbf{\emph{y}}^{\pm}$ as
\begin{equation} \label{Sjrelation}
S_{j}(\textbf{\emph{x}}^{+}+\nu\textbf{\emph{s}}) = \sum_{l=0}^{[j/m]} \frac{a_{m}^l}{l!}
S_{j-lm}(\textbf{\emph{y}}^{+}+\nu\textbf{\emph{s}}), \quad
S_{j}(\textbf{\emph{x}}^{-}+\nu\textbf{\emph{s}}^*) = \sum_{l=0}^{[j/m]} \frac{(a_{m}^*)^l}{l!} S_{j-lm}(\textbf{\emph{y}}^{-}+\nu\textbf{\emph{s}}^*).
\end{equation}
The reader is reminded that the notation of $[a]$ represents the largest integer less than or equal to $a$. Using these relations, we express matrix elements of $\Phi$ and $\Psi$ in Eq. (\ref{3Nby3Ndet2}) through Schur polynomials $S_{j}(\textbf{\emph{y}}^{+}+\nu\textbf{\emph{s}})$, $S_{j}(\textbf{\emph{y}}^{-}+\nu\textbf{\emph{s}}^*)$, and powers of $a_{m}$ and $a_{m}^*$.

Next, we perform row operations to the $\Phi$ matrix in order to remove certain power terms of $a_m$. For this purpose, we notice that when $m=3j+1$ $(j\ge 1)$, coefficients of the highest $a_m$ power terms in $\Phi$'s first column are proportional to
\[ \label{sequenceManaQ}
\hat{S}_{2}, \hat{S}_{5}, \cdots, \hat{S}_{3j-1}, \hat{S}_{1}, \hat{S}_{4}, \cdots, \hat{S}_{3j-2}, \hat{S}_{0}, \hat{S}_{3}, \cdots, \hat{S}_{3j},
\]
and repeating, where $\hat{S}_{j}\equiv S_j(\textbf{\emph{y}}^{+}+\nu\textbf{\emph{s}})$. When $m=3j+2$ $(j\ge 0)$, these coefficients of the highest $a_m$ power terms in $\Phi$'s first column are proportional to
\[ \label{sequenceManaQ2}
\hat{S}_{2}, \hat{S}_{5}, \cdots, \hat{S}_{3j-1}, \hat{S}_{0}, \hat{S}_{3}, \cdots, \hat{S}_{3j}, \hat{S}_{1}, \hat{S}_{4}, \cdots, \hat{S}_{3j+1},
\]
and repeating. In the second and higher columns of $\Phi$, elements are of the same form as those in the first column, except that the index $j$ of every $\hat{S}_{j}$ in them decreases by one with each higher column, and $\hat{S}_{j}\equiv 0$ for $j<0$. Using the first $m$ rows, we perform row operations to remove the highest powers of $a_m$ from the second $m$ rows, leaving the second-highest power terms of $a_m$ with coefficients proportional to $\hat{S}_{j+m}$, where $\hat{S}_{j}$ is the highest $a_m$-power coefficient of each element just being removed. Then, we use the first $m$ rows and the resulting second $m$ rows to eliminate the highest and second-highest power terms of $a_m$ from the third $m$ rows, leaving the third-highest power terms of $a_m$ with coefficients proportional to $\hat{S}_{j+2m}$ in them. This process is continued to all later rows of $\Phi$. Similar column operations are also applied to the matrix $\Psi$ in Eq.~(\ref{3Nby3Ndet2}).

After these row and column operations, we then keep only the highest remaining power of $a_m$ in each matrix element of $\Phi$ and the highest remaining power of $a_m^*$ in each matrix element of $\Psi$. Using these manipulations and the sequence structures in Eqs.~(\ref{sequenceManaQ}) and (\ref{sequenceManaQ2}), for $m \hspace{0.1cm} \mbox{mod} \hspace{0.1cm} 3 =1$ and $2$ respectively, we find that $\sigma_{n,k}^{(Q)}$ in (\ref{3Nby3Ndet2}) is asymptotically reduced to
\[ \label{3Nby3Ndet3}
\sigma_{n,k}^{(Q)}=\widehat{\beta} \hspace{0.06cm} |a_{m}|^{\widehat{K}}
\left|\begin{array}{cc}
\textbf{O}_{(N_{1Q}+N_{2Q})\times (N_{1Q}+N_{2Q})} & \widehat{\Phi}_{(N_{1Q}+N_{2Q})\times \widehat{N}} \\
-\widehat{\Psi}_{\widehat{N}\times (N_{1Q}+N_{2Q})} & \textbf{I}_{\widehat{N}\times \widehat{N}} \end{array}\right| \left[1+O\left(a_{m}^{-1}\right)\right],
\]
where $\widehat{\beta}$ is an $(m,N)$-dependent nonzero constant, $\widehat{K}$ is an $(m,N)$-dependent positive integer, $(N_{1Q}, N_{2Q})$ are nonnegative integers given in Theorem~1, $\widehat{N}=\max(3N_{1Q}, 3N_{2Q}-1)$,
\begin{eqnarray}
&& \widehat{\Phi}=\left(\begin{array}{c} \widehat{\Phi}^{(1)}_{N_{1Q}\times \widehat{N}} \\ \vspace{-0.3cm} \\
\widehat{\Phi}^{(2)}_{N_{2Q}\times \widehat{N}}\end{array}\right), \quad
\widehat{\Psi}=\left(\widehat{\Psi}^{(1)}_{\widehat{N} \times N_{1Q}} \hspace{0.1cm}
\widehat{\Psi}^{(2)}_{\widehat{N} \times N_{2Q}}\right), \label{widehatPhiPsi} \\
&& \widehat{\Phi}_{i,j}^{(I)}=(h_0)^{-(j-1)} S_{3i-I}\left[\textbf{\emph{y}}^{+}(n,k) + (j-1+\nu_0) \textbf{\emph{s}}\right], \\
&& \widehat{\Psi}_{i,j}^{(J)}=(h_0^*)^{-(i-1)} S_{3j-J}\left[\textbf{\emph{y}}^{-}(n,k) + (i-1+\nu_0) \textbf{\emph{s}}^*\right],  \label{widehatPhiPsi3}
\end{eqnarray}
and $\nu_0=N-N_{1Q}-N_{2Q}$. Since the constant factor $\widehat{\beta} \hspace{0.06cm} |a_{m}|^{\widehat{K}}$ in (\ref{3Nby3Ndet3}) does not affect the Manakov solution and can be dropped, the remaining determinant in (\ref{3Nby3Ndet3}) can be rewritten as
\[ \label{cubicrwstype3c}
\sigma_{n,k}^{(Q)}=
\det \left(
\begin{array}{cc}
  \sigma^{[1,1]}_{n,k} & \sigma^{[1,2]}_{n,k} \\
  \sigma^{[2,1]}_{n,k} & \sigma^{[2,2]}_{n,k}
\end{array}
\right) \left[1+O\left(a_{m}^{-1}\right)\right],
\]
\[
\sigma^{[I, J]}_{n,k}=
\left(
\phi_{3i-I, \, 3j-J}^{(n,k, \hspace{0.04cm} I, J)}
\right)_{1\leq i \leq N_{IQ}, \, 1\leq j \leq N_{JQ}},
\]
where the matrix elements in $\sigma^{[I, J]}_{n,k}$ are defined by
\[ \label{mij149}
\phi_{i,j}^{(n,k,I,J)}=\sum_{\nu=0}^{\min(i,j)} \left[ \frac{|p_{1}|^2 }{(p_{0}+p_{0}^*)^2}  \right]^{\nu} \hspace{0.06cm} S_{i-\nu}\left(\textbf{\emph{y}}^{+}(n,k) + \nu_0 \textbf{\emph{s}}+
\nu \textbf{\emph{s}}\right)  \hspace{0.06cm} S_{j-\nu}\left(\textbf{\emph{y}}^{-}(n,k) + \nu_0 \textbf{\emph{s}}^* + \nu \textbf{\emph{s}}^*\right).
\]
The largest index $j$ of $S_j$ involved in the above reduced solution is $\max(3N_{1Q}-1, 3N_{2Q}-2)$. It is easy to see from Theorem~1 that $\max(3N_{1Q}-1, 3N_{2Q}-2)<m$. Thus, the above solution only depends on $S_j$ polynomials with $j<m$, and hence only depends on $y_j^\pm(n,k)$ with $j<m$. From the definition (\ref{defypm}), we see that $y_j^\pm(n,k)=x_j^\pm(n,k)$ when $j<m$. This means that in Eq.~(\ref{mij149}), $\textbf{\emph{y}}^{\pm}(n,k)$ can be replaced by $\textbf{\emph{x}}^{\pm}(n,k)$. Finally, we lump each constant $\nu_0 s_j$ into $a_j$ of $x_j^+(n, k)$, and similarly lump each $\nu_0 s_j^*$ into $a_j^*$ of $x_j^{-}(n, k)$. When $j \hspace{0.1cm} \mbox{mod} \hspace{0.1cm} 3 =0$,
$x_j^\pm(n, k)=0$ per Theorem 3 and does not contain $a_j$. In such a case, we just lump $\nu_0 s_j$ into $x_j^+(n, k)$ and $\nu_0 s_j^*$ into $x_j^{-}(n, k)$, which eventually can be eliminated from the solution for the same reason we did in Eq.~(\ref{defxrpm}) of Theorem 3. After these treatments,
the above determinant in (\ref{cubicrwstype3c}) becomes a $\left(N_{1Q}, N_{2Q}\right)$-th order Manakov rogue wave $[u_{1,N_{1Q}, N_{2Q}}(x,t), u_{2,N_{1Q}, N_{2Q}}(x,t)]$ as given in Theorem~3, whose internal parameters
$(\hat{a}_{1, 1}, \hat{a}_{2,1}, \hat{a}_{4,1}, \hat{a}_{5,1}, \dots, \hat{a}_{3N_{1Q}-1,\hspace{0.05cm} 1})$ and $(\hat{a}_{1, 2}, \hat{a}_{2,2}, \hat{a}_{4,2}, \hat{a}_{5,2}, \dots, \hat{a}_{3N_{2Q}-2, \hspace{0.05cm} 2})$ are related to those in the original rogue wave as
\[  \label{ahata}
\hat{a}_{j,1}=\hat{a}_{j,2}=a_{j}+\nu_0 s_j, \quad j=1, 2, 4, 5, \cdots,
\]
which is the same as the relation (\ref{ahat1a}) in Theorem 5. The error of this lower-order rogue wave approximation is $O(|a_{m}|^{-1})$ in view of Eq.~(\ref{cubicrwstype3c}). This completes the proof of Theorem~5 for the inner region.

We would like to add that, for $s_j$ values we have numerically obtained, we find that $s_{j}=0$ for $j \hspace{0.07cm} \mbox{mod}\hspace{0.07cm} 3 \neq 0$, see Eq.~(\ref{svalueMan}). If this pattern holds in general, then the above relation (\ref{ahata}) would simplify to $\hat{a}_{j,1}=\hat{a}_{j,2}=a_{j}$. But since we have not proved $s_{j}=0$ for all $j \hspace{0.07cm} \mbox{mod}\hspace{0.07cm} 3 \neq 0$, we still keep the $\nu_0 s_j$ term in (\ref{ahata}) and Eq.~(\ref{ahat1a}) of Theorem 5 out of caution.

\vspace{0.1cm}
\textbf{Proof of Theorem 6.} \hspace{0.05cm} The proof of Theorem 6 for R-type Manakov rogue waves is very similar to that for Theorem 5. For that reason, we will only list the differences here.

In the outer region, due to the different matrix indices in Eq.~(\ref{sigma2}) for R-type rogue waves, the corresponding polynomials whose roots give leading-order locations of fundamental rogue waves are naturally R-type Okamoto hierarchy polynomials $R_{N}^{[m]}(z)$. The remaining difference is the calculation of the next-order position shift, i.e., the formula for $\Delta_R$ in Eqs.~(\ref{x0ManaR})-(\ref{t0ManaR}). Repeating earlier calculations for the different R-type matrix indices, we can easily find that
\[ \label{DeltaRform}
\Delta_{R}=\left(\frac{p_2}{p_1}z_0
+\frac{{\rm{i}}p_1^2}{2\hspace{0.04cm} \Re(p_0)}\frac{\Im\left(z_0 \hspace{0.03cm} e^{i\varphi_m}/p_1\right)}{e^{i\varphi_m}}\right)
\frac{\sum_{j=1}^{N} \det_{1 \leq i \leq N} \left[p^{[m]}_{3i-1-1}(z_0),  \cdots, p^{[m]}_{3i-j-1-2}(z_0), \cdots, p^{[m]}_{3i-N-1}(z_0)\right]}{\left[R_{N}^{[m]}\right]'(z_0)}.
\]
Here, the determinant inside the summation of the above formula is the determinant of $R_{N}^{[m]}(z_0)$, i.e., $\det_{1 \leq i, j \leq N} \left[p^{[m]}_{3i-j-1}(z_0)\right]$, except that indices of its $j$-th column are reduced by two.

In the inner region, where $x^2+t^2=O(1)$, we also rewrite the $\sigma_{n,k}^{(R)}$ determinant (\ref{sigma2}) into a $4N\times 4N$ determinant, and then use relations (\ref{Sjrelation}) to rewrite every matrix element of $\Phi$ and $\Psi$ into powers of $a_m$ and $a_m^*$ respectively. For R-type rogue waves, when $m=3j+1$ $(j\ge 1)$, coefficients of the highest $a_m$ power terms in $\Phi$'s first column are proportional to
\[ \label{sequenceManaR}
\hat{S}_{1}, \hat{S}_{4}, \cdots, \hat{S}_{3j-2}, \hat{S}_{0}, \hat{S}_{3}, \cdots, \hat{S}_{3j}, \hat{S}_{2}, \hat{S}_{5}, \cdots, \hat{S}_{3j-1},
\]
and repeating, and when $m=3j+2$ $(j\ge 0)$, these coefficients are proportional to
\[ \label{sequenceManaR2}
\hat{S}_{1}, \hat{S}_{4}, \cdots, \hat{S}_{3j+1}, \hat{S}_{2}, \hat{S}_{5}, \cdots, \hat{S}_{3j-1}, \hat{S}_{0}, \hat{S}_{3}, \cdots, \hat{S}_{3j},
\]
and repeating. Using these sequence structures and performing the same row and column operations as described earlier to remove certain high powers of $a_m$ in the $\Phi$ and $\Psi$ matrices, we find that $\sigma_{n,k}^{(R)}$ can be asymptotically reduced to (\ref{3Nby3Ndet3})-(\ref{widehatPhiPsi3}), except that $(N_{1Q}, N_{2Q})$ are replaced by $(N_{1R}, N_{2R})$ as given in Theorem~2, and $(\widehat{\beta}, \widehat{K})$ are different constants. The rest of the proof is the same as before, and Theorem 6 is then proved.

\vspace{0.1cm}
\textbf{Proof of Theorem 7.} \hspace{0.05cm} The proof of Theorem 7 for Q-type rogue patterns in the three-wave interaction system is very similar to that of Theorem 5 for the Manakov system. In the outer region where $\sqrt{x^2+t^2}=O\left(|a_{m}|^{1/m}\right)$ at large $|a_m|$,
\begin{eqnarray}\label{Skxnasym9}
&& S_{j}(\textbf{\emph{x}}^{+}(n,k) +\nu \textbf{\emph{s}}) =
S_{j}(x_{1}^{+}, x_{2}^{+}, \nu s_{3}, x_{4}^{+}, x_{5}^{+}, \nu s_{6}, \cdots,x_{m}^{+}
+\nu s_m,\cdots)
\sim S_j(\textbf{v}),
\end{eqnarray}
where
\[  \label{vdef9}
\textbf{v}=\left[ (\alpha_1-\beta_1)x+ (c_1\beta_1-c_2\alpha_1)t, \hspace{0.05cm} 0, \cdots, 0, a_{m}, 0, \cdots\right].
\]
Following similar calculations as in the proof of Theorem 5, we find that the highest power term of $a_{m}$ in $\sigma_{n,k}^{(Q)}$ of the three-wave system is
\begin{equation} \label{sigmanmax9}
\sigma_{n,k}^{(Q)} \sim |\alpha|^2 \hspace{0.05cm} |a_{m}|^{2N(N+1)/m} \left|Q_{N}^{[m]}(z)\right|^2,
\end{equation}
where $\alpha=(h_0)^{N(N-1)/2}c_N^{-1}$, $h_0=p_1/(p_0+p_0^*)$, and
\[
z=a_m^{-1/m}\left[(\alpha_1-\beta_1)x+ (c_1\beta_1-c_2\alpha_1)t\right].
\]
Thus, the solution $[u_{1,N, 0}(x,t), u_{2,N, 0}(x,t), u_{3,N, 0}(x,t)]$ is on a constant-amplitude continuous-wave background (\ref{BoundaryCond}), except at or near $(x, t)$ locations $\left(\tilde{x}_{0}, \tilde{t}_{0}\right)$ where
\[  \label{z0def9}
z_0=a_m^{-1/m}\left[(\alpha_1-\beta_1)\tilde{x}_0+ (c_1\beta_1-c_2\alpha_1)\tilde{t}_0\right]
\]
is a root of the polynomial $Q_{N}^{[m]}(z)$, and such $\left(\tilde{x}_{0}, \tilde{t}_{0}\right)$ locations are the leading-order terms of $(\hat{x}_0, \hat{t}_0)$ in Eq.~(\ref{x0t03waveQ}) of Theorem~7.

We can further show that, when $(x, t)$ is in the neighborhood of each of the $\left(\tilde{x}_{0}, \tilde{t}_{0}\right)$ locations given by Eq.~(\ref{z0def9}), the Q-type three-wave rogue solution approaches a fundamental rogue wave that is located within $O(1)$ distance from $\left(\tilde{x}_{0}, \tilde{t}_{0}\right)$. For this purpose, we denote
\[
\hat{x}_2^+(x,t) = \left( \alpha_{2} - \beta_{2} \right) x +\left( c_{1}\beta_{2}-c_{2}\alpha_{2} \right)t,
\]
which are the dominant terms of $x_2^+(x,t)$ in Eq.~(\ref{defxrp2}) with the index `$I$' removed, when $(x, t)$ is in the outer region. Then, for $(x, t)$ in the neighborhood of $\left(\tilde{x}_{0}, \tilde{t}_{0}\right)$, we have a more refined asymptotics for $S_j(\textbf{\emph{x}}^{+}(n,k)+\nu \textbf{\emph{s}})$ as
\begin{eqnarray}\label{Skasym2b}
&& S_{j}(\textbf{\emph{x}}^{+}(n,k) +\nu \textbf{\emph{s}}) =\left[ S_j(\hat{\textbf{v}})+ \hat{x}_2^+(\tilde{x}_0,\tilde{t}_0) S_{j-2}(\hat{\textbf{v}})\right] \left[1+O\left(a_{m}^{-2/m}\right)\right], \quad \quad |a_{m}| \gg 1, \\
&&\hat{\textbf{v}}=\left(x_1^+, 0, \cdots, 0, a_{m}, 0, \cdots\right), \\
&&x_1^+=\left( \alpha_{1} - \beta_{1} \right) x +\left( c_{1}\beta_{1}-c_{2}\alpha_{1} \right)t + n \theta_{1} + k \lambda_{1}.
\end{eqnarray}
Here, the normalization of $a_1=0$ has been utilized. Next, we again rewrite $\sigma_{n,k}^{(Q)}$ in Eq. (\ref{sigma1}) as (\ref{sigmanLap}). Then the contribution to the first determinant in Eq.~(\ref{sigmanLap}) from the first index choice of $\nu_j=j-1$ can be similarly calculated as
\[ \label{tildext09}
\alpha \hspace{0.06cm} a_{m}^{[N(N+1)-1]/m}\left[
(\alpha_1-\beta_1)(x-\tilde{x}_0)+ (c_1\beta_1-c_2\alpha_1)(t-\tilde{t}_0)
+n \theta_1+ k \lambda_1+\hat{\Delta}_Q \right]\left[Q_{N}^{[m]}\right]'(z_0)
\left[1+O\left(a_m^{-1/m}\right)\right],
\]
where
\[
\hat{\Delta}_{Q}=\frac{\hat{x}_2^+(\tilde{x}_0,\tilde{t}_0)}{a_m^{1/m}}\frac{
\sum_{j=1}^{N} \det_{1 \leq i \leq N} \left[p^{[m]}_{3i-1}(z_0),  \cdots, p^{[m]}_{3i-j-2}(z_0), \cdots, p^{[m]}_{3i-N}(z_0)\right]}{\left[Q_{N}^{[m]}\right]'(z_0)}.
\]
As in the definition of $\Delta_Q$ in Eq.~(\ref{DeltaQform}), the determinant inside the summation of the above $\hat{\Delta}_{Q}$ formula is the determinant of $Q_{N}^{[m]}(z_0)$, i.e., $\det_{1 \leq i, j \leq N} \left[p^{[m]}_{3i-j}(z_0)\right]$, except that indices of its $j$-th column are reduced by two. Denoting $\varphi_m\equiv \mbox{arg}\left(a_m^{1/m}\right)$ as before, and using the $(\tilde{x}_0,\tilde{t}_0)$ expressions as obtained from Eq.~(\ref{z0def9}), we get
\[
\frac{\hat{x}_2^+(\tilde{x}_0,\tilde{t}_0)}{a_m^{1/m}}=e^{-{\rm{i}}\varphi_m}\left[
(\alpha_2-\beta_2)\frac{\Im  \left[ \frac{z_{0} e^{{\rm{i}}\varphi_m}}{c_{1} \beta_{1}- c_{2}\alpha_{1}} \right]}{\Im  \left[ \frac{\alpha_{1}-\beta_{1}}{c_{1} \beta_{1}- c_{2}\alpha_{1}} \right]}
+(c_1\beta_2-c_2\alpha_2)\frac{\Im  \left[ \frac{z_{0}  e^{{\rm{i}}\varphi_m}}{\alpha_{1}-\beta_{1}} \right]}{\Im  \left[ \frac{c_{1} \beta_{1}- c_{2}\alpha_{1}}{\alpha_{1}-\beta_{1}} \right]}\right].
\]
Thus,
\begin{eqnarray}
\hat{\Delta}_{Q}&=&e^{-{\rm{i}}\varphi_m}\left[
(\alpha_2-\beta_2)\frac{\Im  \left[ \frac{z_{0} e^{{\rm{i}}\varphi_m}}{c_{1} \beta_{1}- c_{2}\alpha_{1}} \right]}{\Im  \left[ \frac{\alpha_{1}-\beta_{1}}{c_{1} \beta_{1}- c_{2}\alpha_{1}} \right]}
+(c_1\beta_2-c_2\alpha_2)\frac{\Im  \left[ \frac{z_{0}  e^{{\rm{i}}\varphi_m}}{\alpha_{1}-\beta_{1}} \right]}{\Im  \left[ \frac{c_{1} \beta_{1}- c_{2}\alpha_{1}}{\alpha_{1}-\beta_{1}} \right]}\right] \times  \nonumber \\
&& \frac{\sum_{j=1}^{N} \det_{1 \leq i \leq N} \left[p^{[m]}_{3i-1}(z_0),  \cdots, p^{[m]}_{3i-j-2}(z_0), \cdots, p^{[m]}_{3i-N}(z_0)\right]}{\left[Q_{N}^{[m]}\right]'(z_0)}, \label{DeltaQhatform}
\end{eqnarray}
which is an $O(1)$ quantity. Absorbing this $\hat{\Delta}_Q$ term into $(\tilde{x}_0, \tilde{t}_0)$ in Eq.~(\ref{tildext09}), we find that the contribution to the first determinant in Eq.~(\ref{sigmanLap}) under the first index choice of $\nu_j=j-1$ is
\[ \label{tildext029}
\alpha \hspace{0.06cm} a_{m}^{[N(N+1)-1]/m}\left[(\alpha_1-\beta_1)(x-\hat{x}_0)+ (c_1\beta_1-c_2\alpha_1)(t-\hat{t}_0)+n \theta_1+ k \lambda_1\right]\left[Q_{N}^{[m]}\right]'(z_0)
\left[1+O\left(a_m^{-1/m}\right)\right],
\]
where $(\hat{x}_0, \hat{t}_0)$ are given in Eq.~(\ref{x0t03waveQ}) of Theorem 7. The contribution to the first determinant in Eq.~(\ref{sigmanLap}) under the second index choice of $\nu=(0, 1, \cdots, N-2, N)$ can be found the same as that in Eq.~(\ref{Phinnxtnew2}). Using these results and similar ones for the second determinant in Eq.~(\ref{sigmanLap}), we find that
\begin{eqnarray}
&& \sigma_{n,k}^{(Q)}(x,t) = |\alpha|^2 \hspace{0.06cm} \left|\left[Q_{N}^{[m]}\right]'(z_0)\right|^2 |a_{m}|^{[N(N+1)-1]/m}\times  \left\{ \hspace{0.08cm} \left[ (\alpha_1-\beta_1)(x-\hat{x}_0)+ (c_1\beta_1-c_2\alpha_1)(t-\hat{t}_0)+n \theta_1+ k \lambda_1\right]\times   \right. \nonumber \\
&& \hspace{0.4cm} \left.
\left[ (\alpha_1^*-\beta_1^*)(x-\hat{x}_0)+ (c_1\beta_1^*-c_2\alpha_1^*)(t-\hat{t}_0)- n \theta_1^* - k \lambda_1^*\right]+ |h_0|^2
\right\} \left[1+O\left(a_{m}^{-1/m}\right)\right]. \hspace{1cm}
\end{eqnarray}
This expression of $\sigma_{n,k}^{(Q)}(x,t)$,
combined with Eqs.~(\ref{Schpolysolu1b})-(\ref{SchpolysolufN2}), gives a fundamental rogue wave
of the three-wave system as given in Theorem~7, and the error of this prediction is $O\left(a_{m}^{-1/m}\right)$.

In the inner region, where $x^2+t^2=O(1)$, the proof of Theorem 7 is identical to that for Theorem 5. The reason is that Manakov rogue waves and three-wave ones in Theorems 3 and 4 have the same solution structures, except for minor differences in the $\textbf{\emph{x}}^{\pm}$ vectors, but the proof of Theorem 5 for the inner region does not rely on the contents of the $\textbf{\emph{x}}^{\pm}$ vectors. Theorem 7 is then proved.

\vspace{0.1cm}
\textbf{Proof of Theorem 8.} In the outer region of R-type rogue waves in the three-wave system, following procedures very similar to that in the proof of Theorem 7, we can show that the solution separates into $M_{R}$ fundamental rogue waves, whose positions are given by Eq.~(\ref{x0t03waveR}), with
\begin{eqnarray}
\hat{\Delta}_{R}&=&e^{-{\rm{i}}\varphi_m}\left[
(\alpha_2-\beta_2)\frac{\Im  \left[ \frac{z_{0} e^{{\rm{i}}\varphi_m}}{c_{1} \beta_{1}- c_{2}\alpha_{1}} \right]}{\Im  \left[ \frac{\alpha_{1}-\beta_{1}}{c_{1} \beta_{1}- c_{2}\alpha_{1}} \right]}
+(c_1\beta_2-c_2\alpha_2)\frac{\Im  \left[ \frac{z_{0}  e^{{\rm{i}}\varphi_m}}{\alpha_{1}-\beta_{1}} \right]}{\Im  \left[ \frac{c_{1} \beta_{1}- c_{2}\alpha_{1}}{\alpha_{1}-\beta_{1}} \right]}\right] \times  \nonumber \\
&& \frac{\sum_{j=1}^{N} \det_{1 \leq i \leq N} \left[p^{[m]}_{3i-1-1}(z_0),  \cdots, p^{[m]}_{3i-j-1-2}(z_0), \cdots, p^{[m]}_{3i-N-1}(z_0)\right]}{\left[R_{N}^{[m]}\right]'(z_0)}. \label{DeltaRhatform}
\end{eqnarray}
As in the definition of $\Delta_R$ in Eq.~(\ref{DeltaRform}), the determinant inside the summation of the above $\hat{\Delta}_{R}$ formula is the determinant of $R_{N}^{[m]}(z_0)$, i.e., $\det_{1 \leq i, j \leq N} \left[p^{[m]}_{3i-j-1}(z_0)\right]$, except that indices of its $j$-th column are reduced by two. The error of this fundamental rogue wave approximation is $O(|a_{m}|^{-1/m})$. The proof for the inner region is identical to that in the proof of Theorem 6.

\section{Conclusion}
In this article, we have reported new types of rogue patterns associated with Okamoto polynomial hierarchies in the Manakov and three-wave-interaction systems. These rogue patterns exhibit new shapes such as double-triangles, rhombuses, and squares, and they arise when one of the internal free parameters in the rogue wave solutions gets large. The shapes of these patterns are analytically predicted from root structures of Okamoto-hierarchy polynomials through a mapping, which is linear to the leading order but nonlinear to the next order. Due to the nonlinear part of the mapping, rogue patterns are often deformed, sometimes strongly deformed, from Okamoto root structures. Our analytical predictions of rogue patterns have been compared to true solutions, and excellent agreement has been observed, even when rogue patterns are strongly deformed from Okamoto root structures.

To put these results in perspective, let us recall rogue patterns associated with the Yablonskii-Vorob'ev polynomial hierarchy, which we reported earlier in \cite{NLSRWs2021,Yanguniversal}. In those cases, rogue patterns were linear transformations of Yablonskii-Vorob'ev root structures, even when the next-order correction term was included. As a consequence, rogue patterns in \cite{NLSRWs2021,Yanguniversal} were very recognizable from Yablonskii-Vorob'ev root structures. In the current Okamoto case, the associated rogue patterns are deformed from Okamoto root structures, since the mapping between them is nonlinear when the next-order correction term is accounted for. These shape deformations may make the rogue pattern less recognizable from Okamoto root structures, unless the underlying parameter is very large so that the next-order nonlinear correction in the mapping becomes insignificant.

Although we have only demonstrated these Okamoto-hierarchy-related rogue patterns in the Manakov and three-wave-interaction systems, these patterns will definitely also arise in other integrable systems, as long as such systems admit rogue waves whose $\tau$-function determinants can be expressed through Schur polynomials with index jumps of 3. In the Darboux transformation framework, it means that these rogue waves should come from the underlying scattering matrix having a triple eigenvalue. Many other integrable systems admit such rogue waves, such as the coupled Hirota equations \cite{ChenYong2014} and the two-component long-wave-short-wave resonant interaction system \cite{RaoHe2022}. Thus, rogue patterns we reported in this article will arise in all such systems and are universal as well.

\section*{Acknowledgments}
The work of J.Y. was supported in part by the National Science Foundation (USA) under award number DMS-1910282.

\begin{center}
\textbf{Appendix A}
\end{center}

In this appendix, we prove Theorems 1 and 2 regarding roots of Okamoto polynomial hierarchies
$Q_{N}^{[m]}(z)$ and $R_{N}^{[m]}(z)$. The two proofs are similar. Thus, we will only present the proof for Theorem 1 below.

First, we derive the multiplicity of the zero root in $Q_{N}^{[m]}(z)$. For this purpose, we define the Schur polynomial $S^{[m]}_j(z; a)$ as
\begin{equation} \label{Skza}
\sum_{j=0}^{\infty}S^{[m]}_j(z; a) \epsilon^j
=\exp\left[z\epsilon +a \hspace{0.04cm} \epsilon^{m}\right],
\end{equation}
where $a$ is a constant, and $S^{[m]}_j(z; a)\equiv 0$ when $j<0$. Through these Schur polynomials $S^{[m]}_j(z; a)$, we define the following polynomials
\begin{eqnarray} \label{PNza}
&& \widehat{Q}^{[m]}_{N} (z; a) = c_{N}\left| \begin{array}{cccc}
         S^{[m]}_{2}(z;a) & S^{[m]}_{1}(z; a) & \cdots &  S^{[m]}_{3-N}(z; a) \\
         S^{[m]}_{5}(z;a) & S^{[m]}_{4}(z; a) & \cdots &  S^{[m]}_{6-N}(z; a) \\
        \vdots& \vdots & \vdots & \vdots \\
         S^{[m]}_{3N-1}(z;a) & S^{[m]}_{3N-2}(z; a) & \cdots &  S^{[m]}_{2N}(z; a)
       \end{array}\right|.
\end{eqnarray}
It is easy to see that $S^{[m]}_j(z; a)$ is related to the polynomial $p_{j}^{[m]}(z)$ in Eq.~(\ref{pkmz}) as
\begin{equation}
S^{[m]}_j(z; a)=a^{j/m}p_{j}^{[m]}(\hat{z}),  \quad \hat{z}=a^{-1/m}z.
\end{equation}
Thus, the polynomial $\widehat{Q}^{[m]}_{N} (z; a)$ is related to the Okamoto-hierarchy polynomial $Q_{N}^{[m]}(z)$ in Eq.~(\ref{HeiTypePoly2}) as
\[\label{homogeneity1}
\widehat{Q}^{[m]}_{N} (z; a)=a^{N(N+1)/m} Q_{N}^{[m]}(\hat{z}).
\]
This equation tells us that every term in the polynomial $\widehat{Q}^{[m]}_{N} (z; a)$ is a constant multiple of $z^ia^j$, where $i+mj=N(N+1)$. Thus, to determine the multiplicity of the zero root $\hat{z}=0$ in $Q_{N}^{[m]}(\hat{z})$, which is the exponent $i$ of the lowest power of $\hat{z}$ in $Q_{N}^{[m]}(\hat{z})$, we need to determine the term $z^ia^j$ in $\widehat{Q}^{[m]}_{N} (z; a)$ where the power $j$ of $a$ is the highest. To do so, we first expand $S^{[m]}_j(z; a)$ into powers of $a$ as
\[ \label{Sjzarelation}
S^{[m]}_j(z; a)=\sum_{l=0}^{[j/m]} \frac{a^l}{l! (j-lm)!}z^{j-lm}.
\]
This expansion can be obtained by splitting the right side of Eq. (\ref{Skza}) into the product of two exponentials and expanding both exponentials into Taylor series of $\epsilon$, then collecting terms of power $\epsilon^j$ in that product and equating them to $S^{[m]}_j(z; a)$. Using the above relation, we can express matrix elements in the determinant (\ref{PNza}) for $\widehat{Q}^{[m]}_{N} (z; a)$ through powers of $z$ and $a$. Notice that when $m=3j+1$ $(j\ge 1)$, coefficients of the highest $a$ power terms in the first column of that determinant are proportional to
\[ \label{sequenceQ}
z^{2}, z^{5}, \cdots, z^{3j-1}, z^1, z^4, \cdots, z^{3j-2}, z^0, z^{3}, \cdots, z^{3j},
\]
and repeating, and when $m=3j+2$ $(j\ge 0)$, these coefficients are proportional to
\[ \label{sequenceQ2}
z^{2}, z^{5}, \cdots, z^{3j-1}, z^{0}, z^{3}, \cdots, z^{3j}, z^{1}, z^{4}, \cdots, z^{3j+1},
\]
and repeating. In the second and higher columns of $\widehat{Q}^{[m]}_{N} (z; a)$, elements are the same as those in the first column, except that the power $j$ of every $z^{j}$ in them decreases by one with each higher column, and $z^{j}\equiv 0$ for $j<0$.

To obtain the highest power term of $a$ in the determinant $\widehat{Q}^{[m]}_{N} (z; a)$, we perform row operations to this determinant to remove certain power terms of $a$. Specifically, using the first $m$ rows, we perform row operations to remove the highest powers of $a$ from the second $m$ rows, leaving the second-highest power terms of $a$ with coefficients proportional to $z^{j+m}$, where $z^{j}$ is the highest $a$-power coefficient of each element just being removed. Then, we use the first $m$ rows and the resulting second $m$ rows to eliminate the highest and second-highest power terms of $a$ from the third $m$ rows, leaving the third-highest power terms of $a$ with coefficients proportional to $z^{j+2m}$ in them. This process is continued to all later rows of $\widehat{Q}^{[m]}_{N} (z; a)$.

After these row operations, we then keep only the highest remaining power term of $a$ in each matrix element of $\widehat{Q}^{[m]}_{N} (z; a)$. This reduced determinant will be the $z^ia^j$ term in $\widehat{Q}^{[m]}_{N} (z; a)$ where the power $j$ of $a$ is the highest, and the corresponding power $i$ of $z^i$ in this term will be the multiplicity of the zero root in $Q_{N}^{[m]}(\hat{z})$. Recalling
the sequence structures in Eqs.~(\ref{sequenceQ}) and (\ref{sequenceQ2}), for $m \hspace{0.1cm} \mbox{mod} \hspace{0.1cm} 3 =1$ and $2$ respectively, we can readily calculate this reduced determinant and find its $z^i$ term, where $i$ is equal to the quantity $N_Q$ given in Eq.~(\ref{defNQ}) of Theorem 1.

One may notice the close resemblance between the above derivation for the zero root's multiplicity in the Okamoto hierarchy polynomial $Q_{N}^{[m]}(z)$, and the proof of Theorem 5 for the reduced Q-type rogue wave in the inner region. Indeed, these two seemingly very different topics are actually closely related.

Now, we prove the factorization formula (\ref{QNmform}) in Theorem 1. The definition (\ref{pkmz}) of the polynomial $p_{j}^{[m]}(z)$ implies the symmetry
\[
p_{j}^{[m]}(\omega z)=\omega^{j}p_{j}^{[m]}(z),
\]
where $\omega$ is any one of the $m$-th root of 1, i.e., $\omega^m=1$. This symmetry of $p_{j}^{[m]}(z)$ leads to the symmetry of $Q_{N}^{[m]}(z)$ as
\[ \label{QNmzsym}
Q_{N}^{[m]}(\omega z)=\omega^{N(N+1)}Q_{N}^{[m]}(z).
\]
Since we have just established that the multiplicity of the zero root in $Q_{N}^{[m]}(z)$ is $N_Q$, we can write
\[
Q_{N}^{[m]}(z)=z^{N_Q}q_{N}^{[m]}(z),
\]
where $q_{N}^{[m]}(z)$ is a polynomial of $z$ with a nonzero constant term. The symmetry (\ref{QNmzsym}) of the polynomial $Q_{N}^{[m]}(z)$ induces a symmetry for $q_{N}^{[m]}(z)$ as
\[
q_{N}^{[m]}(\omega z)=\omega^{N(N+1)-N_Q}q_{N}^{[m]}(z).
\]
It is easy to check that $N(N+1)-N_Q$ is a multiple of $m$. Hence, $\omega^{N(N+1)-N_Q}=1$, and consequently,
\[
q_{N}^{[m]}(\omega z)=q_{N}^{[m]}(z).
\]
This symmetry of $q_{N}^{[m]}(z)$ dictates that $q_{N}^{[m]}(z)$ can only be a polynomial of $\zeta\equiv z^m$. The form (\ref{QNmform}) of the polynomial $Q_{N}^{[m]}(z)$ is then proved.

\begin{center}
\textbf{Appendix B}
\end{center}

In this appendix, we derive Manakov rogue waves presented in Theorem 3. This derivation is an extension of the earlier derivation in \cite{Yanguniversal} for a simpler type of Manakov rogue waves.

Under the transformation
\[\label{BilinearTrans}
u_{1}(x,t)=\rho_{1}\frac{g}{f} e^{{\rm{i}} (k_{1}x + \omega_{1} t)},\  \ \ u_{2}(x,t)=\rho_{2}\frac{h}{f} e^{{\rm{i}} (k_{2}x + \omega_{2} t)},
\]
where $f$ is real and $(g, h)$ complex, the Manakov system (\ref{ManakModel}) can be converted into the following bilinear equations,
\begin{equation}
\begin{array}{ll}
\left(D_{x}^2 +  \epsilon_{1} \rho_{1}^2 + \epsilon_{1} \rho_{1}^2 \right)f \cdot f =
\epsilon_{1} \rho_{1}^2 g g^*+ \epsilon_{2} \rho_{2}^2 h h^*,\\
\left(i D_{t}+ D_{x}^2 +2i k_{1} D_{x}\right)g \cdot f =0,  \\
\left(i D_{t}+ D_{x}^2 +2i k_{2} D_{x}\right)h \cdot f =0.
\end{array}\label{BiManakov}
\end{equation}
This bilinear system can be reduced from the following higher-dimensional bilinear system in the 2-component Kadomtsev-Petviashvili (KP) hierarchy \cite{OhtaYangWang2011},
\begin{equation}
\begin{array}{ll}
(\frac{1}{2}D_xD_r-1)\tau_{n,k}\cdot\tau_{n,k}=-\tau_{n+1,k}\hspace{0.05cm}\tau_{n-1,k},\\
(D_x^2-D_y+2aD_x)\tau_{n+1,k}\cdot\tau_{n,k}=0,\\
(\frac{1}{2}D_xD_s-1)\tau_{n,k}\cdot\tau_{n,k}=-\tau_{n,k+1}\hspace{0.05cm}\tau_{n,k-1},\\
(D_x^2-D_y+2bD_x)\tau_{n,k+1}\cdot\tau_{n,k}=0,
\end{array} \label{HDManakov}
\end{equation}
where $n, k$ are integers, $\tau_{n,k}$ is a function of four independent variables $(x,y,r,s)$,
$a={\rm{i}}k_{1}$, and $b={\rm{i}}k_{2}$.
For rogue waves, the solution $\tau_{n,k}$ to these higher-dimensional bilinear equations is chosen as  \cite{Yanguniversal}
\begin{equation} \label{tauMana}
\tau_{n,k}=\det_{1\le \nu, \mu \le N}\Big(\phi_{i_\nu, j_\mu}^{(n,k)}\Big),
\end{equation}
where $(i_1, i_2, \cdots, i_N)$ and $(j_1, j_2, \cdots, j_N)$ are arbitrary sequences of indices,
the matrix element $\phi_{ij}^{(n,k)}$ is defined as
\[
\phi_{ij}^{(n,k)}=\mathcal{A}_i \mathcal{B}_{j}\phi^{(n,k)}, \label{phiijnk}
\]
\[
\phi^{(n,k)}=\frac{(p+1)(q+1)}{2(p+q)}\left(-\frac{p-a}{q+a}\right)^{n} \left(-\frac{p-b}{q+b}\right)^{k} e^{\xi+\eta},   \label{phinkMana}
\]
\[
\xi=px+p^2y+\frac{1}{p-a}r+\frac{1}{p-b}s+\xi_{0}(p),
\]
\[
\eta=qx-q^2y+\frac{1}{q+a}r+\frac{1}{q+b}s+\eta_{0}(q),
\]
\[
\mathcal{A}_{i}=\frac{1}{ i !}\left[f_{1}(p)\partial_{p}\right]^{i}, \quad
\mathcal{B}_{j}=\frac{1}{ j !}\left[f_{2}(q)\partial_{q}\right]^{j},  \label{AiBj}
\]
$p, q$ are arbitrary complex constants, and $\xi_{0}(p), \eta_{0}(q), f_1(p), f_2(q)$ are arbitrary functions of $p$ and $q$, respectively.

To reduce the higher-dimensional bilinear system (\ref{HDManakov}) to the original system (\ref{BiManakov}), we need to set
\[
f=\tau_{0,0}, \quad g=\tau_{1,0}, \quad h=\tau_{0,1}, \quad y={{\rm i}}t,
\]
impose the dimension reduction condition
\[ \label{DimMana}
\left(2\partial_x +\epsilon_1\rho_1^2\partial_r+\epsilon_2\rho_2^2\partial_s\right)\tau_{n,k}=C\hspace{0.05cm} \tau_{n,k},
\]
where $C$ is some constant, and impose the conjugation condition
\[ \label{ConjugateMana}
\tau_{-n,-k}=\tau^*_{n,k}.
\]
These two reductions proceed the same way as in Ref.~\cite{YangYang3waves} for rogue waves in the three-wave resonant interaction system.

First, we consider the dimension reduction condition (\ref{DimMana}). Here,
\[
\left(2\partial_x +\epsilon_1\rho_1^2\partial_r+\epsilon_2\rho_2^2\partial_s\right)\phi_{ij}^{(n,k)}=
\mathcal{A}_i \mathcal{B}_{j} \left[\mathcal{F}_{1}(p)+\mathcal{F}_{2}(q)\right]\phi^{(n,k)},
\]
where
\[\label{Q1polynomialb}
\mathcal{F}_{1}(p)= \frac{\epsilon_{1}\rho_{1}^2}{p-a} + \frac{\epsilon_{2}\rho_{2}^2}{p-b}+ 2p,
\]
and $\mathcal{F}_{2}(q)$ is the above $\mathcal{F}_{1}(p)$ function with $p$ switching to $q$ and $(a,b)$ switching to $(-a, -b)$. Suppose the algebraic equation
$\mathcal{F}'_{1}(p)= 0$ admits a non-imaginary double root $p_0$, which happens under conditions
(\ref{CubicCondition}), and the corresponding root $p_0$ is given in Eq.~(\ref{p0Mana}). In this case, the dimension reduction condition (\ref{DimMana}) would be satisfied if we choose $f_1(p)$ to satisfy the functional condition
\[ \label{3rdordlinode}
\left( f_{1}(p) \partial_{p} \right)^{3} \mathcal{F}_{1}(p) = \mathcal{F}_{1}(p),
\]
choose $f_2(q)$ to satisfy a similar condition except to change the index above from 1 to 2 and change $p$ to $q$, and choose the $\tau_{n,k}$ determinant (\ref{tauMana}) as
\[ \label{cubicrwstype3p}
\tau_{n,k}=\det \left(
\begin{array}{cc}
  \tau^{[1,1]}_{n,k} & \tau^{[1,2]}_{n,k} \\
  \tau^{[2,1]}_{n,k} & \tau^{[2,2]}_{n,k}
\end{array} \right),
\]
where
\[ \label{Blockmatrixp}
\tau^{[I, J]}_{n,k}=\mbox{mat}_{1\leq i \le N_{I}, 1\leq j\leq N_{J}} \left(\left.
\phi_{3i-I, \, 3j-J}^{(n,k)}\right|_{p=p_{0}, \hspace{0.06cm} q=q_{0}, \hspace{0.06cm} \xi_0=\xi_{0I}, \hspace{0.06cm} \eta_0=\eta_{0J}}\right), \quad 1\leq I,J \leq 2,
\]
$\phi_{i,j}^{(n,k)}$ is given by Eqs. (\ref{phiijnk})-(\ref{AiBj}), $q_0=p_0^*$, and $N_1, N_2$ are non-negative integers. The reason for this can be found in \cite{YangYang3waves}.

Regarding the conjugation condition (\ref{ConjugateMana}), it can be satisfied when we require
$\eta_{0,I}=\xi_{0,I}^*$.

To introduce free parameters into these solutions, we set
\[ \label{defxi0I2}
\xi_{0,I}=\sum _{r=1}^\infty a_{r,I}  \ln^r \mathcal{W}_{1}(p), \quad I=1, 2,
\]
where $\mathcal{W}_{1}(p)$ is defined through $f_1(p)=\mathcal{W}_{1}(p)/\mathcal{W}'_{1}(p)$, and $a_{r,I}$ are free complex constants.

Next, we remove the differential operators in the matrix elements (\ref{phiijnk}) and derive more explicit expressions of rogue waves through Schur polynomials. This derivation is very similar to that we did in Ref.~\cite{YangYang3waves} for the three-wave system. In fact, this derivation is a bit simpler now due to our introduction of the extra factor $(p+1)(q+1)/2$ in Eq.~(\ref{phinkMana}). Combining these steps, we obtain the Manakov rogue wave expressions given in Theorem 3, except that
the definitions for $x_{r,I}^{+}(n,k)$ and $x_{r,J}^{-}(n,k)$ are as given in Eqs.~(\ref{defxrp})-(\ref{defxrm}) for all $r$ indices, including those where $r \hspace{0.08cm} \mbox{mod} \hspace{0.05cm} 3 = 0$.

But those $x_{r,I}^{+}(n,k)$ and $x_{r,J}^{-}(n,k)$ with $r \hspace{0.08cm} \mbox{mod} \hspace{0.05cm} 3 = 0$ can be removed from the solution. This can be done by using a technique similar to that employed in Appendix A of Ref.~\cite{NLSRWs2021}. After this simplification, Theorem 3 is then proved.

\end{document}